\documentclass[12pt,preprint]{aastex}
\usepackage{natbib,amsmath}

\begin{document}

\def\zmedian{2.44} % Median redshift of the stack
\def\blls{-0.6^{+0.2}_{-0.3}} % slope in the LLS regime
\def\zbest{1.86} % Best estimate of z912
\def\cmfpval{150 - 300} % Crude estimate of the MFP (UPDATE)
\def\bmfpval{252 \pm 52} % Bootstrap estimate of the MFP (UPDATE)
\def\bmmfpval{243 (252)} % Bootstrap median (mean) estimate of the MFP (UPDATE)
\def\zllval{1.8 -- 2.0} % UPDATE
\def\loxval{0.29 \pm 0.05}  % Weighted mean of ACS/WFC3 + Ribaudao at
                            % z=2-2.5
\def\dlalox{0.05}  % UPDATE
\def\acs{ACS/PR200L}
\def\wfc{WFC3/UVIS-G280}
\def\hwfc{WFC3/UVIS-G280}
\def\hacs{\ihst/ACS}
\def\nwfc{53}
\def\mhmpc{h^{-1}_{72} \, \rm Mpc}
\def\mkppo{\tilde \kappa_{912}(z_{\rm stack})}
\def\mgtL{\gamma_{\tau}}
\def\gtL{$\gamma_{\tau}$}
\def\ftelf{$f_\lambda^{\rm Telfer}$}
\def\mftelf{f_\lambda^{\rm Telfer}}
\def\ihst{{\it HST}}
\def\beama{BEAM~A}
\def\beamb{BEAM~B}
\def\beamc{BEAM~C}
\def\wfcteam{WFC3~team}
\def\axeteam{aXe~team}
\def\drp{WFC3+G280-DRP}
\def\hub{h_{72}^{-1}}
\def\dat{$\delta \alpha_{\rm T}$}
\def\mdat{{\delta \alpha_{\rm T}}}
\def\umfp{{\hub \, \rm Mpc}}
\def\mzstack{z_{\rm stack}}
\def\zstack{$z_{\rm stack}$}
\def\zabs{$z_{\rm abs}$}
\def\mzabs{z_{\rm abs}}
\def\mnull{\nu_{\rm 912}}
\def\nnull{$\nu_{\rm 912}$}
\def\intl{\int\limits}
\def\expval{XXX}
\def\nstatqso{193}
\def\maxoff{0.4}
\def\clls{1.9 \pm 0.2}
\def\alls{5.2 \pm 1.5}
\def\blls{-0.9^{+0.4}_{-0.05}}
\def\cmma{\;\;\; ,}
\def\perd{\;\;\; .}
\def\ltk{\left [ \,}
\def\ltp{\left ( \,}
\def\ltb{\left \{ \,}
\def\rtk{\, \right  ] }
\def\rtp{\, \right  ) }
\def\rtb{\, \right \} }
\def\sci#1{{\; \times \; 10^{#1}}}
\def \rAA {\rm \AA}
\def \zlya {$z_{\rm Ly\alpha}$}
\def \mzlya {z_{\rm Ly\alpha}}
\def \zlyb {$z_{\rm Ly\beta}$}
\def \mzlyb {z_{\rm Ly\beta}}
\def \zem {$z_{\rm em}$}
\def \mzem {z_{\rm em}}
\def \mzlls {z_{\rm LLS}}
\def \zlls {$z_{\rm LLS}$}
\def \mzll {z_{\rm 912}}
\def \zll {$z_{\rm 912}$}
\def \ozll {$z_{\rm 912}^{\tau=1}$}
\def \mzend {z_{\rm end}}
\def \zend {$z_{\rm end}$}
\def \zstart {$z_{\rm start}$}
\def \mzstart {z_{\rm start}}
\def\smm{\sum\limits}
\def \mlrest  {\lambda_{\rm r}}
\def \lrest  {$\lambda_{\rm r}$}
\def\mkconst{\kappa_{\mzq}}
\def\mztkll{{\tilde\kappa}_{912}}
\def\ztkll{${\tilde\kappa}_{912}$}
\def\kll{$\kappa_{\rm LL}$}
\def\mkll{\kappa_{\rm LL}}
\def \lll  {$\lambda_{\rm 912}$}
\def \mlll  {\lambda_{\rm 912}}
\def \mtll  {\tau^{\rm LL}_{\rm 912}}
\def \tll  {$\tau^{\rm LL}_{\rm 912}$}
\def \mavtll  {<\mtll>}
\def \tigm  {$\tau_{\rm IGM}$}
\def \mtigm  {\tau_{\rm IGM}}
\def \tlim  {$\tau_{\rm limit}$}
\def \mtlim  {\tau_{\rm limit}}
\def \cmm  {cm$^{-2}$}
\def \cmmm {cm$^{-3}$}
\def \kms  {km~s$^{-1}$}
\def \mkms  {{\rm km~s^{-1}}}
\def \lyaf {Ly$\alpha$ forest}
\def \Lya  {Ly$\alpha$}
\def \lya  {Ly$\alpha$}
\def \mlya  {Ly\alpha}
\def \Lyb  {Ly$\beta$}
\def \lyb  {Ly$\beta$}
\def \lyg  {Ly$\gamma$}
\def \lyd  {Ly$\delta$}
\def \ly5  {Ly-5}
\def \ly6  {Ly-6}
\def \ly7  {Ly-7}
\def \nhi  {$N_{\rm HI}$}
\def \mnhi  {N_{\rm HI}}
\def \lnhi {$\log N_{\rm HI}$}
\def \mlnhi {\log N_{\rm HI}}
\def \etal {\textit{et al.}}
\def \ob {$\Omega_b$}
\def \obh {$\Omega_bh^{-2}$}
\def \om {$\Omega_m$}
\def \ol {$\Omega_{\Lambda}$}
\def \gz {$g(z)$}
\def \mgz {g(z)}
\def \lyaf {Lyman--$\alpha$ forest}
\def \fnz {$f(\mnhi,z)$}
\def \fnhi {$f(\mnhi,X)$}
\def \mfnhi {f(\mnhi,X)}
\def \myfnhi {f_{\rm Ly\alpha}(\mnhi,X)}
\def \yfnhi {$f_{\rm Ly\alpha}(\mnhi,X)$}
\def \lfnhi {$f_{\rm LLS}(\mnhi,X)$}
\def \pfnhi {$f_{\rm pLLS}(\mnhi,X)$}
\def \dfnhi {$f_{\rm DLA}(\mnhi,X)$}
\def \sfnhi {$f_{\rm SLLS}(\mnhi,X)$}
\def \mfnlls {f(\mnhi=10^{17.5}\cm{-2}, X)} 
\def \mlfnhi {f_{\rm LLS}(\mnhi,X)}
\def \mdfnhi {f_{\rm DLA}(\mnhi,X)}
\def \msfnhi {f_{\rm SLLS}(\mnhi,X)}
\def \ztot {$\Delta z_{\rm TOT}$}
\def \mztot {\Delta z_{\rm TOT}}
\def \mlplls {\ell_{\rm{PLLS}}(z)}
\def \lplls {$\ell_{\rm{PLLS}}(z)$}
\def \mlzlls {\ell_{\rm{LLS}}(z)}
\def \lzlls {$\ell_{\rm{LLS}}(z)$}
\def \mllls {\ell_{\rm{LLS}}(X)}
\def \llls {$\ell_{\rm{LLS}}(X)$}
\def \ldla {$\ell_{\rm{DLA}}(X)$}
\def \lslls{$\ell_{\rm{SLLS}}(X)$}
\def \mlslls{\ell_{\rm{SLLS}}(X)}
\def \mlox{\ell(X)}
\def \lox{$\ell(X)$}
\def \loz{$\ell(z)$}
\def \mloz{\ell(z)}
\def \tlox{$\ell(X)_{\tau \ge 2}$}
\def \flox{$\ell(X)_{\tau \ge 0.5}$}
\def \mtlox{\ell(X)_{\tau \ge 2}}
\def \nlls {$\mathcal{N}_{\rm LLS}$}
\def \mnlls {\mathcal{N}_{\rm LLS}}
\def \slls {$\sigma_{\rm LLS}$}
\def \mslls {\sigma_{\rm LLS}}
\def \drlls {$\Delta r_{\rm LLS}$}
\def \mdrlls {\Delta r_{\rm LLS}}
\def \mlmfp {\lambda_{\rm mfp}^{912}}
\def \lmfp {$\lambda_{\rm mfp}^{912}$}
\def \dbeta {$\Delta \beta_{\rm LLS}$}
\def \mdbeta {\Delta \beta_{\rm LLS}}
\def \mbplls {\beta_{\rm pLLS}}
\def \bplls {$\beta_{\rm pLLS}$}
\def \btlls {$\beta_{\rm LLS}$}
\def \mbtlls {\beta_{\rm LLS}}
\def \mbtlya {\beta_{\rm Ly\alpha}}
\def \teff {$\tau_{\rm eff}^{\rm LL}$}
\def \mteff {\tau_{\rm eff}^{\rm LL}}
\def \mtmtl {\tau_{\rm eff}^{\rm metals}}
\def \tmtl {$\tau_{\rm eff}^{\rm metals}$}
\def \mtlyman {\tau_{\rm eff}^{\rm Lyman}}
\def \tlyman {$\tau_{\rm eff}^{\rm Lyman}$}
\def \tlya {$\tau_{\rm eff}^{\rm Ly\alpha}$}
\def \mtlya {\tau_{\rm eff}^{\rm Ly\alpha}}
\def \mnmin {\mnhi^{\rm min}}
\def \nmin {$\mnhi^{\rm min}$}
\def \O {${\mathcal O}(N,X)$}
\newcommand{\cm}[1]{\, {\rm cm^{#1}}}
\newcommand{\fobs}[1]{\bar f^{\rm obs}_{#1}}
\newcommand{\fint}[1]{\bar f^{\rm SED}_{#1}}
\def \snrlim {SNR$_{lim}$}
\def\mglls {\gamma_{\rm LLS}}
\def\mavgt {<\mtll>}
\def\dldN {$\Delta \mlox / \Delta \log \mnhi$}  

\title{The HST/ACS+WFC3 Survey for Lyman Limit Systems II: Science}

\author{
John M. O'Meara\altaffilmark{1},
J. Xavier Prochaska\altaffilmark{2},
Gabor Worseck\altaffilmark{2},
Hsiao-Wen Chen\altaffilmark{3}, \&
Piero Madau\altaffilmark{2}
}
\altaffiltext{1}{Department of Chemistry and Physics, Saint Michael's College.
One Winooski Park, Colchester, VT 05439}
\altaffiltext{2}{Department of Astronomy and Astrophysics, UCO/Lick
  Observatory, University of California, 1156 High Street, Santa Cruz,
  CA 95064}
\altaffiltext{3}{Department of Astronomy and Astrophysics, 
University of Chicago, 640 S. Ellis Ave, Chicago, IL 60637}

\begin{abstract}
We present the first science results from our Hubble Space Telescope
Survey for Lyman limit absorption systems (LLS) using the low
dispersion spectroscopic modes of the Advanced Camera for Surveys and the
Wide Field Camera 3.  Through an analysis of 71 quasars, we determine the 
incidence frequency of LLS per unit redshift and per unit path length,
\loz\ and \lox\ respectively, over the redshift range $1 < z< 2.6$, 
and find a weighted mean of \lox $=0.29 \pm 0.05$ for $2.0 < z < 2.5$
through a joint analysis of our sample and that of Ribaudo \etal\
(2011).  Through stacked spectrum analysis, we determine a median
(mean) value of the mean free path to ionizing radiation
at $z=2.4$ of $\mlmfp = \bmmfpval \mhmpc$, with an error on the mean value of
$\pm 43 \mhmpc$. We also
re-evaluate the estimates of \lmfp\ from \cite{pwo09} and place
constraints on the evolution of \lmfp\ with redshift, including 
an estimate of the ``breakthrough'' redshift of $z = 1.6$
 
Consistent
with results at higher $z$, we find that a significant fraction
of the opacity for absorption of ionizing photons comes from systems
with \nhi $ \le 10^{17.5}$\cmm\ with a value for the total Lyman opacity of
$\mtlyman = 0.40 \pm 0.15$.  Finally, we determine that at
minimum, a 5-parameter (4 power-law) model is needed to describe the
column density distribution function \fnhi\ at $z\sim 2.4$,  find
that \fnhi\ undergoes no significant change in shape between $z \sim
2.4$ and $z \sim 3.7$, and provide our best fit model for \fnhi.
\end{abstract}

\section{Introduction}
The importance of Lyman limit systems (LLS), those quasar
absorption line systems with \nhi\ $>10^{17.2}$ \cmm\ has been well
understood. Studies of LLS at redshifts $z > 2.6$ stretch back many
decades, with the LLS being some of the first absorption line systems
studied quantitatively \cite{tytler82}.  Through to the present day,
ground based surveys of the LLS span the full range of observability,
$2.6 > z > 6$ (e.g. \cite{ssb}, \cite{storrie94}, \cite{peroux01}, 
\cite{songaila10}, \cite{pow10}).  The 1990s also introduced studies
of the LLS from space (e.g. \cite{key_lls}), with the largest datasets
being provided by the the Faint Object Spectrograph and Space
Telescope Imaging Spectrograph onboard the Hubble Space Telescope 
spanning the range $0.3 < z < 2.6$ (see
\cite{ribaudo11} for a summary).  The space-based studies peak in
sensitivity at $z\sim 1$, declining to both higher and lower redshifts.
Nearly all early studies of the LLS provided a general picture of a
rapidly evolving population with an incidence frequency described by a
$(1+z)^\gamma$ power law, and $\gamma \simeq 1.5$.  Early studies also
provided constraints on the column density distribution function, 
describing it as a single power law over the full range in LLS hydrogen
column density (e.g. \cite{tytler87}; \cite{lzt91}; \cite{pwr93}),
while more recent studies (e.g. \cite{peroux03}; \cite{opb+07};
\cite{pow10}) have considered more complicated descriptions.

The optical depths from LLS at 1 Rydberg, 
combined with their increased frequency
per unit redshift compared to the higher \nhi\ damped Lyman alpha
systems (DLAs;  \nhi\ $\ge 10^{20.3}$\cmm), suggest that they strongly
influence, if not dominate the
attenuation of photons emitted by galaxies and quasars, and
thus set the intensity of the extragalactic UV background, and
determine the mean free path of ionizing photons in the intergalactic
medium (IGM; e.g.,\cite{me03}; \cite{pwo09}, hereafter PWO09;
\cite{dww08}; \cite{calverley11};\cite{McQuinn11}; \cite{hm12}). Thus, LLS 
play a crucial role in the cosmological history of interplay between
radiation and baryons.

Both the low density, highly ionized \lyaf\ ( \nhi\ $<10^{17.2}$\cmm\ )
 and the high density, predominantly neutral
DLAs have been well studied and constrained in part because of their relative
 ease of observation.  The \lyaf\ is ubiquitous in quasar spectra, 
 is visible from the ground at $z > 1.6$, and is relatively easy to
 constrain in terms of \nhi\ owing to its lack of saturation in the
 Lyman series. One must go to higher order Lyman series transitions to
 determine the \nhi\ as \nhi\ increases, but this can be achieved with
 high S/N, high-resolution data \citep[e.g.][]{rudie12}.  
The DLA show strong damping wing features in
 their \lya\ line, allowing for easy determination of their \nhi\ even
 in low-resolution data \citep[e.g.][]{wgp05}.  
By contrast, the LLS, although frequently observed, are by comparison
poorly constrained,
 as both the \lya\ line and the Lyman break must be well covered in
 the spectrum to constrain the \nhi\ \citep[e.g.][]{ppo+10}.  
At the low end of the LLS H~I
 column density range, significant wavelength coverage at rest
 wavelengths in the LLS frame of $\lambda < 912$\AA\ are required to
 accurately determine the \nhi\ through the shape of the recovery in
 flux.  These complications place stronger
 constraints on the redshift range where large statistical surveys for
 LLS can be done from the ground, namely $z>2.5$.

Nevertheless, our motivation for studying the LLS is clear.  A
complete census of absorption across the full range of \nhi\ is
required to understand the properties of the IGM at a given redshift,
and the evolution in redshift of that census provides tests of models of
structure formation, evolution in the UV background, and the
interplay between galaxies and their environment.  At $z\sim 3.7$,
\cite{pow10} (hereafter POW10) analyzed spectra from the SDSS DR7
catalogue to establish the incidence frequency of LLS and to
characterize a number of biases inherent in LLS analysis at any
redshift. PWO09 also use the SDSS DR7 catalogue, and produce a
measurement of the mean free path to ionizing radiation, \lmfp, over
the same redshift range in POW10.  Combined with the incidence
frequency measurements and measurements of the \lyaf, LLS, and DLA in the
literature, POW10 determined the column density distribution function
\fnhi\ at $z = 3.7$.  This \fnhi\ underscores the importance of the LLS
on our complete understanding of absorbers in the universe, as it
showed that a number of strong inflections in \fnhi\ exist at this
redshift.  Recent developments in theoretical models and simulations  
of the LLS regime \citep[e.g.][]{altay11}
have made significant improvements in
reproducing these inflections with the inclusion of self shielding.
The path forward is now clear: to move beyond a simple counting of the LLS
 and to extend the knowledge of \fnhi\ and
\lmfp\ to both higher and lower redshifts to provide a determination their
evolution with cosmic time, and to better understand their physical
nature.

At lower redshifts ($z<2.5$), a determination of \fnhi\ and \lmfp\ which
includes the contribution from LLS is made difficult by the simple
fact that one must go to space based telescopes to obtain coverage of
the Lyman break at rest wavelength $\lambda = 912$\AA.  The region $1
< z < 2.5$ is particularly difficult, as most space-based instruments
have poor NUV throughput at high resolution. Previous studies of LLS at $z <
2.5$ offered differing results regarding the evolution of the incidence
frequency, \loz\, of the lls.  \cite{lzt91} argued for a fairly
constant \loz\, with rapid evolution at higher $z$, whereas
\cite{key_lls} and \cite{storrie94} argued for a single power law
evolution over the entire range $0 < z < 4$. 
Recently, \cite{ribaudo11} performed an {\it HST} archival study of
the LLS frequency distribution at $z < 2.5$, with the bulk of their
statistical power at $0.75 < z<1.5$.  They find that the LLS frequency
with redshift \loz\ is well described by a power-law \loz$\propto
(1+z)^\gamma$ with $\gamma = 1.33 \pm 0.61$ at $z < 2.6$.
They place constraints on \fnhi\ and \lmfp\ (finding that the former
is incompatible with a single power-law), but do so
largely by relying on measurements at different redshifts than the
bulk of their sample for the higher column density LLS and the DLA.

In this paper, we present the first set of scientific results from our
campaign to better explore the LLS at redshifts where our
knowledge of optically thick systems is  poorly constrained,
namely $1< z < 2.6$.  Our data sample comes from a
survey performed with the \textit{Hubble Space Telescope} using the
low-resolution NUV spectroscopic modes of the \textit{Advanced Camera
  for Surveys} and the \textit{Wide Field Camera 3} instruments.  The
survey is described in the first paper in this series \cite{omeara11}, 
which we refer to hereafter as Paper~I.  This paper is
organized as follows: Section \ref{sec:quasars} summarizes the
results of Paper~I, Section \ref{sec:spectral} describes how we model
LLS absorption in our sample, Section \ref{sec:stats} describes our
statistical analysis and results for the incidence frequency of the
LLS, Section \ref{sec:stack} describes our analysis of stacked
spectra and the determination of the mean free path of ionizing
photons under specific assumptions regarding the stacked QSO SED and
any intrinsic inflections it might have, Section \ref{sec:fn} presents our analysis of the column density
distribution function, and Section \ref{sec:discuss} offers a summary and
discussion.

Unless stated otherwise, all results in this paper assume a
``standard'' Lambda+CDM cosmology with  $H_0 = 72 \mkms \, \rm
Mpc^{-1}$, $\Omega_m = 0.26$,  and $\Omega_\Lambda = 0.74$.

\section{Summary of Quasar Selection and \ihst\ Datasets}
\label{sec:quasars}

In Paper~I, we reported in detail the quasar sample and  \ihst\
observations for our survey.  
We also described the reduction procedures and presented the
extracted 1D spectra that form the basis of the following analysis.
This section provides a brief summary of Paper~I.

For our \ihst\ snapshot programs in Cycles 15 and 17, we selected 100
quasars at $\mzem \approx  2.5$ from the SDSS Data Release 5 based
primarily on their optical photometry.  Specifically, we restricted
the list to quasars with $g < 18.5$\,mag, $2.30 < \mzem < 2.60$,
and spectra without very strong associated absorption. 
Because the SDSS quasar spectroscopic sample is based primarily 
on optical color-selection \citep{rfn+02}, it is possible it may be
biased relative to a complete quasar sample \citep[e.g.][]{wp11}.
Indeed, PWO09 demonstrated that the cohort of SDSS quasars 
at $z \approx 3.5$ are biased {\it toward} sightlines with strong
Lyman limit absorption in the $u$-band.  For our selection criteria,
the presence of an intervening LLS should have negligible effect on
the quasar colors.  Nevertheless, quasars at $z \sim 2.5$ have a typical
color that is sufficiently close to the stellar locus that the SDSS
team chose selection criteria which favor UV-excess quasars.
\cite{wp11} have demonstrated that this biases the spectroscopic
sample at $z \sim 2.6$ to have bluer spectral energy distributions (SEDs) than a
complete sample (see their Figure~16).  We return to this issue later
in the paper.

From this list (see Table~1 of
Paper~I),  18 quasars were queued in
Cycle~15 to observe with the ACS/PR200L prism and \nwfc\ quasars with the
WFC3/UVIS-G280 grism in Cycle~17.  We developed customized software to
extract, wavelength calibrate, and flux the data to produce a fully
calibrated 1D spectrum for each quasar.  
The \acs\ spectra cover observed wavelengths $\lambda = 1500-5000$\,\AA\
and the \wfc\ spectra span roughly $\lambda = 2000-6000$\,\AA. 
Our spectra have relatively high S/N per pixel down to $\lambda
\approx 1800$\,\AA\ for the ACS dataset and $\lambda \approx
2000$\,\AA\ for the \wfc\ dataset.  This reflects the
relatively high UV fluxes of the quasars and the low
dispersion of the spectrometers: at $\lambda = 2500$\AA\ the spectra
have FWHM~$\approx 60$\AA.
Uncertainties in the wavelength calibration are on the order of
2~pixels, in the form of a rigid shift in pixel space.
Table~\ref{tab:quasars} summarizes the quasars comprising the survey.

%%%%%%%%%%%%%%%%%%%%%
\section{Fitting for Lyman Limit Absorption}
\label{sec:spectral}

The principle goal of our \ihst\ program was to survey $z \sim 2.5$
quasars for Lyman limit absorption at $z \lesssim 2$.  At these
redshifts, the Lyman limit lies below Earth's
atmospheric cutoff and one requires space-borne, UV spectroscopy.  Because the
Lyman limit is a continuum opacity, its analysis does not require high
spectral resolution.  Instead, one prefers well-fluxed spectra with
high S/N, and, ideally, quasars without complex continua.
Our dataset nicely achieves these criteria (Paper~I).

Our approach to the Lyman limit analysis is straightforward.  First,
we must estimate the observed flux of each source in the
absence of Lyman continuum opacity.  This quantity is not the
intrinsic flux of the quasar; the \lya\ forest scatters light at
rest wavelength $\mlrest < 1215$\AA\ beginning with \lya\ opacity and eventually
including the full Lyman series.  Because of the low spectral
resolution of these spectra,  we cannot identify individual absorption
lines and therefore observe the intrinsic quasar flux attenuated by a
relatively smooth Lyman series opacity (see $\S$~\ref{sec:formalism}
for a full description).  Therefore, our approach was to fit a template
spectrum to the flux at $\mlrest \approx 950-1150$\AA.  We have found
that the data are reasonably well described by  
the \cite{telfer02} radio-quiet quasar template spectrum
\ftelf\ shifted to the observer frame and 
allowing for a scaled normalization $C$ and power-law tilt $\alpha$, i.e.,

\begin{equation}
f_\lambda^{\rm conti} (\mlrest < 1200{\rm \AA}) = C \, \mftelf \, \ltp
\frac{\lambda_{\rm obs}}{2500 \rm \AA} \rtp^\alpha
\perd
\label{eqn:qso_conti}
\end{equation}
We emphasize that even though the Telfer spectrum is intended to
represent the mean intrinsic SED of $z \sim 1$ quasars,
$f_\lambda^{\rm conti}$ represents the intrinsic flux of each of our
quasars attenuated by Lyman series opacity.
The values for $C$ and $\alpha$ were determined using a custom,
interactive GUI that allows one to visually compare a model continuum
with each quasar spectrum.  
In general, the data at $\lambda_{\rm
  rest} = 950-1150$\AA\ in the QSO rest-frame constrain the $C$ and
$\alpha$ parameters
with reasonable confidence ($\approx 10-20\%$ uncertainty).
The analysis on each quasar was done independently by JMO and JXP, and
we report our preferred models in Table~\ref{tab:quasars}.
We then extrapolate this model to $\mlrest < 950$\AA.

With $f_{\lambda}^{\rm conti}$ estimated as above, we proceeded to fit any
significant and sudden drop in the flux below $\lambda_{\rm rest} =
912$\,\AA\ as Lyman limit opacity.  Specifically, we model the
observed flux as the continuum flux described above modulated
by one or more ``systems'' yielding appreciable Lyman limit opacity.
Each system is characterized by a redshift $z_{\rm abs}$ and an
optical depth at the Lyman limit $\mtll \approx \mnhi/10^{17.19}
\cm{-2}$. In the following, we refer to systems with $\mtll \ge 2$ as
Lyman limit systems (LLSs) and systems with $\mtll < 2$ as partial
Lyman limit systems (pLLSs\footnote{We note the difference between
  pLLS as defined 
  here, and PLLS, which is a ``proximate'' LLS, i.e. one within 3000
  \kms\ of the QSO redshift.}), 
although we occasionally use LLS to refer
to all systems exhibiting detectable Lyman limit opacity.
The resulting model is

\begin{equation}
f^{\rm model}_\lambda = f^{\rm conti}_\lambda \, \exp \ltk - \smm_i^N
\tau_\lambda^{{\rm LL},i} \rtk
\label{eqn:model}
\end{equation}
where the sum is over all absorbers identified along the sightline
($z_{\rm abs} \le z_{\rm em}$) and $\tau_\lambda$ is the opacity in the rest
frame of the absorber \citep{verner96}.  

In a minority of cases ($\sim 30\%$), the
observed spectra are well described by one or zero Lyman limit
systems. For the remainder of cases, however, the data
are best, and very well, 
modeled with multiple LLS with \tll\ varying from 0.2 to 2.
As with the continuum estimation, the modeling was performed by hand using a
custom GUI that accounts for the spectral resolution of the \ihst\
data.  
We also experimented with $\chi^2$-minimization algorithms, but these
gave unrealistically small statistical errors ($\approx 1-2\%$) owing
to the high $S/N$ of the spectra (and high reduced $\chi^2$ values
because the models did not include \lya\ forest absorption).  
Indeed, the results are dominated by the systematic error of
continuum placement and, to a lesser extent, line-blending by the IGM.
For these reasons, we have proceed with a by-hand approach and
estimate the uncertainties through inter-comparisons amongst multiple
authors, and further with the results from a Monte Carlo analysis.

Each spectrum was modeled independently by JMO and JXP and then
the authors compared these results and agreed, along with a third
author, GW, on a solution for cases in dispute.
For the systems with $\mtll > 2$ listed in
Table~\ref{tab:wfc3_survey}, the two authors had agreed in 35/39 cases
there was a
$\mtll >2$ system within $\delta z = 0.05$ of the redshift listed in
the table.  
Of the 4 disparate cases, most had $\mtll \approx 2$.
In these cases, the redshifts have an RMS of $\approx
0.01$.  For the systems in the table with $0.5 < \mtll < 2$, both
authors agreed on a system within $\delta z = 0.05$ in 20/27 cases.
The RMS in the optical depths between the two sets of models is
$\approx 0.05$ and the RMS in redshifts is $\approx 0.02$.

To gauge our ability to successfully identify LLS absorption, we
performed an analysis on 100 mock quasar sightlines.  The
sightlines were generated by drawing randomly from Monte Carlo
absorption line lists, each created using the best-fit column density
distribution function described in \ref{sec:fn_fit}.  Absorption lines
were generated from these lists, convolved to the WFC3 resolution, 
and were placed on absorption-free QSO
spectra with varying power law tilts and signal to noise ratios chosen
to best mimic the data variety of our sample.  
The spectra were then surveyed by one author (JMO) for LLS absorption using the same
interface as with the real data.  The results of this analysis are as
follows:
When we impose the survey definition criteria given in
\ref{sec:survey_def}, the mock LLS spectra contain a total of 91
$\mtll > 0.5$ absorbers in the survey path, with 72 systems having
$\mtll > 2$, 11 having $1 < \mtll \le 2$, and 8 with $0.5 < \mtll \le 1.0$.
JMO identified 76 systems having
$\mtll > 2$, 9 having $1 < \mtll \le 2$, and 10 with $0.5 < \mtll \le 1.0$.
The primary nature of the discrepancy between the input mock LLS
sample and those recovered dealt with systems very near to the optical
depth boundary of each bin.  For example, 
three of the four $\mtll > 2$ discrepant
cases, the input mock spectra had $17.4 < $\lnhi $\le 17.5$, but were
given a column density of either \lnhi$ = 17.5$ or \lnhi$ = 17.55$
Likewise, one of the two  $1 < \mtll \le 2$ discrepancies and two of
the three $0.5 < \mtll \le 1.0$ cases stemmed from the mock LLS lying
within  \lnhi $= 0.05$ of the boundary between bins in $\mtll$ .
In one case, a single $\mtll > 2$ system was used to describe two mock
LLS with $\mtll > 1$ which lied within $\delta z =0.1$ of each other.
In only a single case was a completely incorrect system used to model
the input mock spectrum, where a spurious $\mtll \sim 1$ system was
included.  Although not included in any of our statistical analyses
below, $>50\%$ of all $0.2 < \mtll < 0.5$ absorption was correctly
identified, confirming our ability to be sensitive to such absorption.

\begin{figure}
\includegraphics[width=5in]{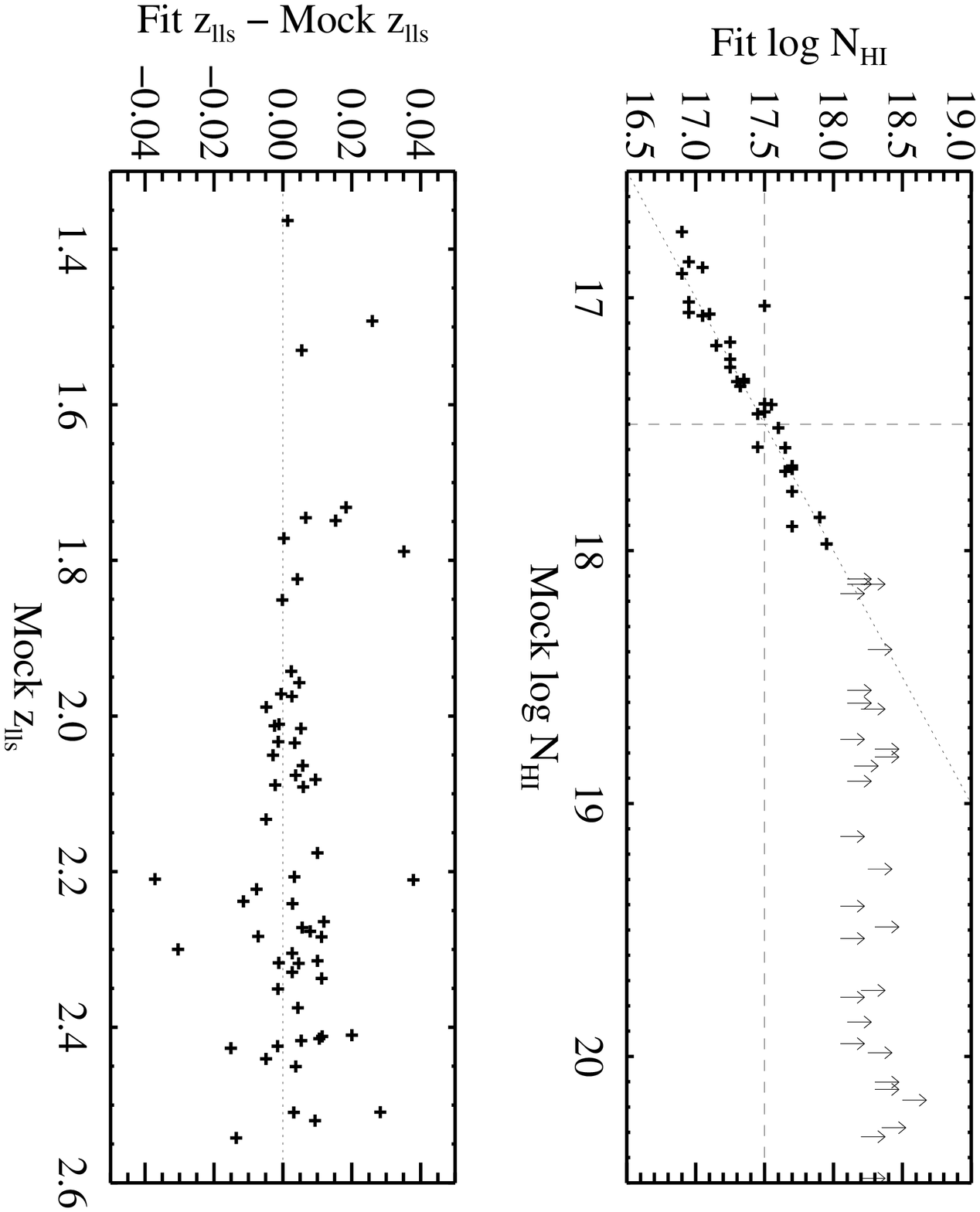}
\caption{Comparisons between fit and input values for \lnhi\ (upper panel) and
  \zlls\ (lower panel) for a mock sample of spectra designed to mimic
  the WFC3 data.  The horizontal and vertical dashed lines in the
  upper panel correspond to $\mtll=2$.  For most $\mtll > 2 $ LLS, we
  can only obtain a lower limit on \lnhi\ and denote those systems
  with arrows.}
\label{fig:mock_compare}
\end{figure}

Figure \ref{fig:mock_compare} gives a further exploration of the range
of uncertainties in our data.  In the upper panel, we show the
estimates of \lnhi\  in LLS within the range \lnhi$ >16.9$
derived from fits to the mock spectra compared to their input values.
We choose this range to fully explore the range of values for which we
can make more precise statements about \nhi, namely the statistical survey range $0.5 <
\mtll < 2.0$ with additional measurements at higher optical
depths when the data allows for an \nhi\ do be determined in
 the cases where we can observe the recovery in flux below the Lyman limit
(although we note that all $\mtll \ge 2$ absoprtion
is lumped into a single bin for our statistical analysis).  We find
overall good agreement between the fit values and the input mock values for
\nhi\ over this range with a mean difference between fit and input
of $-0.02$, and $\sigma = 0.12$.  In the lower panel of Figure
\ref{fig:mock_compare}, we explore our ability to properly determine
the redshift of the LLS for the $\mtll > 0.5$ LLS.  We find a mean difference between fit
and input $z_{\rm{lls}}$ of $-0.002$ and $\sigma=0.014$, again a good agreement.

\begin{figure}
\includegraphics[width=5in]{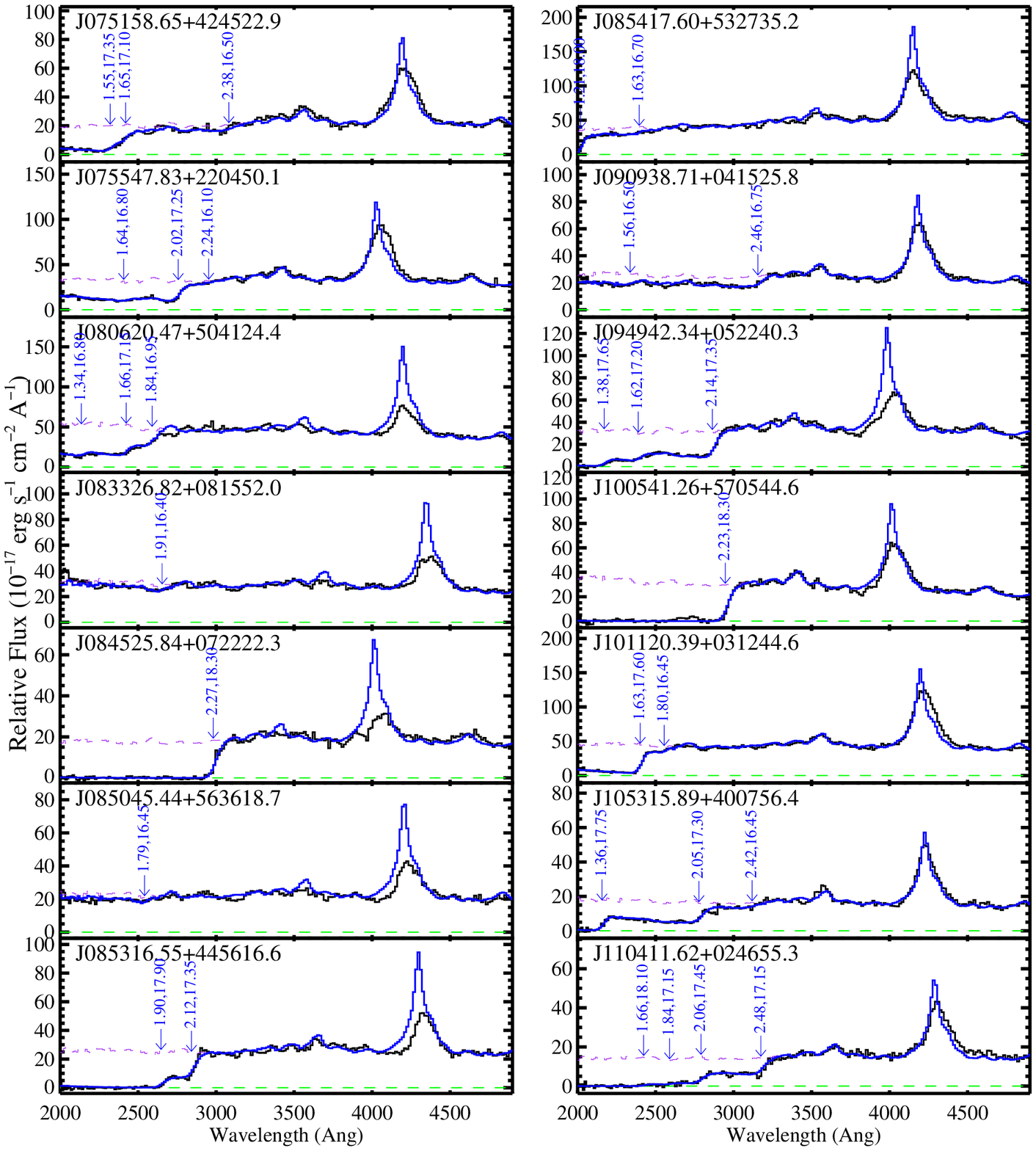}
\caption{\wfc\ quasar spectra (black histogram) and the models (blue
  solid line) we have generated to
  assess Lyman limit absorption along the sightlines.  The arrows mark
  the redshift and \nhi\ values for each modelled LLS.  The dashed
  purple line traces our estimate of the continuum without LL
  absorption but includes Lyman series opacity.
}
\label{fig:wfc3_models}
\end{figure}

These comparisons provide an estimate of the systematic uncertainty
related to the LLS modeling.
In addition, we estimate the systematic uncertainties of continuum
uncertainty and system blending to be 
1\%\ ($\approx 1000 \mkms$) for the redshift
estimation 
and $\approx 15\%$ for the optical depth measurements.\footnote{All 
models assume $b=25 \mkms$;  this has no bearing on our results given
the low spectral resolution.}
Figures~\ref{fig:wfc3_models} and \ref{fig:acs_models} present
the spectra and adopted models for \wfc\ and \acs\ respectively, 
and Table~\ref{tab:models} tabulates the results.
%%%%%%%%%%%%%%%%%%%%%
\section{The Statistics of LLS at $z \approx 2$}
\label{sec:stats}

\subsection{Survey Definition}
\label{sec:survey_def}

From the spectral analysis of the previous section, it is relatively
straightforward to perform a statistical analysis of Lyman limit
absorption.  A fundamental description of IGM absorption systems is the
\nhi\ frequency distribution, \fnhi.  Akin to a luminosity function,
\fnhi\ gives the number of absorbers with column density \nhi\ per
$d\mnhi$ interval and per path length $dX$, where 

\begin{equation}
dX = \frac{H_0}{H(z)}(1+z)^2 dz  \perd
\label{eqn:dX}
\end{equation}
The absorption path length was introduced to yield a constant \fnhi\
for systems that arise from discrete absorbers (e.g.\ galaxies)
with a constant product of comoving number density and physical size
\citep{bp69}. 
Standard practice is to survey quasars for absorbers as a function of
\ion{H}{1} column density and estimate \fnhi\ from simple counting
statistics.

Owing to uncertainties in the
continuum placement and limitations in S/N limit, analysis of
our quasar spectra has a limiting sensitivity to
Lyman limit absorption of $\mtll \gtrsim 0.2$ or $\mnhi \gtrsim
10^{16.5} \cm{-2}$. However, 
as explained below, we only include those systems with $\mtll > 0.5$
in our statistical exploration of incidence frequency and column
density distribution.  Furthermore, it is impossible to yield any
estimate of \tll\ beyond a lower limit
for systems with $\mtll > 4$ because all
such systems have zero measurable flux below the Lyman
limit.  Although we can use the recovery in flux below the Lyman
limit to obtain \nhi\ for LLS with $2 < \mtll < 4$, such cases require
significant spectral coverage which is not always available in our
data.  As a result, our statstical sample will consder all systems with $\mtll >
2$ together as a single population.
\footnote{One can improve the estimate for \nhi\ from the Lyman
  series, e.g.\ the damping wings of \lya, but our spectral resolution
  precludes such analysis.}  
In short, the analysis has sensitivity to \tll\ over approximately 
one order of magnitude and many systems only yield a lower limit to
\tll.   Therefore, one may estimate \fnhi\ directly in only a narrow
range of \ion{H}{1} column densities. 
For these reasons, studies of Lyman limit absorption have provided
rather limited direct constraints on \fnhi.

Instead, researchers have focused primarily on the zeroth moment of
\fnhi, i.e.\ the incidence of Lyman limit absorption per unit
path length 

\begin{equation}
\mlox_{\tau \ge \mtlim} = \intl_{N_{\rm HI, limit}}^{\infty} \mfnhi \, d\mnhi
\label{eqn:lox}
\end{equation}
or per unit redshift\footnote{Also commonly referred
  to as $d\mathcal{N}/dX$ and $d\mathcal{N}/dz$.} 
\loz. 
The incidence of Lyman limit systems is measured from a limiting
optical depth \tlim\ corresponding to a limiting \ion{H}{1} column
density $N_{\rm HI, limit}$,
generally set by the spectral quality.

In principle, the \loz\ quantity is a direct observable.
Standard practice is to estimate it from the ratio of the number of
systems detected within a redshift interval \nlls\ to the total
redshift path surveyed $\Delta z$:

\begin{equation}
\mloz = \frac{\mnlls}{\Delta z} \perd
\label{eqn:loz}
\end{equation}
Observationally, there are several factors that complicate such an LLS
survey (see e.g. Appendix C in \cite{tytler87}). 
First,  it is very challenging to successfully identify
multiple LLSs that have small separations in redshift space.
  One is limited by the spectral
resolution, the S/N of the data,  the precise characteristics of
the systems, and the density of the \lya\ forest at the wavelengths of
interest.  For our data, we cannot resolve two LLS located within
$|\delta v| < 10,000 \, \mkms$ ($\delta z \approx 0.1$) 
and therefore consider all such complexes as a single LLS with a
summed optical depth.   
Second, the presence of a partial LLS reduces the quasar flux and the
resultant spectral S/N but may not prohibit the search for an
additional LLS.  As described in detail in POW10, one must
establish a strict and proper criterion for the survey path or the
results will suffer from a ``pLLS bias''.  

To this end, we establish these criteria for the survey:

\begin{enumerate}
\item The ending redshift \zend\ is set to be $3000 \mkms$ blueward of
  the quasar emission redshift.  This is established in part to
  account for the relatively large uncertainty in quasar redshifts and
  also to minimize any biases from gas in the environment including
  the quasar. 
\item If there are no LLS detected, the starting redshift 
  $\mzstart = \rm (2200 \, \AA / 911.7641 \, \AA) - 1 = 1.4$ for the
  WFC3 data and $\mzstart = 1.2$ for the ACS data.  This
  provides enough spectral coverage ($\approx 200$\,\AA) to
  confidently identify an LLS system before the spectra deteriorate in
  quality.
\item For the WFC3 sample, identification of an LLS 
  with $\mtll \ge 0.5$ and $z=\mzabs$
  precludes the
  search for any LLS with $z < \mzabs$ and $\mtll \le 1$.
\item For  he WFC3 sample, 
  identification of an LLS with $\mtll \ge 1$  and $z=\mzabs$ 
  precludes the search for any LLS with $z < \mzabs$ and $\mtll < 2$ .
\item In both samples, identification of an LLS with $\mtll \ge 2$  and $z=\mzabs$ 
precludes the search for any other LLS with $z < \mzabs$.Because of potential sky subtraction complications
 (see Paper~I), we have only performed a survey
for $\mtll \ge 2 $ LLS in the \acs\ spectra. 
\end{enumerate}
Together, the final four criteria establish \zstart\ values for a
range of limiting optical depth. 

With these survey criteria, it is straightforward to evaluate the
survey path of our dataset.  This is described in Figure~\ref{fig:goz},
which presents the so-called sensitivity function $g(z)$ for the
survey, i.e.\ the number of unique quasar spectra at a given redshift
where one can search for an LLS to a given optical depth limit.  The
survey path is similarly summarized in Tables~\ref{tab:wfc3_survey}
and \ref{tab:acs_survey} where we
list the start and ending redshift for the LLS survey of each
QSO.  We note that the sensitivity function is comparable to that from
\cite{ribaudo11} at the low ($z\sim 1.5$) redshift end of our
sample, increasing to nearly twice their value at higher redshifts,
making the two samples highly complementary.
None of the quasars that they surveyed are in our sample.

\subsubsection{Results}
\label{sec:stat_results}

One may now simply evaluate Equation~\ref{eqn:loz} to derive estimates
for the incidence of LLS absorption at $z<2$, as a function of
limiting optical depth.  Figure~\ref{fig:loz} presents the results for
the \wfc\ survey for $\mtll \ge 0.5, 1$ and 2, and for the \acs\
survey for $\mtll \ge 2$ LLS.  We present the results for two redshift
intervals, $z=[1.2, 2.0)$ and $[2.0, 2.6)$.  As one predicts for an
expanding universe, we observe a decreasing incidence between the two
redshift bins for all samples.  Overlayed on Figure \ref{fig:loz} are
the results from \citet{songaila10} for $\mtll \ge 1$ determined over
the redshift range $0 < z < 6$.  Our results are consistent with
theirs and those of \citet{ribaudo11}, but we note that our survey
provides better sensitivity for the epochs $1.5 < z < 2.5$.

Equation \ref{eqn:dX}, along with the transformation
\begin{equation}
\ell(X) dX = \ell(z) dz
\label{eqn:lxlz}
\end{equation}
and the cosmological evolution of the Hubble constant
\begin{equation}
H(z) = H_{0} \ltk \Omega_{\Lambda} +
\Omega_{m}\left(1+z\right)^3 \rtk^{1/2}
\label{eqn:hoz}
\end{equation}
allows for the determination of $\ell(X)$ from our sample.  We use the
same bins in redshift and sub-sampling in $\mtll$ as for \loz.  The
results for \lox\ are shown in Figure \ref{fig:lox}.  Also shown in
Figure \ref{fig:lox} is the single power law fit for $\mtll \ge 2$ LLS
from \cite{ribaudo11} which combines their analysis of $z<2.5$ HST
spectra and the $z > 3.5$ SDSS results from POW10.
Our results generally agree with those of \cite{ribaudo11},
namely that at redshifts $z<2$, $\ell(X)$ evolves only very weakly (in
fact, the \wfc\ results are more consistent with no evolution),
transitioning to a sharper evolution with $z$ at $z>2$.  We further
concur with their observation that the $2.5 < z < 3.5$ span must be
fully explored to best understand this transition in evolution of $\ell(X)$.
Additionally, Ribaudo \etal\ (2011) use their results to constrain the
differential column density distribution, \fnhi.
We address this
quantity below, using the additional information gained from an
analysis of stacked spectra.  

We now place our results for
$\ell(X)$ from \wfc\ alongside those obtained from the SLLS 
\citep{opb+07} and the DLA \citep{pw09} in Figure~\ref{fig:complox}.  
Quantitatively, we find that the SLLSs and DLAs (i.e.\ systems with
$\mnhi \ge 10^{19} \cm{-2}$) contribute a significant fraction of
\tlox. By inference, this implies that there are relatively few LLS
with $\mnhi = 10^{17.5} - 10^{19} \cm{-2}$ and therefore that \fnhi\
is shallow at these column densities.  We return to this issue in
$\S$~\ref{sec:fn_highest}.
We also note with interest the relative contributions to \lox\
that each bracket in $\mtll$ makes.  Specifically, we see that the
integral contribution to \lox\ of the LLS with $0.5 < \mtll < 2$ is
nearly equal in magnitude to that for the LLS with $\mtll > 2$.  
As \lox\ corresponds to the integral of \fnhi\ over \nhi, this implies
a pronounced change in shape in \fnhi\ for $\mtll < 2$.  We
explore this result in greater detail in $\S$~\ref{sec:fn_plls}.
Finally, we combine our results for $\mtll > 2$ for \acs, \wfc, and
the results from the archival study of \cite{ribaudo11}.  The
weighted mean for these results at $z\sim [2,2.5]$ is $\mtlox = \loxval$.
We adopt this measurement in all analysis that follows.
Table~\ref{tab:incidence} summarizes all of these results.

\subsection{Systematic Errors in the Incidence Frequency}
To assess the effects of uncertainty in the \nhi\ and $z_{\rm abs}$
values for the LLS, we performed a Monte Carlo analysis where we
randomly modified each system with a normal deviate and assuming
$\sigma_{\mnhi} = 0.05$\,dex and $\sigma_{\rm z} = 0.02$ (see
$\S$~\ref{sec:spectral}).  We then recalculated \loz\ 1000 times for
each redshift bin.
We find that none of the trials have an \loz\
value that lies outside the $85\%$ confidence level given by Poisson
uncertainty.  Furthermore, we explored the effects of proximity to the
quasar by varying our upper wavelength cut.  This simultaneously
addresses issues associated either with inaccurate quasar emission redshifts,
$z_{qso}$, or with the possibility that quasars arise in overdense
regions, which might effect our statistics.  To do so, we expanded the 
proximity region from 3000 \kms\ blueward of the quasar emission
redshift to 10000 \kms, and find our results to be essentially unchanged.

In \cite{pow10} we identified and described at length a systematic
bias to LLS surveys associated with pLLS.  This bias occurs when one
truncates the search path for an LLS owing to the presence of one or
more pLLS, as we have done in this paper for systems with $\mtll <2$.
It is possible, therefore, that we have over-estimated the incidence
of systems with $\mtll \approx 1$.  The pLLS bias, however, is
proportional to the incidence of Lyman limit absorption which at these
redshifts is small.  Indeed, we estimate that this is a $<10\%$ effect
and therefore within the statistical error associated with our small
sample size.

The other, and unavoidable, bias described by \cite{pow10} is related
to the blending of Lyman limit systems. One or more systems localized
to $\delta z \sim 0.1$ will be unresolved by our spectral resolution
and therefore treated as a single LLS.  This raises the incidence of
$\mtll > 2$ systems at the expense of pLLS.  Again, this bias is
proportional to the overall incidence of LLS and at these redshifts
the effects will be small \citep[$<10\%$; Figure~7 of][]{pow10}.

%%%%%%%%%%%%%%%%%%%%%%%%%%%%%%%%%%%%%%%%%%%%%%%%%%%%%%%%%%
%%%%%%%%%%%%%%%%%%%%%%%%%%%%%%%%%%%%%%%%%%%%%%%%%%%%%%%%%%
%%%%%%%%%%%%%%%%%%%%%%%%%%%%%%%%%%%%%%%%%%%%%%%%%%%%%%%%%%
\section{Stacked Spectrum Analysis}
\label{sec:stack}

In this section we generate an average quasar spectrum from the \hwfc\
spectra and then analyze this dataset to estimate the mean free path
to ionizing radiation \lmfp, following PWO09. We begin with
a review and expansion of their formalism.

\subsection{Formalism}
\label{sec:formalism}

In PWO09 we introduced a new technique for constraining the mean free
path to ionizing radiation, \lmfp.  The approach analyzes a
`stacked' quasar spectrum, constructed from a strict average of a
cohort of quasars with common emission redshift, at wavelengths
blueward of the Lyman limit.  This stacked spectrum, by definition,
yields the average observed flux $\fobs{\lambda}$ of quasars at a given redshift.  For
wavelengths redward of \lya, the flux is dominated by the spectral
energy distribution (SED) of quasars, with only a minor attenuation
from the average metal-line absorption of gas in the IGM,
$\mtmtl(z)$.  Blueward of \lya, e.g.\ at $\approx 1125$\AA, 
the observed flux $\fobs{1125}$ is the
product of the average intrinsic quasar SED $\fint{1125}$ and the average absorption
from the \lya\ forest\footnote{Observationally, it is common practice
  to assess \tlya\ to a limiting \ion{H}{1} column density, e.g. to
  exclude DLAs or even LLSs.}
$\exp(-\mtlya)$ at the redshift $\mzlya =
1125\rAA \, (1+\mzem)/1215.67\rAA - 1$,

\begin{equation}
\fobs{1125}(\mzem) = \fint{1125} \, \exp \ltk - \mtlya(\mzlya)
\rtk \perd
\label{eqn:Lya}
\end{equation}
As one progresses to shorter wavelengths (lower \zlya), \tlya\ is
observed to decrease because the incidence of \lya\ lines decreases
\citep[e.g.][]{rau98}. However, one eventually incurs additional opacity
from \lyb\  from gas at redshift $\mzlyb = \lambda/1025.72\rAA - 1$, and
then \lyg\ opacity, and eventually the full Lyman series,

\begin{equation}
\mtlyman(\mzem,\lambda) = \smm_{n=2}^\infty \tau_{\rm eff}^{n}(z_n)  \cmma
\label{eqn:tau_Lyman}
\end{equation}
with $z_n = \lambda(1+\mzem)/\lambda_n - 1$ and $n=2,3,4,...$
corresponding to \lya, \lyb, \lyg, etc \citep[e.g.][]{madau96}.

At $\mlrest < 912$\AA, the emitted photons also experience the continuum opacity of Lyman
limit absorption $\mteff(\mzem,\mzll)$ with
\begin{equation}
\mzll \equiv \frac{\lambda}{\mlll} \, (1+\mzem) -1  \;\;\; .
\end{equation}
This quantity is related to
the mean free path \lmfp, defined to be the distance that a packet of photons travels before
suffering an $\rm e^{-1}$ attenuation (on average).  
Specifically, we define \lmfp\ to be the {\it physical} distance from \zem\ to the
redshift $\mzll^{\rm mfp}$ where $\mteff(\mzll^{\rm mfp},\mzem) = 1$.  
Note that with this definition, \lmfp\ applies to photons with energies
greater than 1\,Ryd at \zem.
The mean free path to ionizing photons establishes the mean intensity
of the extragalactic UV background 
\citep[EUVB;][]{hm96,meiksin09} and, in essence, defines the epoch of Hydrogen
reionization (i.e.\ \lmfp\ has negligible value in a neutral universe).  

Estimations of \teff\ (and thereby \lmfp) have 
been made previously by integrating evaluations of the \ion{H}{1} frequency
distribution \fnz\ of absorption systems in the IGM
\citep[e.g.][]{mhr99,flh+08}:

\begin{equation}
\mteff(z_{912},\mzem) = \intl_{z_{912}}^{\mzem} \intl_0^\infty f(\mnhi,z')
   \lbrace 1 - \exp \ltk - \mnhi \sigma_{\rm ph}(z') \rtk \rbrace d\mnhi dz' 
\label{eqn:teff}
\end{equation}
where $\sigma_{\rm ph}$ is the photoionization cross-section evaluated
at the photon frequency.   This approach is fraught with great
uncertainty because: (i) the frequency of absorbers with roughly unit optical depth
at the Lyman limit $\mtll \lesssim 1$ is very difficult to ascertain;
and (ii) the incidence of optically thick absorbers ($\mtll > 2$)
suffers from significant systematic uncertainty (POW10). 

In PWO09 we defined a new formalism to evaluate \teff\ from stacked
quasar spectra.  
We introduced an effective opacity from Lyman limits \kll\ which
evolves with redshift and frequency as

\begin{equation}
\mkll(z,\nu) = \mztkll(z) \ltp \frac{\nu}{\nu_{912}} \rtp^{-2.75} = 
  \mztkll(z) \ltp \frac{1+z}{1+z_{912}} \rtp^{-2.75} \cmma
\label{eqn:kLL}
\end{equation}
where the second term accounts for the
photoionization cross-section\footnote{In PWO09 we assumed a
  $\nu^{-3}$ dependence which is not as accurate.  Furthermore, there
  is essentially no frequency dependence on the effective opacity for
  systems with very high \nhi, which our approach ignores.} 
(accurate to within $1\%$ to $\mlrest \approx 600$\AA).
One may relate this opacity to the optical depth through the standard
definition\footnote{This formalism wrongly ignores stimulated
  emission, but that has a negligible effect on our model.}
\begin{equation}
\mteff(r,\nu) = \intl_0^r \mkll(r',\nu) \, dr' \cmma
\label{eqn:odepth}
\end{equation}
and relate distance to redshift via cosmology

\begin{equation}
\frac{dr}{dz} = \frac{c}{(1+z) H(z)} 
  = \frac{c/H_0}{(1+z) \sqrt{\Omega_{\rm m} (1+z)^3 + \Omega_\Lambda}} \perd
\label{eqn:drdz}
\end{equation}
In PWO09 we performed this analysis at $z \sim 4$ where the Hubble
parameter $H(z)$ is dominated by matter and one can approximate $dr/dz
\propto (1+z)^{-5/2}$.  
At the redshifts of interest here ($z \lesssim 2$), we find that a
power-law representation,

\begin{equation}
\frac{dr}{dz}(z<2.6) \approx \frac{dr}{dz} (z=2.6) \ltk
\frac{1+z}{3.6} \rtk^{-2.315} \cmma
\end{equation}
is accurate to within 1\%\ for $z=1-2.5$. We adopt this
power-law approximation in the following, taking $dr/dz(z=2.6) = 322 \umfp$.
Altogether our expression becomes

\begin{equation}
\mteff(z_{912},\mzem) = 6237 \, {\rm Mpc} \,
(1+z_{912})^{2.75} 
\intl_{z_{912}}^{\mzem} \mztkll(z') \, (1+z')^{-5.065} \, dz' \perd
\label{eqn:newteff}
\end{equation}

Note that this formalism is primarily introduced to parametrically
solve for \lmfp\ by establishing the redshift where $\mteff=1$.
In contrast to the PWO09 analysis of SDSS spectra, the \hwfc\ data
have the wavelength coverage and sensitivity to produce a stacked
spectrum to $\mlrest \approx 600$\AA.  In part, this reflects the 
larger mean free path to ionizing radiation at $z \sim 2$ than $z
\sim 4$.  If one wishes to analyze the stacked spectrum to these
wavelengths, however, it is not safe to assume 
that the
modulations in flux at $\mlrest < 912$\AA\ are due solely to Lyman
limit opacity
(as was done by PWO09\footnote{This
  assumption was justified, in part, because the analysis was
  performed over a very short wavelength (redshift) interval, $\mlrest
  \approx 850\rAA-900$\AA.}).
First, one must consider evolution in the intrinsic SED
$\fint{\lambda}$ over these
wavelengths.  For example, the $z \sim 1$ average quasar spectrum
produced by \cite{telfer02} follows a power-law $\fint{\lambda} \propto
\lambda^{-0.43}$ at $\mlrest < 1200$\AA\ which implies a 20\%\ higher
flux at $\mlrest = 600$\AA\ than at the Lyman limit.
Second, the Lyman series opacity \tlyman\ is certain to decrease with
redshift.  Therefore, the observed flux should increase at
$\mlrest < 900$\AA\ in the absence of Lyman limit opacity. 

Lastly, we also expect the mean free path
to increase with decreasing redshift as the universe expands.  In the
SDSS analysis at $z \sim 4$, the mean free path was so short that we had no
sensitivity to redshift evolution in \kll\ from a single stacked
spectrum (PWO09).  By examining stacked spectra over a (narrow) range of
redshifts, however, we revealed evidence for an increasing \lmfp\ with
decreasing redshift.  In the \hwfc\ stacked spectrum, 
the flux covered includes opacity from absorbers spanning from
$\mzll \approx 1 - 2.5$ and one must
consider explicit redshift evolution in \kll.
We assume the $\mztkll(z)$ opacity term in Equation~\ref{eqn:kLL}
evolves as a power-law
with the scale-factor $a=1/(1+z)$, 

\begin{equation}
\mztkll(z) = \mztkll(\mzstack) \ltk \frac{1+z}{1+\mzstack}
\rtk^{\gamma_\kappa} \cmma
\label{eqn:kLL_z}
\end{equation}
where $\mzstack$ is the median redshift of the quasars
whose spectra are averaged.
This assumed functional form for \ztkll\ is motivated by the fact that the evolution 
should be dominated by expansion of the Universe and the empirical
observation that many aspects of the IGM are well-modeled by power-laws of
the scale factor.  We would reconsider this parameterization if it
provided a poor model for the data.

\subsection{Generating the \hwfc\ Stacked Spectrum}
\label{sec:make_stack}

As described in the previous subsection, constraints on the mean free
path at $z<2.5$ may be derived from analysis of the average quasar
spectrum of $\mzem \approx 2.5$ quasars.  Our \hwfc\ dataset comprises
\nwfc\ quasars with $\mzem = 2.3 - 2.6$ (median of 2.44), each chosen to be free of
significant associated absorption.

We have taken the following procedure to combine the individual
spectra, accounting for the differences in \zem\ and also the
non-uniform dispersion of the \wfc\ grism spectra.

First, we correct for reddening of the quasar flux by adopting the
$E(B-V)$ estimate from \cite{sfd98} and a standard Galactic extinction curve
\cite{fm90}.  In general, this is a small correction because the
median reddening value is $E(B-V) = 0.02$\,mag.  On the other hand,
our spectra cover the so-called 2175\AA\ bump in the extinction curve
which implies an additional correction.
Second, we generate a rest-frame wavelength array with fixed dispersion
$\Delta \lambda$.  The dispersion value was set to be large enough to
include at least one entire pixel from the \wfc\ spectra at rest
wavelengths $\mlrest < 1215$\AA.  Specifically, we adopted
$\Delta\lambda = 6.1897$\AA.  Third, each quasar spectrum was shifted
to the rest-frame and normalized by the median flux from $\mlrest =
1450\rAA-1470$\AA.  Fourth, we assigned the entire flux of each pixel from
the original spectrum to the nearest pixel of the stacked spectrum.
Lastly, all of the flux values in the final spectrum were averaged to
produce the $\bar f_\lambda$ spectrum, normalized to unity at $\lambda
= 1450$\AA.

The resultant \hwfc\ stacked spectrum is shown as the solid black line
in Figure~\ref{fig:wfc3_stack}, and tabulated in Table~\ref{tab:wfc3_stack}. 
To assess sample variance, we generated a set of
stacked spectra with standard bootstrap techniques.  Specifically, we
generated 500 stacked spectra by drawing randomly from the \nwfc\
quasar spectra, allowing for duplications.  The shaded region in
Figure~\ref{fig:wfc3_stack} shows the RMS at each pixel after clipping
any $3 \sigma$ outliers.  To some degree, this bootstrap analysis provides
estimates on the uncertainty in our measurement of the average observed
quasar flux at $z \approx 2.4$, subject to the selection criteria of the
sample.  

Overplotted on the stacked spectrum is the average radio-quiet quasar spectrum
of \cite{telfer02}, produced from a set of $z \sim 1$ quasars observed with
UV spectrometers on \ihst\ and modified as follows.   The
Telfer spectrum has been normalized at $\mlrest = 1450$\AA, smoothed to
the \wfc\ spectral resolution, and rebinned to the dispersion solution
of the \hwfc\ stacked spectrum.  At wavelengths $\mlrest \approx
1300$\AA, the \hwfc\ stacked spectrum has flux that modestly exceeds
the Telfer spectrum.  This suggests that the \hwfc\ stack has a bluer
underlying power-law continuum but it could also result from strong
line-emission (e.g.\ \ion{O}{1}~1302) at those wavelengths.  We 
explore this issue in greater depth in $\S$~\ref{sec:teff_lya}.  

At all wavelengths $\mlrest < 1200$\AA, the Telfer spectrum
exceeds the \hwfc\ stack.  Here, the \wfc\ stacked spectrum includes 
absorption by the intergalactic medium, whereas the Telfer spectrum
has been corrected for IGM absorption (to the best of their abilities).
For wavelengths $\mlrest \approx 1100$\AA,
the offset is $\sim 10\%$ and it increases with decreasing wavelength as
additional terms in the Lyman series contribute
(Equation~\ref{eqn:tau_Lyman}). 
Beyond the Lyman limit, the flux for the \hwfc\ stack drops rapidly
owing to the integrated continuum opacity of the Lyman limit \teff\
(Equation~\ref{eqn:teff}).  
At $\mlrest
\approx 600$\AA, the ratio of the Telfer spectrum to the \hwfc\ stack
implies a total effective optical depth $\tau_{\rm eff}^{\rm TOT}
\sim 1.5$.  We expect the opacity at these wavelengths to be dominated by the
Lyman limit opacity.  We now model this absorption and
thereby place new constraints on the IGM at $z \sim 2$.

%%%%%%
\subsection{Mock Spectra}
\label{sec:mocks}

To better gauge the effects of sample variance, to test our stacking, 
and to examine the evolution of quantities such as \tlyman, we
created a number of sets of mock \wfc\ spectra.  The mocks were
utilized at each stage of the analysis of the real \wfc\ stack
described below, both to develop and test our analysis algorithms, and
to better explore the range of likely errors for each measured
quantity.

The mock spectra were generated from simulated \ion{H}{1} line
distributions based on empirical parameterizations
\citep[e.g.][]{madau96,wp11}. 
Under the implicit assumption that the \lya\ forest is a collection of
Voigt profiles with uncorrelated parameters (redshift $z$, column
density \nhi, Doppler parameter $b$), we populated each
simulated  sightline with absorbers until the \lya\ effective
optical  depth converged to a given value. 
If the \lya\ effective optical depth evolves as 
$\tau_\mathrm{eff}^\mathrm{Ly\alpha}\propto\left(1+z\right)^{\gamma+1}$ 
the line density is $l\propto\left(1+z\right)^\gamma$. At $z<1.5$ we assumed
$\tau_\mathrm{eff}^\mathrm{Ly\alpha}=0.017\left(1+z\right)^{1.20}$
(Kirkman et al. 2007), whereas at $z>1.5$ we incorporated the observed
steepening in the $\tau_\mathrm{eff}^\mathrm{Ly\alpha}$ evolution by
taking $\tau_\mathrm{eff}^\mathrm{Ly\alpha}=0.0062\left(1+z\right)^{3.04}$ 
(Dall'Aglio et al. 2008).

For the Doppler parameters we adopted the single parameter
distribution function by \cite{hr99} %Hui & Rutledge (1999) 
$dn/db\propto b^{-5}\mathrm{exp}\left(-b^4/b_\sigma^4\right)$ 
with $b_\sigma=24$~km\,s$^{-1}$ \citep{kim+01} %(Kim et al. 2001) 
restricted to 10~km\,s$^{-1}\le b<100$~km\,s$^{-1}$.
The main distribution of interest is the column density distribution
for which we assumed a triple power law
$f(\mnhi,\mnhi^\mathrm{min},\mnhi^\mathrm{max})=C_i\times \mnhi^{-\beta_i}$ 
over the range 12$<\log \mnhi <$22.
The constants $C_i$ implicitly depend on the 
$\tau_\mathrm{eff}^\mathrm{Ly\alpha}$ evolution, so we determined them
from 5000 simulated sightlines for each of our sets of slopes $\beta_i$.
By varying these three slopes we recovered different values of the
mean free path 
(Equation~11). In total, we ran 42 different models that resulted 
in a wide range of MFPs at $z=2.4$ from 40~Mpc to the horizon length. 
For simplicity we discarded most of these and considered only 
those with simulated MFPs close to our measured value.

With these Monte-Carlo simulated line lists we then generated mock
 \wfc\ spectra as follows: For each of the 53 quasars in our sample 
we first generated a mock quasar SED by multiplying
the \cite{telfer02} composite with a power law 
$f_\lambda=(\lambda_\mathrm{rest}/2500\mathrm{\AA})^{\alpha}$ with
$\alpha$ drawn from a Gaussian distribution with $\sigma=0.2$.
The mock SED was normalized between 1430 and 1470~\AA.
We then randomly drew one of the 5000 sightline realizations, 
simulated the resolved Lyman series and continuum spectrum at 
$0<z<z_\mathrm{em}$, multiplied it onto the interpolated SED, 
convolved it with the \wfc\ grism line spread function and rebinned 
it to the \wfc\ grism dispersion solution.
Lastly, we added Gaussian noise to the spectra to match the quality of 
our \wfc\ spectra (S/N$\sim 30$ at 1350\AA), taking into account both
the shape of the mock quasar spectrum and the \wfc\ sensitivity
function. 

%%%%%%%%
\subsection{Modeling the Stacked Spectrum}
\label{sec:modeling}

In this subsection, we describe our approach to modeling the stacked
quasar spectrum from the \hwfc\ sample (Figure~\ref{fig:wfc3_stack},
Table~\ref{tab:wfc3_stack}).    
Our scientific emphasis is to place new
constraints on the magnitude and evolution of the mean free path
\lmfp\ at $z \approx 2$. Nevertheless, the data redward of the Lyman
limit offer additional constrains on the intrinsic SED of $z \sim 2.5$
quasars and also the Lyman series opacity at $z \sim 2$.
We consider each of these in turn.

%%%%%%%%%%%%%%%%%%%%
\subsubsection{QSO SED}
\label{sec:qsoSED}
Central to our experiment is the fact that we use quasar light 
to probe the foreground IGM.  To some extent, it is an
observational necessity for revealing the nature of this medium.
Galaxies are too faint for current facilities, especially at $\mlrest
< 2000$\AA, and gamma-ray bursts or supernovae are too faint and/or
fade too rapidly to generate sufficiently large samples, especially
with UV spectroscopy.  

As is apparent from previous work on the rest-frame SED of quasars at
UV wavelengths \cite[e.g.][]{telfer02,vanden01,scott04}, these sources do
not have smoothly varying, intrinsic continua.  Although the underlying
SED is roughly a power-law ($f_\lambda \propto \lambda^\alpha$ with
$\alpha \approx -1.3$ for $\mlrest > 1200$\AA), there are
significant and broad emission lines from the \ion{H}{1} Lyman series
and metal-line transitions.  Furthermore, individual quasars exhibit a
range of power-law slopes and varying strength of line-emission.  
Despite this diversity, it is remarkable that quasars as a
population have a very similar average SED at all redshifts
\citep[e.g.][]{vanden01}.
Outside the \lya\ forest, for example, the SED
in the UV is nearly identical between the $z \sim 1$ Telfer
spectrum and the average spectrum of $z \sim 3$ quasars
\citep{vanden01}.  
There is only a hint of a harder SED for higher $z$ quasars
\citep{telfer02} and also modest differences in the equivalent widths of
high-ionization emission lines \citep{baldwin77}.  
At wavelengths $\mlrest < 1200$\AA, however, \cite{telfer02} report a
break in the quasar SED power-law to $f_\lambda \propto
\lambda^{-0.4}$ that is not evident in other analyses
\citep[e.g.][]{scott04}.  This difference could be related to redshift
evolution and/or the average luminosity of the quasars sampled.
Recently, \cite{shull12} have published a composite spectrum from
high-dispersion {\it HST}/COS observations (with corrected Lyman limit
absorption) and find results that are in good agreement with the
Telfer et al.\ SED.

Because of the remarkable similarity in the average SED of bright
quasars with redshift, we adopt the following assumption for modeling
the \hwfc\ stacked spectrum:  we assume the underlying SED is the
Telfer et al.\ radio-quiet spectrum modulated by a tilted power-law,

\begin{equation}
\fint{\lambda} = C_{\rm T} \, f_\lambda^{\rm
  Telfer} \, \ltp \frac{\lambda}{1450 \rm \AA} \rtp^\mdat \cmma
\label{eqn:intrinsic}
\end{equation}
and a scaling parameterized by $C_{\rm T}$.
Returning to Figure~\ref{fig:wfc3_stack}, we propose that the offset between the
Telfer spectrum and the \hwfc\ stack at $\mlrest \approx 1300$\AA\ may
result from a difference in the average power-law of the two stacked
spectra.  Indeed, the data are well matched if we assume $\mdat
\approx -0.5$.  Interestingly, such a tilt is akin to arguing that
there is no break in the average quasar SED at $\lambda_r \approx
1200$\AA, contrary to the findings of \cite{telfer02} but consistent
with other estimations \citep{scott04}.

The inclusion of a tilt also allows for the fact that \cite{telfer02}
may not have properly or entirely corrected for IGM absorption at
$\lambda < 900$\AA.  For example, the authors did not correct for 
partial LLS, which are expected to be rare at $z<1$ \citep{ribaudo11}
but not negligible.

In the following, we
restrict \dat\ by demanding that the effective \lya\
opacity \tlya\ derived from our stacked spectra matches previous
estimates from analysis of higher spectral-resolution data
\citep{kts+05}.
Similarly, the parameter $C_{\rm T}$ allows for a normalization offset
between the Telfer stack and the \hwfc\ stack.  Although each spectrum
was normalized at $\mlrest \approx 1450$\AA, there may be a modest
difference in the emission-line strength at these wavelengths and
it is non-trivial to precisely measure the flux of each stacked
spectrum at a given wavelength.  Therefore, we allow for a 5\%\
modulation in the normalization.

%%%%%
\subsubsection{Constraints on the Intrinsic SED from \tlya}
\label{sec:teff_lya}

As noted in the previous sub-section, we have significant reason to believe that the average
intrinsic quasar SED for our \wfc\ sample does not follow the standard
Telfer quasar spectrum.  This is suggested by the offset at
$\lambda_{\rm r} \approx 1300$\AA, but is even more apparent in the \lya\
forest (Figure~\ref{fig:wfc3_stack}).  At $\lambda_{\rm r} \approx
1100$\AA, for example, the Telfer spectrum lies only a few percent
above the stacked spectrum which would imply almost zero effective
\lya\ opacity at $z \approx 2$.  This contradicts our knowledge of
the IGM at these redshifts, as informed by high S/N, high
spectral-resolution observations \citep[e.g.][]{kts+05,kbv+07}.
Indeed, we now invert the problem to constrain the average quasar SED
of our quasar sample, using previous estimates of \tlya. 

Previous authors have estimated the effective opacity from the \lya\
forest using higher spectral resolution data of $z \sim 2.5$ quasars
\citep{kts+05,kbv+07}.  The standard approach is to estimate the
intrinsic quasar continuum $\fint{ }$ from the data directly and then measure the
average absorption,

\begin{equation}
D_A \equiv 1 - f^{\rm obs}/f^{\rm SED} \cmma
\label{eqn:DA}
\end{equation}
between the \lya\ and \lyb\ emission lines corresponding to a
redshift slightly less then \zem.
Depending on the author, estimations for $D_A$ may include only the
low density \lya\ forest, all \lya\ lines (i.e.\ including strong
absorbers like damped \lya\ systems), and/or metal-line absorption.
Because our \wfc\ stacked spectrum includes the opacity from all of
these sources, we wish to compare against a total estimate $D_A^T$.
\cite{tytler04} report a total $D_A^T$ value at $z=2.14$ of $0.18 \pm
0.04$.  Analysis of SDSS quasar spectra provides a similar value
\citep{dww08}. 

In Figure~\ref{fig:DA} we present a series of SED models for our
intrinsic spectrum comparing to a range of tilts and normalizations
(\dat, $C_T$) applied to the Telfer spectrum.  In each case, we have
demanded that the average opacity at $\lambda_{\rm r} = [1080,
1140]$\AA\ fall within the $1\sigma$ interval given by $D_A^T$.  
Restricting $C_T$ to $\pm 5\%$, we find \dat\ values ranging from
$\approx -0.2$ to $-1$.  It is very unlikely that the unaltered Telfer
spectrum provides a good description of the intrinsic SED for
our \wfc3\ cohort.  Instead, 
the data favor a bluer SED at $\lambda_{\rm r} <
1200$\AA.  In fact, the preferred tilt for $C_T = 1$ of $\mdat \approx
-0.5$ nearly corresponds to maintaining the $f_\lambda \propto
\lambda^{-1.3}$ power-law that is observed for quasars at $\lambda_{\rm
  r} > 1200$\AA\ \citep{telfer02,vanden01}.  While it is possible that a
portion of this tilt relates to error in the fluxing of the \wfc3\
spectra (Paper~I), it is very unlikely to be entirely explained by
such systematic effects.  Instead, we conclude that the \wfc\ quasar
cohort has a harder SED than the $z \sim 1$, Telfer et al.\
radio-quiet sample.  
In addition to the implications for our analysis of the IGM, a harder
SED would imply a higher emissivity of ionizing photons from $z \sim 2$
quasars.  In turn, it would increase the quasar contribution to the
extragalactic UV background at these redshifts.  

The bluer SED relative to the Telfer analysis may also arise from the
color-selection of SDSS quasar candidates.  At $z \sim 2.5$, the
average quasar color lies near the stellar locus
\citep[e.g.][]{richards06} and therefore, the SDSS team weighted
their targeting algorithms toward quasars with UV-excess (i.e.\ bluer
SED).  Indeed, \cite{wp11} have estimated that SDSS quasars with
spectroscopic redshift $z \approx 2.5$ would have $\mdat \approx -0.3$
(see their Figure 16).  
We encourage additional analysis of quasars at $z > 2$ to further
explore this issue.

Although the above analysis prefers a hard SED, 
we emphasize that a tilt $\mdat \approx -1$ is not well
supported by the individual spectra.  None of the quasars exhibit such
a blue SED at $\lambda \ll 1200$\AA.  For example, the quasars
J083326+081552
and J121519+424851, which appear to have minimal Lyman limit absorption from
the IGM, are better described by $\mdat \approx -0.5$ at such wavelengths.  
Therefore, in the following, we 
restrict the analysis to SED models with $\mdat = [-0.8, -0.2]$.

\subsubsection{Lyman Series Opacity}
\label{sec:teff_lyman}

Although the \lya\ forest is characterized by a series of narrow
($\delta v \le 50\mkms$) and stochastically distributed
absorption-lines, the \hwfc\ stacked spectrum exhibits no discrete
absorption features.  This follows from the low spectral-resolution of
the individual \wfc\ spectra and also the effects of spectral
stacking.  
As detailed in
$\S$~\ref{sec:formalism}, the Lyman series opacity should increase
with decreasing rest-wavelength as additional transitions contribute
(Equation~\ref{eqn:tau_Lyman}).
Below the Lyman limit, however, the
total Lyman opacity \tlyman\ is then expected to
decline with decreasing redshift (i.e.\ as the universe expands).  

One may relate the effective opacity at a given Lyman transition to
the frequency distribution of IGM absorption lines $f(\mnhi,b,z)$ as, 

\begin{equation}
\tau_{\rm eff}^n = \int\int\int f(\mnhi,b,z) \exp(-\tau_\nu^n) d\mnhi \,
db \, dz \cmma
\end{equation}
where $\tau_\nu^n$ is the line opacity of transition $n$ and is a
function of $\mnhi, b$, and $z$.  Because $f(\mnhi)$ is estimated to
decline at least as steeply as $\mnhi^{-1.5}$, one predicts the
effective opacity of a given Lyman transition to be dominated by the 
the minimum column density for line saturation.  This means that
\tlya\ is set by lines with $\mnhi \approx 10^{14} \cm{-2}$
whereas the higher order transitions are defined by lines with
$\mnhi \approx 10^{17} \cm{-2}$.  
Indeed, for reasonable \fnhi\ distributions we estimate that 
roughly half of the total opacity derives from the first few
transitions ($n=2-5$, i.e.\ \lya-\lyd) 
with the remainder contributed by higher order transitions.

To estimate the \tlyman\ opacity from the frequency distribution,
therefore, one requires: an accurate description over many orders of
magnitude in \nhi, an assessment of the Doppler parameter distribution,  
and also an accurate estimation of their
evolution with redshift.  While there has been some analysis on
$f(N,b,z)$ 
at $z \sim 2$, the constraints are quite limited. 
Given this uncertainty, we have made our own estimate of
\tlyman\ from the \wfc\ stacked spectrum, corresponding to $z=z_{\rm
  stack}=2.44$ .  In
Figure~\ref{fig:wfc3_tlyman} we estimate \tlyman\ allowing for the
range of intrinsic quasar SEDs described in the previous subsection.
The top panel shows the recovered \tlyman\ values as a function of
tilt $\delta\alpha_{\rm T}$ in the SED.  The bottom panel shows the 
range of allowed SEDs compared against the \wfc\ stacked spectrum. 
We also present the unaltered Telfer spectrum (purple, dotted line) to
further emphasize that this SED is inconsistent with the observations.
If we assume that $\mdat = -0.8$ to $-0.2$ at 95\%\ c.l., then this
implies $\mtlyman = 0.40 \pm 0.15$ at similar confidence.  There will
be additional uncertainty from sample variance, but we estimate that
this contributes less than the uncertainty related to the quasar SED.

This estimate for \tlyman\ may be crudely compared against the
incidence of $\tau > 1$ LLS derived from our spectra.  Specifically,
one may assume that $\approx 50\%$ of the opacity is contributed by the higher
order lines and then compare this value to an
estimate for \tlyman\ for LLS from an evaluation of equation~\ref{eqn:teff}. 
Taking $\mnhi \ge 10^{17.19} \cm{-2}$, $\ell(z) = 1.5$ for $z=2.0$,
$b_{\rm eff} = 35 \,\mkms$, and assuming $f(N,z) \propto N^{-1.5}
(1+z)^{1.33}$ yields $\tau_{\rm eff} = 0.06$ for the transitions
$n=5-30$.   This value is considerably lower than the \tlyman\ value
derived from above, and it suggests that systems with $\mnhi <
10^{17.2} \cm{-2}$ dominate the Lyman series opacity.  This conclusion
is consistent with our inferences of a steep $f(N)$ distribution at
$\mnhi \approx 10^{16} \cm{-2}$
\citep[$\S$~\ref{sec:stats},\ref{sec:fn};  see also][]{ribaudo11}.

In the following, we model the \wfc\ stacked spectrum with a range of
\tlyman\ values.  In fact, we take exactly the \tlyman\ value required
to reproduce the observed flux at $\mlrest = 912$\AA\ for a given
quasar SED and then allow for a 10\%\ variation which significantly
exceeds what can be measured from our high $S/N$ stack and therefore
is a conservative range.  
For $\lambda_{r} < 912$\AA, we further assume that
\tlyman\ decreases as a $(1+z)^{\gamma_\tau}$ power-law, 

\begin{equation}
\mtlyman(\lambda_{\rm r}) = \mtlyman(\lambda_{\rm r}=912{\rm \AA}) \, 
  \ltp \frac{1+z_{912}}{1+\mzem} \rtp^{\mgtL}  \perd
\end{equation}
This is justified by the fact that \fnz\ is generally well-described
by such a power-law.  On the other hand, different \nhi\ regimes and
different redshift ranges may be described by different \gtL\ 
values \citep{janknecht06}.  We proceed under the expectation that the
total \tlyman\ is nevertheless well described by a power-law over the
redshift interval relevant to our analysis $\mzll \approx 1.5$ to 2.4.
Specifically, we allow \gtL\ to range from 1.2 to 1.8 which spans the
range of values estimated for the \lya\ forest \citep{janknecht06} and LLS
\citep[$\S$~\ref{sec:stats};][]{ribaudo11} at $z \lesssim 2$.

\subsubsection{Lyman Limit Opacity}
\label{sec:LLOpacity}

Following the formalism presented in $\S$~\ref{sec:formalism},
we characterize the Lyman limit opacity \kll\ as given by
equations~\ref{eqn:kLL} and \ref{eqn:kLL_z}.
This is a two parameter model that sets the normalization at $z =
z_{\rm stack}$ and allows for redshift evolution.
In the following, we will demand that $\gamma_\kappa > 0$, i.e. that
the \lmfp\ increases\footnote{Formally,
  $\gamma_\kappa = 0$ also implies an increasing \lmfp\ because of the
  expanding universe.} 
with decreasing redshift as expected from the
decreasing incidence of LLS
($\S$~\ref{sec:stat_results}).

\subsection{Constraining the Mean Free Path}
\label{sec:MFP}
 We now proceed to compare
a suite of models against the observed \hwfc\ stacked spectrum to
estimate the mean free path \lmfp.
We proceed with standard $\chi^2$ analysis where the uncertainty at
each pixel in the stacked spectrum is estimated from a bootstrap
analysis ($\S$~\ref{sec:make_stack}).  This yields a `best' estimate
for \lmfp.  In turn, we provide an estimate of the uncertainty in this
quantity by repeating such analysis on a set of stacked spectra
generated with standard bootstrap techniques. 
There are several issues to note regarding this approach.
First, the bootstrap analysis of the stacked spectrum provides an estimate of the RMS at each
pixel in the stack, but we caution that the PDF need not follow a true
Gaussian.  On the other hand, we find our results are relatively
insensitive to what we assume for the error in the stacked spectrum.
%{\bf [Test this]}
Second, the bootstrap analysis may not properly reflect the full
uncertainty in the stacked spectrum related to sample variance, i.e.,
the stacked spectrum was derived from a total of 53 quasars.  
Lastly, this approach ignores the fact that
the data at $\mlrest < 912$\AA\ are highly correlated.  This
correlation occurs because of the nature of Lyman limit absorption;  
a system at
$z=z_{912}$ attenuates the flux at all wavelengths $\mlrest < 912{\rm
  \AA} (1+z_{912})/(1+\mzem)$.  Therefore, one predicts (and observes)
a monotonic decrease in the stacked spectrum for $\mlrest < 912$\AA.
The first two issues described above suggest an underestimate of the
uncertainty while the last point may lead to an overestimate.

Our model has six parameters: two for the quasar SED (\dat,
$C_{\rm T}$), two to model the Lyman series opacity
(\tlyman($z=\mzem$), \gtL), and two to model the Lyman
limit opacity ($\mkppo, \gamma_\kappa$).
In the previous subsection we imposed constraints on the 
parameters as summarized in Table~\ref{tab:best_mfp}.  
We then constructed a $\chi^2$ grid in this six dimensional parameter
space for rest wavelengths $\mlrest = 700.0 - 911.76$\AA.
Although the stacked spectrum extends to $\mlrest = 600$\AA,
uncertainty in the quasar SED, the evolution of \tlyman , and sample
variance are much greater at these wavelengths.
Furthermore, our principle goal is to estimate \lmfp\ which we find
occurs at $\mlrest \approx 800$\AA.  These issues motivated our decision to
terminate the quantitative comparison at $\mlrest = 700$\AA.

Figure~\ref{fig:wfc3_kappa} shows the results for the best-fit model
and the suite of models with $\Delta \chi^2 \equiv \chi^2_\nu -
\chi^2_{\nu, \rm min} < 0.5$.  
This choice for a limiting $\Delta \chi^2$ is primarily illustrative;
it does not impact our uncertainty estimate for \lmfp.
The top panel displays the range of allowed $\gamma_\kappa$ and $\mkppo$ values;
there is an obvious degeneracy between these two parameters.
In short, the data permit a more opaque universe at $z=2.4$ that
rapidly evolves to a lower opacity or a less opaque universe that
evolves more slowly.

Despite this degeneracy in $\mkppo$ and $\gamma_\kappa$, the results imply
a relatively tight constraint on \lmfp.  In
Figure~\ref{fig:wfc3_z912}, we plot the \ozll\ values for the full set
of allowed models as a function of the SED tilt \dat\ and redshift
evolution of the Lyman series opacity $\gamma_\tau$.  We find that
\ozll\ values from \zllval\ which for our adopted $\Lambda$CDM
cosmology implies $\mlmfp = \cmfpval \mhmpc$.
Again, we consider these estimates to be largely illustrative, not
quantitative;  systematic effects and sample variance dominate the
uncertainty. 

One approach to assessing uncertainty related to sample variance is to
repeat the above analysis for a suite of stacked spectra generated by
standard bootstrap techniques.  Figure~\ref{fig:mfp_boot} presents the
\lmfp\ values for each of 500~realizations of the stacked spectrum
using the same formalism described above.  We recover a median (mean)
mean free path of $\mlmfp = \bmmfpval \mhmpc$.
We consider this value to be our best estimate for \lmfp\ at
$z=2.44$.  This same analysis suggests an uncertainty in \lmfp\ of
$\approx 20\%$.  This should be considered a minimum estimate for the
uncertainty.
We also considered the systematic error reddening corrections might have on our
stack analysis
by turning off the de-reddening and re-computing \lmfp.  The results
of this change are within our estimates on the error for \lmfp, with the implied
value approximately 15\,Mpc lower than our fiducial value.  
The set of best-fit parameters are presented in Table~\ref{tab:best_mfp}.

Consider, further, the sources of error contributing to the \lmfp\
measurement.  Our analysis only considers
the stacked \wfc\ spectrum at rest wavelengths $\mlrest = 700 - 910$\AA\
where one observes a rapid ($>60\%$) decline in the relative flux with
decreasing $\mlrest$.  Our model of the relative flux consists of three
independent effects: 
(i) a reduction in the flux due to the effective
opacity of Lyman limit absorption \teff; 
(ii) an increase in the flux due to decreasing Lyman series opacity
\tlyman;
and 
(iii) changes in the flux owing to relative changes in the QSO SED.
We emphasize that the latter two effects are expected to be small
($<10\%$ in relative flux), in particular because the analysis is
performed on such a narrow range of \lrest.  For example, if we ignore
variations in \tlyman, we recover the same \lmfp\ value to within
3\%.  Similarly, varying the tilt in the QSO SED modifies \lmfp\ by
only several percent.
These two aspects of the model, however, do work together in that they both
give higher relative fluxes at lower \lrest.   We find
that adopting a redder QSO SED (e.g.\ the Telfer spectrum 
without any tilt) and no \tlyman\ evolution would yield 
an  $\approx 10\%$ higher \lmfp\ value than our favored value.  Even
redder (i.e. \dat $>0$) tilts have been considered, and would
significantly increase our estimate of \lmfp.  These tilts are highly
disfavored, however, as they summarily fail to reproduce the
constraints provided by \tlya. We cannot reject the possibility that
the underlying QSO SED could have additional strong inflections near
rest frame wavelengths of 912\AA, thus adding significant additional
uncertainties to our estimate of \lmfp, but we see no indications of
such features in other QSO SED studies \citep{telfer02,scott04,shull12}.
We conclude, therefore, that uncertainties in these aspects of the
modeling impose an $\approx 10\%$ systematic error in the \lmfp\ analysis.

Presently, uncertainty in the estimated \lmfp\ value is dominated by
sample variation.  As indicated in Figure~\ref{fig:wfc3_stack}, the
stacked spectrum has an $\approx 20\%$ scatter at $\mlrest \approx
800$\AA.  This dominates the uncertainty in the results
(Figure~\ref{fig:mfp_boot}).   We estimate that one would have to
increase the sample size by at least a factor of four before uncertainties
related to the QSO SED or \tlyman\ would contribute substantially.
Lastly, we comment that extending the \lmfp\ analysis to lower
redshifts ($z<1.5$) where the \lmfp\ value is presumably much larger
will prove progressively more difficult.  As the Lyman
limit opacity has a weaker impact on the relative flux, 
uncertainties in the QSO SED and/or \tlyman\ could dominate the
analysis.   

\subsection{Comparisons with other estimates of \lmfp\ at $z\sim 2.4$}
Our bootstrap estimate of \lmfp$= \bmfpval \mhmpc$ can now be compared with other
estimates derived through different methods at this redshift, namely
through application of Equation \ref{eqn:teff} after adopting an
estimate for \fnhi\.  
The exact approach to incorporating systems with $\mtll \approx 1$ can 
dramatically affect the results, and lend to very 
different estimates of \lmfp\ when using this
methodology.  \cite{flh+08} arrive at a value of \lmfp$=163 \mhmpc$
for $z=2.4$,
but establish the normalization of \fnhi\ by considering only those
LLS with $\mtll>1$. \cite{songaila10} provide new observations and
analyses of LLS at $z > 4$, along with low redshift GALEX LLS
observations to adopt a single power-law form for \fnhi\ in the LLS
regime, and the same methodology as \cite{flh+08} to arrive at  
\lmfp$=172 \mhmpc$ at $z=2.4$.  \cite{wp11} employ the \lmfp\ from
PWO09, plus new constraints on \fnhi\ at $14.5 < $\lnhi$< 19.0$ to
determine a multiple power-law \fnhi, and arrive at a higher
value of \lmfp$=220 \mhmpc$ at $z=2.4$.  Finally, \cite{hm12} employ
the \fnhi\ from POW10 to arrive at \lmfp$=185 \mhmpc$ at $z=2.4$, with
a very steep redshift dependence at these redshifts (by $z=2$, \lmfp
has increased to $242 \mhmpc$).
Clearly, the derived \lmfp\ values from \fnhi\ estimations are very
sensitive to the poorly constrained regime at $\mtll \approx 1$.
The methodology presented here has the advantage that it is wholly
independent of \fnhi\ and the results for \lmfp\ can then be used to
offer constraints on \fnhi , as in the next section.

\section{Constraints on $f(N)$ at $z \sim 2.2$}
\label{sec:fn}
We now turn our focus to an exploration of the column density
distribution function \fnhi.  
Our analysis of $\ell(X)$ for the LLS along with our stacked
spectrum analysis of \lmfp\ allow us to place constraints on \fnhi\
over many decades in H~I column density.  We have already seen in
section \ref{sec:stat_results} that we expect significant (and
multiple) deviations from a single power-law for \fnhi\ in order to
match constraints from the \lya\ forest on the low H~I column density 
end, and the SLLS and DLA on the high H~I column density column end.  
We address the problem in stages, from high to low \nhi\ before
performing a full and simultaneous fit to all of the observational
constraints at $z \sim 2$. 

\subsection{Constraints on \fnhi\ for $\mnhi \gtrsim 10^{18} \cm{-2}$}
\label{sec:fn_highest}

Our survey of Lyman limit systems provides an integral constraint on
the \nhi\ frequency distribution at high values
(equation~\ref{eqn:lox}; POW10).  Specifically, we observe an
incidence of $\tau > 2$ LLS at $z\approx 2$ of $\mtlox = \loxval$
corresponding to systems with $\mnhi \ge 10^{17.5} \cm{-2}$.
This integral constraint may then be compared against the observed
incidence of super Lyman limit systems (SLLS; $10^{19} \cm{-2} < \mnhi
< 10^{20.3} \cm{-2}$) and damped \lya\ systems (DLAs; $\mnhi \ge
10^{20.3} \cm{-2}$) that have been estimated from surveys analyzing
the damping wing of the \lya\ transition
\citep{peroux05,phw05,opb+07,pw09}.
If one assumes a functional form for \fnhi\ at $\mnhi < 10^{19}
\cm{-2}$, e.g.\ a single power-law, then we may directly constrain its
parameters as follows.

Figure~\ref{fig:complox} presents our estimates for \tlox\ compared
against the incidence of SLLS and DLAs, in cumulative form.  For the
DLAs we have assumed $\mlox_{\rm DLA} = \dlalox$ from the SDSS survey
of \cite{pw09}, derived from the redshift interval $z = [2.2,
2.4]$.  We note that 
systematic error associated with the SDSS spectral coverage
affects \ldla\ at these redshifts
\citep{np+09}, but the DLAs have a sufficiently small
contribution that we may neglect this issue.
We also adopt the \fnhi\ distribution
for DLAs from \cite{pw09} over the same redshift interval.  This is
plotted in Figure~\ref{fig:fn_high}.  For the SLLS, we adopt the value 
of $\mlox_{\rm SLLS} = 0.13$ given by \citep{opb+07}, which they
derive for SLLS with \lnhi $\ge 19.0$ and $1.7 < z < 3$.

Together, Figure \ref{fig:complox} shows that the 
SLLS and DLAs contribute $\approx 50-80\%$ of \tlox. 
Even if we assume the upper end of our \tlox\ estimate,\footnote{We
  further note that \cite{ribaudo11} reported $\mtlox = 0.27 \pm 0.09$
  from their survey of the {\it HST} archive.}
it is evident that systems with $\mnhi = 10^{17.5} \cm{-2} - 10^{19}
\cm{-2}$ have a relatively modest contribution.  
In turn, this implies a shallow \fnhi\ for
$\mnhi< 10^{19} \cm{-2}$.  Figure~\ref{fig:fn_high} shows the range of
constraints for a single power-law covering that \nhi\ interval,

\begin{equation}
f( 10^{17.5}\cm{-2} \le \mnhi < 10^{19} \cm{-2}; X) =
  k_{\rm LLS} \mnhi^{\beta_{\rm LLS}} \cmma
\label{eqn:power_LLS}
\end{equation}
which assumes 
(i) $\mtlox = \loxval$; 
(ii) \ldla=0.05;
(iii) $\log f(\mnhi=10^{19}\cm{-2},X) = -20.2 \pm 0.2$.
The latter constraint follows
from $\ell(X)_{\rm SLLS} = 0.13 \pm 0.04$ and 
$\beta_{\rm SLLS} = -1.2 \pm 0.2$.
We estimate $\log k_{\rm LLS} = -9.2$ and $\beta_{\rm LLS} = \blls$,
and note that the two parameters are highly correlated.

The derived slope for \fnhi\ in the LLS regime is very shallow;
$\beta_{\rm LLS} > -1$ implies the universe exhibits greater
cross-section to gas with $\mnhi = 10^{19} \cm{-2}$ than $10^{18}
\cm{-2}$, which seems unlikely.  However, a similar result was derived (in nearly identical fashion) for LLS at $z \sim 4$
by POW10.  At both epochs, the relatively low incidence of $\tau > 2$
LLS implies a flattening of \fnhi\ at $\mnhi \approx 10^{18} \cm{-2}$
that we speculate is associated with transitioning from an optically
thick to an optically thin regime.  
We return to this issue in $\S$~\ref{sec:discuss}.

%%%
\subsection{Constraints on \fnhi\ for $\mnhi \approx 10^{17} \cm{-2}$}
\label{sec:fn_plls}

In Section~\ref{sec:stats} we presented results on the incidence of
LLS for several limiting opacities at the Lyman limit: $\tau_{\rm limit} =
0.5, 1$ and 2.  While these measurements are not independent\footnote{We
  further remind the reader that \flox\ was required to exceed \tlox\
  in the analysis.  While this must be true, it further emphasizes
  that the two results are highly correlated.}
(they were derived from the same sightlines and the samples overlap),
one may compare the results to offer an estimate for \fnhi\ in the
interval $\mnhi = 10^{16.9} \cm{-2} - 10^{17.5} \cm{-2}$
corresponding to $\mtll = 0.5 - 2$.  

Once again, we parameterize \fnhi\ as a single power-law, 

\begin{equation}
f( 10^{16.9}\cm{-2} \le \mnhi < 10^{17.5} \cm{-2}; X) =
  k_{\rm pLLS} \mnhi^{\beta_{\rm pLLS}} \perd
\label{eqn:power_pLLS}
\end{equation}
To estimate both $\beta_{\rm pLLS}$ and $k_{\rm pLLS}$, we must impose
an additional constraint on \fnhi\ from the LLS analysis of the
previous sub-section.  Specifically, we consider the range of allowed values
for $f(\mnhi=10^{17.5}\cm{-2}, X)$ which Figure~\ref{fig:fn_high}
demonstrates could range from $\approx 10^{-18.5} - 10^{-20} \cm{2}$. 
In Figure~\ref{fig:fn_pLLS}a, we plot the offset in incidence $\Delta
\mlox$ from the $\mtlim > 2$ measurement for $\mtlim > 1$ and $\mtlim
> 0.5$.
Overplotted on these offsets are a series of curves for
the predicted offsets, as a function of \tlim, for a range of $\mfnlls$ 
and $\beta_{\rm pLLS}$ values.  Models with $\mfnlls < 10^{-19} \cm{2}$
predict too few pLLS for $\mbplls > -2$.  Indeed, only
for $\mfnlls = 10^{-18} \cm{2}$ is a power-law
with $\mbplls \approx -1$ permitted in the pLLS regime.

This issue is further illustrated in Figure~\ref{fig:fn_pLLS}b where
we indicate the allowed values for \bplls\ (i.e.\ in agreement within
$1\sigma$ of both $\Delta \mlox$ values) as a function of
$\mfnlls$.  We conclude that \pfnhi\ must be steeper than $\beta_{\rm
  pLLS} = -1.5$ unless $\mfnlls > 10^{-18.7} \cm{2}$.
A steepening of \fnhi\ in the pLLS regime has been previously reported
by \cite{ribaudo11} based on their LLS analysis.
Similarly, POW10 argued for a steepening of \fnhi\ at $\mnhi < 10^{17}
\cm{-2}$ in order reproduce the mean free path estimates of
PWO09.   We discuss this result further in
$\S$~\ref{sec:discuss}.

While a steep \fnhi\ is permitted physically, it is quite surprising
given the very shallow \fnhi\ estimated for the LLS systems in the
previous sub-section.  Indeed, at its extreme, the results imply a
shift in slope $\Delta \beta_{\rm LLS} \equiv \beta_{\rm LLS} - \beta_{\rm
  pLLS}$ of unity or greater over an \nhi\ interval of $\approx
1$\,dex. This point is further illustrated in
Figure~\ref{fig:fn_dbeta} where we estimate \dbeta\ taking into
account the \lox\ measurements and constraints on the SLLS (the DLA
results are inconsequential).
For our best estimate of \tlox=0.3, we recover $\mdbeta>1$, i.e.\ a
rapid steepening in \fnhi.  Only at the upper end of permissible
values for \tlox\ may \dbeta\ be small.  We further emphasize that any
new measurements of \lslls\ or \ldla\ that gave higher values would
further increase \dbeta.

%%%
\subsection{Constraints on \fnhi\ for $\mnhi < 10^{17} \cm{-2}$}
\label{sec:fn_low}

We complete our piece-meal examination of \fnhi\ by imposing
constraints from the mean free path analysis of $\S$~\ref{sec:stack}.
Our evaluation of \lmfp\ yields an additional integral constraint on
\fnhi\ as described in equation~\ref{eqn:teff}.  Unlike \lox, the
\lmfp\ constraint includes redshift evolution in the \nhi\
frequency distribution.  Our results
on \lox\ suggest very weak redshift evolution for \fnhi\ at $z<2.5$
\citep[Figure~\ref{fig:lox}; see also][]{ribaudo11}, at least for
systems with $\mnhi \approx 10^{17} \cm{-2}$.  Furthermore, our
estimate of the mean free path is sufficiently short that we are only
sensitive to \fnhi\ for a small redshift interval at $z \approx 2$.
We proceed, therefore, by assuming no evolution in \fnhi.

Figure~\ref{fig:cumul_tau} shows calculations of the effective Lyman
limit opacity \teff\ for a series of \fnhi\ models, identical for
$\mnhi > 10^{17.5} \cm{-2}$ but with strict power-laws at lower \nhi\
with $\mbplls = [-1, -1.5, -2, -2.5]$.  In each case, we estimate
\teff\ from $z=\zbest$ to $z=\zmedian$ which is the redshift interval
for which our mean free path analysis yielded $\mteff = 1$.    The red
shaded region indicates the $1\sigma$ uncertainty in this estimate.

For the favored values of \lmfp\ and for the \fnhi\ estimates at
$\mnhi > 10^{17.5} \cm{-2}$, we find that LLSs with $\tau > 2$
contribute approximately half of the required optical depth.  Indeed,
unless we adopt the largest \lmfp\ allowed by the data and the highest
incidence of LLS then systems with $\mnhi < 10^{17.5} \cm{-2}$ must
make a significant contribution to \teff.  Figure~\ref{fig:cumul_tau}
indicates that $\mbplls \ll -1$; in fact $\mbplls < -2$ is preferred
for our central values.  This offers (nearly) independent evidence
that \fnhi\ steepens significantly in the pLLS regime.

\subsection{Joint Constraints on a Complete \fnhi\ Model}
\label{sec:fn_fit}

As the previous subsections reveal, 
our estimates of the mean free path \lmfp\ and the incidence of LLS
\lox\ provide new constraints on the \nhi\ frequency distribution
\fnhi\ at $z \sim 2.5$, especially for $\mnhi \approx 10^{15} - 10^{18} \cm{-2}$.
By combining these measurements with other evaluations and
constraints on \fnhi\ from the literature, we may infer the \nhi\ distribution
across $\approx 10$ orders of magnitude. 
This evaluation informs astrophysical quantities sensitive to
properties of the IGM \citep[e.g.\ the EUVB;][]{hm12} and provides
insight into the physical origin of absorption line systems.
At the most basic level, we test whether a single \fnhi\ model can
reproduce all of the observational constraints.  Any significant
inconsistency would stress a fundamental flaw in our standard methods
for studying the IGM. 

Before proceeding, it is important to emphasize what may be a subtle
distinction between two conceptualizations of \fnhi:
(1) an analysis of the \fnhi\ distribution that one uses to estimate
quantities like \tlya\ and \lmfp\
(Equations~\ref{eqn:tau_Lyman},\ref{eqn:teff}) and 
(2) an observational \fnhi\ derived through line-fitting analysis of
absorption-line spectra.  At larger \nhi\ values, where the systems
are rare and line-blending is uncommon, the two types of \fnhi\ should
equate.  At low \nhi, however, the two may diverge.  Consider, for
example, a mock spectrum generated from an analytic \fnhi\ following
the procedures described in Section~\ref{sec:mocks}.  One could then
perform a line-profile analysis of the mock spectrum and would
certainty find fewer low \nhi\ lines than inputted.  
Ultimately, neither type of \fnhi\ actually follows the physical
description of the IGM suggested by cosmological simulations, i.e. the
concept of an undulating density field.  In this respect, any estimate
of \fnhi\ has limited physical significance.
We proceed, nevertheless, with the primary goal of estimating an
analytic-oriented \fnhi\ distribution.

Absent a physically motivated model for \fnhi, we assume that it
follows a series of power-laws, monotonically decreasing in value with
increasing \nhi\ but 
discontinuous in the first derivative.  
This approach is partly motivated by previous estimations of \fnhi\
over modest \nhi\ intervals, but it primarily reflects our ignorance of
the underlying distribution function.  Analytically, a power-law
approach has the advantage of being robust to non-physical
deviations, e.g., significant wiggles that a spline or higher-order
polynomial may introduce.  
We normalize \fnhi\ at $\mnhi = 10^{12} \cm{-2}$
with the value $10^{k_{12}}$. 
We do not evaluate \fnhi\ for $\mnhi < 10^{12} \cm{-2}$
because there are very poor constraints at these column densities;
for example, gas with such low column densities should contribute less than
5\%\ to the total effective \lya\ opacity.\footnote{Of course, such gas
  may trace the majority of the volume of the universe and could
  contribute a significant fraction of the mass density.}
We then describe \fnhi\ as a series of broken
power-laws with slope $\beta$ in each segment defined by
a series of \nhi\ `pivots'. For example, $\beta_{20.3}$ defines the slope from 
$\mnhi = 10^{20.3} \cm{-2}$ to the next \nhi\ pivot.  
Initially, we also allowed for redshift evolution in \fnhi, in particular
to model \lmfp.  We found,
however, that the data offered only very weak constraints.
In the following, we assume a $(1+z)^{1.5}$ evolution in \fnhi, which
is roughly consistent with the observed evolution in \tlya\ at these redshifts
\citep{kts+05}.

In addition to the observational constraints imposed by the results of
this paper (\lmfp, \lox), we also include
(1) evaluations of \fnhi\ estimated from the line-fitting survey of
\cite{kim02} but re-evaluated to our own choice of \nhi\ binning and
to our assumed cosmology.  We restrict the evaluations to $\mnhi < 10^{14.5}
\cm{-2}$ to maintain significant statistical power;
(2) measurements of \fnhi\ for SLLS from \cite{opb+07};
(3) measurements of \fnhi\ for DLAs from \cite{pw09};
and
(4) an estimate of \tlya\ (from $D_A$) at $z=2.4$ \citep{kts+05}.
Table~\ref{tab:fn_constraints} lists these constraints and we discuss
a few modifications to the reported values and errors below.

From the empirical constraints, we construct deviates between
observation $y$ and model $m$, $d \equiv (y-m)/\sigma$, with $\sigma$
the 1-sigma uncertainty.  This includes the integral
constraints on \fnhi: \tlya, \lox\, and \lmfp.  For all of these, we
assume Gaussian errors which is not formally correct but offers a fair
approximation.  For the case of \tlya, \cite{kts+05} estimated an error of
only 3\%\ but we adopt a more conservative error of 10\%\ to account for
systematic uncertainty and the fact that our model
assumes a fixed Doppler parameter for the \lya\ forest of
$b=24\mkms$.  We adopt a model with the same constraint as the 
\cite{kts+05} analysis that was restricted to systems
with $\mnhi <10^{17.2} \cm{-2}$, but which includes metal line
absorption.
For the highest \nhi\ values (constrained by the DLAs), we calculate
the deviate from $\mnhi = 10^{21.5} - 10^{22.3} \cm{-2}$ and consider
this as a single constraint.
Lastly, we optimized our parametric model with the software package
MPFIT with initial guesses estimated through a by-eye comparison to the
various constraints.

It is illustrative to consider a series of \fnhi\ models with
increasing complexity, i.e.\ a increasing number of power-law
segments,
to establish the simplest model which satisfies all constraints.
Following the results of the previous sub-section, we model \fnhi\ at
$\mnhi > 10^{18} \cm{-2}$ with a series of 3 power-laws, defined by
\nhi\ pivots at
$\log \mnhi = 20.3$ and 21.5\,dex.  We then optimized a 4-parameter
model for \fnhi\ under the full set of observational constraints.
This model assumes a single power-law with slope $\beta_{12}$ from
$\mnhi = 10^{12} - 10^{20.3} \cm{-2}$.  Not surprisingly, the model
yields very poor results with a reduced chi-squared
$\chi^2_\nu > 10$.
The model cannot match the shallow slope suggested by \lox\ 
(Figure~\ref{fig:fn_high}) with the steep slope implied by 
\lmfp\ (Figure~\ref{fig:cumul_tau}) and the \lya\ forest observations.
In short, the data require a break in \fnhi\ below $\mnhi = 10^{18} \cm{-2}$.  

We next evaluated a set of 5-parameter models (4 power-laws) with an
additional \nhi\ pivot at $\approx 10^{17.5} \cm{-2}$.  To our
surprise, we found this model well-reproduced ($\chi^2_\nu \approx 1$)
the \tlya, \lmfp, \lox,
and SLLS/DLA constraints provided $\beta_{12} \approx -1.65$
(Figure~\ref{fig:bothfn_z25}a). This implies
a significant break to a shallower \fnhi\ ($\Delta \beta > 0.6$) for
the LLS.
Perhaps coincidentally, this slope is in good agreement with the shape
of \fnhi\ reported for the $z<1$ universe \citep{pss04,lh07}.
This model also implies that absorption with low \nhi\ values
($<10^{13.5} \cm{-2}$) contribute significantly $(\sim 50\%)$ to the
effective \lya\ opacity.  In turn, the model severely over-predicts the
incidence of such weak absorbers compared to the empirical estimations
of \cite{kim02}.  Those authors reported a more
shallow \fnhi\ at $\mnhi < 10^{14} \cm{-2}$, inconsistent with our
5-parameter model.  On the other hand, such
low \nhi\ absorption is the most subject to uncertainties associated
with S/N, continuum placement, and line-blending.  
We also posit that this difference may be a manifestation of the
distinction between the analytic and observational \fnhi\
distributions noted above.   We proceed to consider a yet more complex
model but note that one is not entirely required.

In an effort to reproduce all of the observational constraints given in
Table~\ref{tab:fn_constraints}, we examined a 6-parameter model (5
power-laws) that includes two \nhi\ pivots between $\log \mnhi = 14$
and 19\,dex.  We experimented with parameterizing these
two \nhi\ pivots and found very weak constraints on their values.
Therefore, we have fixed them at $\mnhi = 10^{14.5} \cm{-2}$ and
$\mnhi = 10^{17.5} \cm{-2}$ and caution that the results should not be
considered unique.  
Figure~\ref{fig:bothfn_z25}b presents the best-fit \fnhi\ distribution
compared against the observational constraints.  
The reduced $\chi^2$ is near unity and is dominated by the \fnhi\
evaluations at low and high \nhi. 
This model shows a shallow distribution function at low \nhi\ which
steepens to $\beta_{14.5} \approx -2$ at   
$\mnhi \gtrsim 10^{14.5} \cm{-2}$, as inferred by \cite{kim02}. 
Again, we conclude that \fnhi\ is steeper than $\beta = -1.6$ at
$\mnhi \approx 10^{15} \cm{-2}$ and then transitions to $\beta \approx
-1$ in the LLS regime.
This is consistent with our inferences based on the \lmfp\ and partial
LLS analyses (Sections~\ref{sec:MFP},\ref{sec:fn_low}).

Despite the relatively low $\chi^2_\nu$ value for this 6-parameter
model, there is significant tension.  A
shallow \fnhi\ at low \nhi\ values requires that a significant
fraction of the \lya\ opacity arise from systems with $\mnhi \approx 10^{14}
\cm{-2}$.  Indeed, a 6-parameter model that pivots at $\mnhi = 10^{14}
\cm{-2}$ instead of $10^{14.5} \cm{-2}$ under-predicts the
\tlya\ value and may be ruled out.  The model shown in
Figure~\ref{fig:bothfn_z25}, meanwhile, over-predicts the \cite{kim02}
estimates of \fnhi\ at $\mnhi > 10^{14} \cm{-2}$.  
As the precision in the observations improves, we
speculate that it may become impossible to generate an \fnhi\
distribution that satisfies all of the constraints.

To summarize our main findings on \fnhi\ at $z \approx 2.4$:

\begin{itemize}
\item All 4-parameter (3 power-law) models for \fnhi\ are ruled out by
  the observations.
\item A 5-parameter model with a single power-law ($\beta \approx
  -1.65$) spanning from $\mnhi = 10^{12} \cm{-2}$ to $\approx
  10^{17.5} \cm{-2}$ well reproduces the \tlya, \lmfp, \lox, and
  SLLS/DLA observations.  This model predicts a large contribution to
  the effective \lya\ opacity from gas with low \nhi\ values and also
  over-predicts the incidence of such gas compared to the
  line-analysis of \cite{kim02}.  Such line-analysis is challenged,
  however, by S/N, continuum placement, and line-blending and we
  recommend further consideration of the systematic uncertainties.
\item To fit the \fnhi\ evaluations from \cite{kim02}, one must
  consider a 6-parameter model (5 power-laws).  This allows for a
  shallow \fnhi\ ($\beta_{12} \approx -1.3$) at $\mnhi \lesssim 10^{14}
  \cm{-2}$ that must break sharply to a steep distribution
  $(\beta_{14.5} \approx -2$) for intermediate \nhi\ values.
\item All of the successful models require a relatively steep ($\beta
  < -1.6$) distribution function for $\mnhi = 10^{15} - 10^{17}
  \cm{-2}$ that then flattens in the LLS regime to $\beta \approx -1$.
\end{itemize}

Table~\ref{tab:corrmat} lists the best-fit values, the $1\sigma$
uncertainties derived from the diagonal elements of the covariance
matrix, and the correlation matrix for the model.  The parameters are
well constrained, with the exception of the slope at $\mnhi >
10^{21.5} \cm{-2}$ which is best considered as an upper limit
\citep[$\beta_{21.5} < -3$;][]{phw05}.
On the other hand, several of the parameters are highly correlated
(e.g.\ $k_{12}$ and $\beta_{12}$) and one cannot assert that our model
is unique.  
We also caution that our analysis ignored the parametric freedom of
pivot placement; their values were motivated by the observations.  Lastly, we
re-emphasize that the broken power-law model is assuredly non-physical
and therefore an incomplete description of the universe.
Nevertheless, the results impose a general description of \fnhi, whose
implications are discussed below.

\section{Discussion}
\label{sec:discuss}
We now consider the larger implications of our results.

\subsection{Evolution in \lox\ and \lmfp}
\label{sec:mfp}

In Figure \ref{fig:mfp_vs_z}, we present our measurement of \lmfp\
along with the results of PWO09 at higher redshift.  Overplotted on
Figure \ref{fig:mfp_vs_z} are curves which describe the evolution in
\lmfp\ with redshift when assume the 5-parameter model functional form
for the IGM at $z=2.4$.  
The different curves correspond to allowing \loz\ to evolve as
$(1+z)^\gamma$, and letting $\gamma$ vary from 1.5 to 2.5.  The curves are
forced to have the same value, and match the PWO09 SDSS result at
$z=4$. The importance of our new measurement for \lmfp\ is immediately obvious:  
without information at lower redshifts, the high-redshift data has no 
discriminatory value in $\gamma$. 
 With the inclusion of the {\it HST} data, however, we can rule out extrema
 in $\gamma$, especially towards the high end.  Specifically, we can
 rule out $\gamma =2.5$.  We note that our results are again
 consistent with those of \cite{ribaudo11}, who favor a value of
 $\gamma = 1.8$.  Their analyses, however, only explicitly determines
 the \lmfp\ for systems with $\tau_{912} > 2$.
Figure \ref{fig:mfp_vs_z} also shows the need for
 further exploration of \lmfp\ at other redshifts, specifically at $z
 < 2$, $z\simeq 3$, and $z \ge 4.5$ to provide a full
 description of the evolution of \lmfp\ over cosmic time.  Songaila \&
 Cowie (2010) present a measurement of \lmfp\ at $5< z < 6$ from ESI
 spectra which indicate a slightly lower value for \lmfp\ than would be
 predicted from lower redshift power-law fits, but we caution that the
 number of spectra in their sample is relatively low, and that their
 method for determining \lmfp\ involves a number of assumptions not
 made in our analysis here.  

The solid gray curve in Figure~\ref{fig:mfp_vs_z} shows the Horizon of
the universe as a function of redshift.  When \lmfp\ exceeds this
radius, ionizing photons are capable of travelling the visible
universe before being attenuated.  In this respect, every source of
ionizing photons can `see' the other.  This epoch is termed the
`breakthrough' redshift \citep[e.g.][]{mm93}.  Given the current
constraints on \lmfp\ and the simple models presented in
Figure~\ref{fig:mfp_vs_z}, our best estimate is $z_{\rm break} = 1.6$
and we require $z_{\rm break} < 2$.  Future observations on the
incidence of optically thick absorption at $z \lesssim 1$ should
firmly establish the value (Howk et al., in prep.).

\subsection{The Inflection in \fnhi\ through the LLS Regime}

One of the primary results of our analysis is that the frequency of
systems with $\mnhi \lesssim 10^{17} \cm{-2}$ is significantly higher
than that predicted from a simple extrapolation of the \fnhi\ distribution estimated for
$\mnhi > 10^{17} \cm{-2}$.  Stated another way, our results demand a
higher frequency of pLLS than one would predict based on the
incidence of LLS.  We have reached this conclusion from two
complementary analysis of our dataset:
(i) the direct counting of pLLS versus LLS (Figure~\ref{fig:fn_pLLS}); and
(ii) our analysis for the total effective Lyman limit opacity of LLS
(Figure~\ref{fig:cumul_tau}).  
Each of these results imply the steepening of \fnhi\ with decreasing
\nhi.  Such a steepening of \fnhi\ at $\mnhi
\lesssim 10^{17.2}$ has been inferred previously at $z=3.7$ (POW10)
based on \lmfp\ analysis and at $z \approx 1.5$ from a direct
survey of pLLS \citep{ribaudo11}.

This result is further illustrated in Figure~\ref{fig:dldN} which
plots the differential contribution of systems to \lox\ per $\Delta
\log \mnhi = 0.5$\,dex interval, \dldN. 
These \dldN\ values were simply calculated from the power-law
description of \fnhi\ shown in Figure~\ref{fig:bothfn_z25}b.  
As with \lox, each evaluation here is proportional to the comoving
number density of sources, $n_c$, that give rise to the absorption times their 
average physical cross-section $\Delta A$ at that range of column densities.
For $\mnhi < 10^{17} \cm{-2}$, the $\Delta \mlox$ values rise steeply
with decreasing \nhi.  At $\mnhi \approx 10^{17.5} \cm{-2}$, however,
the $\Delta \mlox$ values flatten and are nearly constant until $\mnhi
\approx 10^{20.5} \cm{-2}$.  This simply reflects, of course, the
`inflection' in \fnhi\ across the LLS regime.

The flattening of \fnhi\ as \nhi\  decreases from $\approx 10^{20.5}$
has been recognized and discussed previously
\citep{opb+07,pow10} at higher redshifts, and also at $z\sim
0$ \citep{corbelli02}.
This is expected to occur because of the 
transition in gas from a predominantly
neutral state to a predominantly ionized plasma \citep{zheng02}. 
This leads to a significantly lower cross-section for gas with $\mnhi
\le 10^{20} \cm{-2}$ than otherwise and hence an
observed flattening in \dldN.  Detailed radiative transfer
calculations of galaxies in high-resolution cosmological simulations
have confirmed this picture \citep{fg11,fumagalli11a,altay11}.

The steepening of \fnhi\ at $\mnhi \approx 10^{17} \cm{-2}$ with
decreasing \nhi\ suggests a similar `phase transition'.  
Because $\mnhi = 10^{17.2} \cm{-2}$ corresponds to $\mtll = 1$,  
the obvious culprit for an inflection is the transition 
from the optically thin regime to optically thick gas. 
Qualitatively, this may work as follows.
For systems with $\mnhi \ll 10^{17.2} \cm{-2}$, the material is
optically thin and one expects to receive roughly the same
radiation field from the EUVB.  In this case, the distribution of
\nhi\ values that result is dominated by the distribution
of densities within the volume \citep[e.g.][]{schaye01_lya}.  Our results
indicate an \nhi\
distribution characterized by $\mfnhi \propto \mnhi^{-2}$.  In the
absence of radiation transfer effects (i.e.\ the transition to
optically thick gas), we may presume that the \nhi\ distribution would
have shown a similar shape for $\mnhi > 10^{17} \cm{-2}$.  Instead,
as the gas becomes optically thick, the radiation field has been attenuated
and one recovers a higher \ion{H}{1} column density than in the
optically thin limit.  This non-linear
effect naturally leads to the flattening of \fnhi\ at $\mnhi \approx 10^{17}
\cm{-2}$.   A full treatment of these effects within the context of
cosmological simulations is clearly warranted.

\subsection{Evolution in \fnhi}
Buoyed by the relative success of our analysis, we reconsider the
\nhi\ frequency distribution at $z \sim 3.7$, previously evaluated
by \cite{pow10}.  Table~\ref{tab:fn_constraints} lists the
observational constraints.  For the \lya\ forest ($\mnhi = 10^{12} -
10^{14.5} \cm{-2}$), we adopt a constraint on the slope of \fnhi\
based on line-fitting analysis \citep[e.g.][]{kt97,kim02}.  We allow
for significant uncertainty in this constraint since, as explored by
these and other authors, issues such as continuum placement, completeness,
and non-uniqueness of models to address line blending often arise,
especially at higher redshifts. 
The normalization of \fnhi\ at these low column densities is set by the
integrated opacity \tlya.  At $z \sim 4$, there is significant
disagreement in the empirical estimates of \tlya; \cite{fpl+08} report a value
$\mtlya \approx 0.8$ while other authors have published much lower values
\citep[$\mtlya \approx 0.7$][]{kbv+07,dww08}.  The difference in these
values substantially exceeds the estimated uncertainties, statistical and
systematic.  In our primary analysis, we adopt the higher value but
then comment on the implications of lower values.
The other empirical constraints are estimates of \fnhi\ from the SLLS and
DLAs \citep{opb+07,pw09}, and 
\lmfp\ and \lox\ integral constraints (PWO09,POW10).

Table~\ref{tab:corrmat} summarizes the best-fit model for $z=3.7$, error
estimates, and the correlation matrix while Figure~\ref{fig:fn_z37}
presents a comparison of the best-fit model with the data.
Similar to the $z=2.4$ analysis, we find that a six parameter model
(with identical \nhi\ pivots) provides a good description of the
observational constraints at $z=3.7$.  More remarkably, the best-fit
slopes all lie within $1 \sigma$ of the estimates at lower redshift.
The only statistically significant evolution lies in the
normalization; $k_{12}$ is $\approx 0.3$\,dex higher at higher
redshift.

The apparent lack of evolution in the shape of \fnhi\ between
redshifts 2.4 and 3.7 takes on additional interest when we consider
the strong evolution in \lox\ over the same timescale.  As \lox\
relates to the comoving number density of absorbers and their physical
size, we can surmise from our results that although either (or both)
of these two quantities must evolve with time, a change in the column
density distribution of absorbers producing the LLS does not.  This
implies that if it is the change in physical size which is providing
 the bulk of the evolution in \lox, and if the LLS arise in galaxy
 halos, the distribution of the gas within the halos is the same over
 cosmic time.  The same conclusion is reached by \citet{fumagalli11a}
 in their simulations.

\subsection{Implications for the nature of LLS}
Finally, we consider these results in the context of current understanding on
the origin of the LLS.  At low redshift, it has been statistically
established that systems with $\mnhi \gtrsim 10^{14.5} \cm{-2}$ are
associated with the circumgalactic medium (CGM) of galaxies
\citep{pwc+11,rudie12}. 
Specifically, galaxies of all luminosity show a very high covering
fraction of significant \lya\ absorption to impact parameters $r
\approx 300$\,kpc which may be defined as the extended CGM of these
systems.  \cite{pwc+11} demonstrate that $n_c$ for $L>0.01L^*$ galaxies
at $z\sim 0$ is large enough that all of the strong \lya\ absorbers
($>0.3$\AA) are associated with this extended CGM.
At high $z$, one also observes a significantly higher incidence of
\ion{H}{1} absorption in the CGM of $L\approx L^*$ galaxies
\citep{rakic12} and a remarkably high incidence of optically thick gas
in the CGM of galaxies hosting quasars \citep{qpq2}.  One may
infer, therefore, that gas with $\mnhi \gtrsim 10^{15} \cm{-2}$ at 
$z \sim 2$ primarily traces the extended CGM of high $z$ galaxies.
From numerical simulations, one reaches similar conclusions
\citep[e.g.][]{fumagalli11a,vdv11}.   

Recent work in cosmological simulations has also produced \fnhi\ with
numerous inflections.  For example, \citet{altay11} provide a
determination of \fnhi\ at $z=3$ through a combination of line-fitting
at low \nhi\ and projection at \nhi\ in the LLS and DLA regime.  Their
\fnhi\ is  consistent with those presented here at both $z=2.4$ and
$z=3.7$.  
Of particular
interest is their observation that the presence of self shielding
produces the flattening of \fnhi\ in the LLS regime, as discussed
above.  Unfortunately, their results are constrained to a single
redshift.  \citet{vdv11} use the same set of
simulations but focus instead on the gas fraction in halos, and the
future history of LLS gas.  They conclude that the majority of LLS gas
is in cold-mode galaxy halo gas and will be incorporated in those galaxies' ISM
within 1.2 Gyr.  While they do not provide estimates of \fnhi\ at
other $z$, one may interpret their result along with our observed lack
in evolution of \fnhi\ to imply that to maintain
the shape of \fnhi\ at later times in the universe, cold gas must
continue to be placed in galaxy haloes and be observed as LLS.
Finally, \citet{fumagalli11a} also find that the majority of LLS are
cold-mode gas.  Notably, they follow their simulations through the
redshift range $ 1.4 <  z < 4$ and find, in agreement with our
results, that \fnhi\ does not evolve with time.

\section{Summary and Future Work}
We have presented the first scientific results from our \textit{HST} program
to study Lyman limit absorption at redshift $z\simeq 2$.  
We have:
\begin{itemize}
\item Determined \loz\ and \lox\ over the redshift range $1 < z< 2.6$, 
finding a weighted mean of \lox $=0.29 \pm 0.05$ for $2.0 < z < 2.5$.
\item Stacked our WFC3 spectra to find  a median
(mean) value of the mean free path to ionizing radiation
of $\mlmfp = \bmmfpval \mhmpc$ with an error on the mean value of
$\pm 43 \mhmpc$.
\item Shown either that the SED of $z\sim 2$ quasars is harder than those at
    $z\sim 1$ or that there are significant systematics in the SDSS
    selection function.
\item Constrained the evolution of \lmfp\ with redshift, and
    estimating the breakthrough redshift to be $z = 1.6$.  The
    methodology behind this result is subject to
the assumption that there are no significant inflections in the
intrinsic QSO spectrum other than the one required to produce the
observed Lyman alpha forest opacity.  Including further inflections
would produce very large uncertainties in the derived \lmfp.
\item Shown that no 4-parameter (3 power-law) form can
    describe \fnhi\ at $z=2.4$, and provide a best-fit 5-parameter
    model.
\item Shown that \fnhi\ does not vary significantly in shape between
    $z=2.4$ and $z=3.7$.
\end{itemize}

 While the current data sample are more than adequate to constrain the
 incidence of LLS and their contribution to \fnhi\ and \lmfp\ at
 $z\sim 2$, they are of too poor resolution to make any statements on
 metal absorption in the LLS.  Given their increased number over the
 DLA at these (and all) redshifts, the LLS likely contribute to a
 significant fraction of the metal budget.  In the next paper in this
 series, we combine our \textit{HST} results with those from
 Keck+HIRES to address the metallicity of LLS at $z\sim 2$, a quantity
 which is currently poorly constrained.

\acknowledgements
It is our pleasure to thank
 K. Griest, G. Altay, M. Fumagalli, R. Simcoe, N. Lehner, and
C. Howk for numerous helpful discussions that significantly improved
this paper.  We also thank the anonymous referee for their comments.

This work is based on observations made with the NASA/ESA Hubble 
Space Telescope, 
obtained at the Space Telescope Science Institute, which is operated
by the Association of
 Universities for Research in Astronomy, Inc., 
under NASA contract NAS 5-26555. These observations are associated
with programs 10878 and 11594.
JXP also acknowledges support from an NSF CAREER grant (AST-0548180).
Support for programs 10878 and 11594 were provided by NASA through a 
grant from the Space Telescope Science Institute, which is operated by
the Association of Universities for Research in Astronomy, Inc., under
NASA contract NAS 5-26555.

Funding for the SDSS and SDSS-II has been provided by the Alfred
P. Sloan
Foundation, the Participating Institutions, the National Science
Foundation,
 the U.S. Department of Energy, the National Aeronautics and Space
Administration, the Japanese Monbukagakusho, the Max Planck Society,
and the Higher Education Funding Council for England. The SDSS Web
site is http://www.sdss.org/.
The SDSS is managed by the Astrophysical Research Consortium for the
 Participating Institutions. The Participating Institutions are the
 American Museum of Natural History, Astrophysical Institute
 Potsdam, University of Basel, University
of Cambridge, Case Western Reserve University, University of Chicago,
Drexel University, Fermilab, the Institute for Advanced Study, the
Japan Participation Group, Johns Hopkins University, the Joint
Institute
for Nuclear Astrophysics, the Kavli Institute for Particle
Astrophysics
and Cosmology, the Korean Scientist Group, the Chinese Academy of
Sciences (LAMOST), Los Alamos National Laboratory, the
 Max Planck Institute for Astronomy (MPIA), the Max Planck Institute
for Astrophysics (MPA), New Mexico State University, Ohio State
University, University of Pittsburgh, University of Portsmouth,
Princeton University, the United States Naval Observatory, and
the University of Washington.

%\bibliographystyle{/u/xavier/paper/Bibli/apj}
%\bibliography{/u/xavier/paper/Bibli/allrefs}
\bibliographystyle{apj}
\bibliography{jmo_allrefs}

\begin{deluxetable}{ccclccrrcc}
\tablewidth{0pc}
\tablecaption{Quasar Sample and Continuum\label{tab:quasars}}
\tabletypesize{\footnotesize}
\tablehead{\colhead{Object Name} &\colhead{Plate$^a$} & \colhead{Fiber$^a$} &\colhead{\zem} & \colhead{$g$} &
\colhead{C$^b$} & \colhead{$\alpha^b$}}
\startdata
\cutinhead{Quasars Observed with \wfc}
J075547.83+220450.1&1264& 163&2.319&17.58& 92.3&$ 0.4$&\\
J075158.65+424522.9& 434& 555&2.453&17.77& 58.0&$ 0.6$&\\
J084525.84+072222.3&1298& 512&2.307&17.87& 49.8&$ 0.5$&\\
J085045.44+563618.7& 448& 466&2.464&18.05& 65.7&$ 0.3$&\\
J085316.55+445616.6& 897& 392&2.540&18.17& 74.8&$ 0.4$&\\
J085417.60+532735.2& 449& 273&2.418&17.20&113.4&$ 0.9$&\\
J080620.47+504124.4&1780& 181&2.457&17.57&144.1&$ 0.0$&\\
J083326.82+081552.0&1759& 283&2.581&17.80& 90.9&$ 0.1$&\\
J090938.71+041525.8&1193& 296&2.444&18.01& 73.1&$ 0.2$&\\
J094942.34+052240.3& 994& 245&2.282&17.75& 90.9&$ 0.5$&\\
J105315.89+400756.4&1435& 205&2.482&18.15& 49.8&$ 0.2$&\\
J100541.26+570544.6& 558&  30&2.308&18.02& 90.5&$ 0.0$&\\
J101120.39+031244.6& 574& 281&2.458&17.79&127.7&$ 0.3$&\\
J113550.68+460705.0&1442&   1&2.496&17.67& 87.2&$ 0.4$&\\
J114358.52+052445.0& 838& 591&2.561&16.95&190.7&$ 0.3$&\\
J110735.58+642008.7& 596& 385&2.316&17.38&171.9&$ 0.1$&\\
J111928.38+130251.0&1605& 406&2.394&18.01& 56.9&$ 0.5$&\\
J110411.62+024655.3& 508& 520&2.532&18.21& 41.5&$ 0.5$&\\
J124831.65+580928.9&1317& 577&2.599&17.74& 77.9&$ 0.2$&\\
J125345.49+051611.3& 848& 438&2.398&17.99& 49.7&$ 0.4$&\\
J125914.85+672011.8& 495& 494&2.443&17.93& 97.8&$ 0.2$&\\
J121519.42+424851.0&1450& 486&2.310&17.73&124.0&$ 0.2$&\\
J122015.50+460802.4&1371& 327&2.446&18.01& 81.9&$ 0.4$&\\
J122836.05+510746.2& 971& 448&2.450&17.88& 94.6&$ 0.0$&\\
J123515.84+630113.4& 780& 546&2.383&17.38&157.8&$ 0.4$&\\
J133521.96+454238.2&1463&  29&2.452&18.00&114.3&$ 0.0$&\\
J130055.67+055620.5& 849& 330&2.446&17.99& 95.6&$ 0.2$&\\
J130240.16+025457.6& 524& 427&2.414&17.84&117.5&$ 0.2$&\\
J131123.09+453159.4&1375& 459&2.403&18.20& 61.8&$ 0.0$&\\
J131855.75+531207.2&1041& 319&2.321&18.08& 70.3&$ 0.5$&\\
J132312.83+414933.0&1462& 482&2.440&18.06&105.2&$ 0.4$&\\
J132552.17+663405.7& 496& 114&2.511&17.92&105.3&$ 0.3$&\\
J133433.88+035545.2& 853& 220&2.583&18.14& 89.3&$ 0.2$&\\
J133523.82+463742.1&1463& 586&2.474&17.97& 86.7&$ 0.0$&\\
J133646.56+015741.9& 528& 141&2.379&17.94&112.0&$ 0.0$&\\
J134211.98+601525.0& 786& 255&2.399&17.91&101.2&$ 0.1$&\\
J135412.28+542100.1&1323& 213&2.294&17.98& 72.5&$ 0.3$&\\
J135445.66+002050.3& 301& 385&2.504&18.07& 65.6&$ 0.0$&\\
J135831.78+050522.8& 856& 469&2.455&17.17&155.9&$ 0.4$&\\
J141528.47+370621.2&1643& 537&2.374&17.63& 96.3&$ 0.3$&\\
J140028.81+643030.9& 498& 296&2.359&18.09& 56.4&$ 0.5$&\\
J145453.53+032456.8& 588& 103&2.368&17.97& 95.4&$ 0.0$&\\
J153335.82+384301.1&1293&  30&2.529&17.50&128.3&$ 0.2$&\\
J153514.65+483659.7&1167& 471&2.542&17.84&108.0&$ 0.1$&\\
J154042.98+413816.3&1053& 306&2.516&17.46& 89.7&$ 0.0$&\\
J162516.42+294318.3&1421& 463&2.357&17.72& 70.6&$ 0.3$&\\
J161003.54+442353.7& 814& 301&2.588&18.24& 46.7&$ 0.2$&\\
J162548.79+264658.7&1408& 281&2.518&17.34&234.4&$ 0.0$&\\
J165137.52+400218.9& 630& 560&2.343&17.46&109.1&$ 0.6$&\\
J172409.19+531405.5& 359& 189&2.547&17.87& 52.8&$ 0.6$&\\
J211157.78+002457.5& 986& 403&2.325&17.85& 94.0&$ 0.3$&\\
J213629.44+102952.2& 731&  56&2.555&17.98& 59.1&$ 0.2$&\\
J233823.16+150445.2& 747& 493&2.419&17.62&108.9&$ 0.2$&\\
\cutinhead{Quasars Observed with \acs}
J010657.94-085500.1& 659& 420&2.350&18.08&  2.6&$ 0.0$&\\
J023359.71+004938.5& 407& 600&2.522&17.85&  4.0&$ 0.4$&\\
J034024.57-051909.2& 462& 445&2.340&17.95&  3.9&$ 0.0$&\\
J040241.42-064137.9& 464&  32&2.432&17.93&  3.0&$ 0.1$&\\
J084619.53+365836.8& 865& 135&2.335&18.18&  3.3&$ 0.3$&\\
J080413.66+251633.8&1205& 164&2.302&17.90&  1.9&$ 3.0$&\\
J085952.36+350724.7&1211& 506&2.373&18.22&  2.5&$ 0.6$&\\
J092849.24+504930.5& 767&  41&2.349&18.14&  3.1&$ 0.1$&\\
J102900.79+622342.0& 773& 314&2.470&17.78&  5.9&$ 0.6$&\\
J104321.55+624127.3& 773& 177&2.355&17.74&  4.3&$ 2.2$&\\
J114311.61+651513.5& 597& 604&2.392&17.59&  4.7&$ 0.0$&\\
J121944.79+461015.3&1451& 140&2.484&18.02&  2.5&$ 0.1$&\\
J120729.75+042909.9& 842&  68&2.413&18.21&  1.5&$ 0.3$&\\
J145554.30+521828.5&1164& 288&2.348&17.77&  3.9&$ 0.7$&\\
J165914.54+380900.7& 632& 120&2.343&18.12&  3.0&$ 1.2$&\\
J161815.53+370103.9&1056& 623&2.496&17.74&  4.1&$ 0.2$&\\
J161538.88+392051.1&1336& 137&2.315&18.11&  3.0&$ 0.0$&\\
J230011.74-102144.4& 725& 295&2.297&18.26&  3.1&$ 0.0$&\\
\enddata
\tablecomments{Quasars observed as part of our \hwfc\ and \hacs\ survey for LLS absorption (Paper~I).}
\tablenotetext{a}{SDSS-DR7 designations.}
\tablenotetext{b}{Absorbed continuum fitting parameters.  The modeled continuum is of the form: $f^{\rm conti}_\lambda = C \mftelf \, (\lambda / \rm 2500 \, \AA)^\alpha$.We caution the reader
that these models are not meant to precisely describe the intrinsic 
spectral energy distributions of the quasars, but instead describes the flux attenuated by the Lyman series absorption (see the text).}
\end{deluxetable}
 
\clearpage

\begin{deluxetable}{ccc}
\tablewidth{0pc}
\tablecaption{LLS Models \label{tab:models}}
\tabletypesize{\footnotesize}
\tablehead{\colhead{Quasar} &\colhead{$z_{\rm abs}$} & \colhead{\nhi}}
\startdata
\cutinhead{Quasars Observed with \wfc}
J075158.65+424522.9&2.380&16.50\\
&1.654&17.10\\
&1.545&17.35\\
J075547.83+220450.1&2.238&16.10\\
&2.024&17.25\\
&1.640&16.80\\
J080620.47+504124.4&1.840&16.95\\
&1.658&17.15\\
&1.343&16.80\\
&1.080&17.05\\
J083326.82+081552.0&1.911&16.40\\
\cutinhead{Quasars Observed with \acs}
J010657.94-085500.1&2.149&16.40\\
&1.947&18.10\\
J023359.71+004938.5&2.066&16.60\\
&1.393&16.35\\
J034024.57-051909.2&2.065&18.15\\
J040241.42-064137.9&2.160&17.10\\
&1.495&16.85\\
&1.778&16.65\\
\enddata
\tablecomments{[The complete version of this table is in the electronic edition of the Journal.  The printed edition contains only a sample.]}
\end{deluxetable}
 
\clearpage

\begin{deluxetable}{lcccccc}
\tablewidth{0pc}
\tablecaption{\wfc\ LLS STATISTICAL SURVEY\label{tab:wfc3_survey}}
\tabletypesize{\footnotesize}
\tablehead{\colhead{Quasar} &\colhead{\zem} & 
\colhead{$z_{\rm start}^{\tau>2}$} &\colhead{$z_{\rm start}^{\tau>1}$} &\colhead{$z_{\rm start}^{\tau>0.5}$} &
\colhead{\zlls} & \colhead{log \nhi}}
\startdata
J075158.65$+42$4522.9&2.453&1.576&1.576&1.576&1.576&17.60\\
J075547.83$+22$0450.1&2.319&1.400&2.026&2.026&2.026&17.30\\
J080620.47$+50$4124.4&2.457&1.400&1.400&1.840&1.840&16.95\\
J083326.82$+08$1552.0&2.581&1.400&1.400&1.400&\ldots & \ldots \\
J084525.84$+07$2222.3&2.307&2.275&2.275&2.275&2.275&18.45\\
J085045.44$+56$3618.7&2.464&1.400&1.400&1.400&\ldots & \ldots \\
J085316.55$+44$5616.6&2.540&1.903&2.117&2.117&1.903&17.90\\
&&&&&2.117&17.35\\
J085417.60$+53$2735.2&2.418&1.400&1.400&1.400&\ldots & \ldots \\
J090938.71$+04$1525.8&2.444&1.400&1.400&1.400&\ldots & \ldots \\
J094942.34$+05$2240.3&2.282&1.400&2.141&2.141&2.141&17.35\\
J100541.26$+57$0544.6&2.308&2.239&2.239&2.239&2.239&17.75\\
J101120.39$+03$1244.6&2.458&1.634&1.634&1.634&1.634&17.60\\
J105315.89$+40$0756.4&2.482&1.400&2.050&2.050&2.050&17.35\\
J110411.62$+02$4655.3&2.532&1.613&2.062&2.495&1.613&18.15\\
&&&&&2.062&17.40\\
&&&&&2.495&17.15\\
J110735.58$+64$2008.7&2.316&2.093&2.093&2.093&2.093&18.15\\
J111928.38$+13$0251.0&2.394&1.400&2.149&2.149&2.149&17.30\\
J113550.68$+46$0705.0&2.496&2.114&2.114&2.114&2.114&17.55\\
J114358.52$+05$2445.0&2.561&2.139&2.139&2.139&2.139&17.50\\
J121519.42$+42$4851.0&2.310&1.400&1.400&1.400&\ldots & \ldots \\
J122015.50$+46$0802.4&2.446&1.400&1.400&1.891&1.891&16.95\\
J122836.05$+51$0746.2&2.450&1.400&1.400&1.400&\ldots & \ldots \\
J123515.84$+63$0113.4&2.383&2.238&2.238&2.238&2.238&18.50\\
J124831.65$+58$0928.9&2.599&1.400&1.400&2.487&2.487&16.95\\
J125345.49$+05$1611.3&2.398&2.353&2.353&2.353&2.353&18.55\\
J125914.85$+67$2011.8&2.443&2.359&2.359&2.359&2.359&17.95\\
J130055.67$+05$5620.5&2.446&2.058&2.058&2.058&2.058&17.60\\
J130240.16$+02$5457.6&2.414&1.917&1.917&1.917&1.917&18.10\\
J131123.09$+45$3159.4&2.403&1.400&1.400&1.809&1.809&17.05\\
J131855.75$+53$1207.2&2.321&1.585&1.585&2.080&1.585&18.00\\
&&&&&2.080&17.00\\
J132312.83$+41$4933.0&2.440&2.267&2.267&2.267&2.267&18.30\\
J132552.17$+66$3405.7&2.511&1.792&2.373&2.373&1.792&17.60\\
&&&&&2.373&17.30\\
J133433.88$+03$5545.2&2.583&1.400&2.269&2.269&2.269&17.45\\
J133521.96$+45$4238.2&2.452&2.110&2.110&2.110&2.110&17.65\\
J133523.82$+46$3742.1&2.474&2.327&2.327&2.327&2.327&17.80\\
J133646.56$+01$5741.9&2.379&1.400&1.400&1.400&\ldots & \ldots \\
J134211.98$+60$1525.0&2.399&1.995&1.995&1.995&1.995&17.60\\
J135412.28$+54$2100.1&2.294&2.249&2.249&2.249&2.249&17.60\\
J135445.66$+00$2050.3&2.504&1.400&1.400&1.400&\ldots & \ldots \\
J135831.78$+05$0522.8&2.455&1.895&1.895&1.895&1.895&18.15\\
J140028.81$+64$3030.9&2.359&2.232&2.232&2.232&2.232&18.45\\
J141528.47$+37$0621.2&2.374&2.124&2.124&2.124&2.124&18.45\\
J145453.53$+03$2456.8&2.368&1.861&1.861&1.861&1.861&17.50\\
J153335.82$+38$4301.1&2.529&1.815&1.815&1.815&1.815&18.35\\
J153514.65$+48$3659.7&2.542&1.929&1.929&1.929&1.929&18.15\\
J154042.98$+41$3816.3&2.516&2.186&2.186&2.186&2.186&18.50\\
J161003.54$+44$2353.7&2.588&1.400&1.400&2.331&2.331&17.05\\
J162516.42$+29$4318.3&2.357&1.400&1.400&1.400&\ldots & \ldots \\
J162548.79$+26$4658.7&2.518&1.400&1.400&1.400&\ldots & \ldots \\
J165137.52$+40$0218.9&2.343&1.949&1.949&1.949&1.949&18.10\\
J172409.19$+53$1405.5&2.547&2.149&2.149&2.149&2.149&18.00\\
J211157.78$+00$2457.5&2.325&1.884&1.884&1.884&1.884&17.60\\
J213629.44$+10$2952.2&2.555&1.555&1.555&2.500&1.555&17.60\\
&&&&&2.500&17.15\\
J233823.16$+15$0445.2&2.419&2.226&2.226&2.226&2.226&17.75\\
\enddata
\end{deluxetable}

\begin{deluxetable}{lcccccc}
\tablewidth{0pc}
\tablecaption{\acs\ LLS STATISTICAL SURVEY\label{tab:acs_survey}}
\tabletypesize{\footnotesize}
\tablehead{\colhead{Quasar} &\colhead{\zem} & 
\colhead{$z_{\rm start}^{\tau>2}$} &
\colhead{\zlls} & \colhead{\nhi}}
\startdata
J010657.94$-08$5500.1&2.350&1.947&1.947&18.10\\
J023359.71$+00$4938.5&2.522&1.200&\ldots & \ldots \\
J034024.57$-05$1909.2&2.340&2.065&2.065&18.15\\
J040241.42$-06$4137.9&2.432&1.200&\ldots & \ldots \\
J080413.66$+25$1633.8&2.302&1.200&\ldots & \ldots \\
J084619.53$+36$5836.8&2.335&1.200&\ldots & \ldots \\
J085952.36$+35$0724.7&2.373&1.200&\ldots & \ldots \\
J092849.24$+50$4930.5&2.349&1.400&1.400&17.65\\
J102900.79$+62$2342.0&2.470&2.279&2.279&17.55\\
J104321.55$+62$4127.3&2.355&1.200&\ldots & \ldots \\
J114311.61$+65$1513.5&2.392&2.101&2.101&18.20\\
J120729.75$+04$2909.9&2.413&1.908&1.908&17.70\\
J121944.79$+46$1015.3&2.484&2.171&2.171&18.10\\
J145554.30$+52$1828.5&2.348&1.200&\ldots & \ldots \\
J161538.88$+39$2051.1&2.315&2.177&2.177&17.85\\
J161815.53$+37$0103.9&2.496&1.200&\ldots & \ldots \\
J165914.54$+38$0900.7&2.343&1.200&\ldots & \ldots \\
J230011.74$-10$2144.4&2.297&1.558&1.558&17.85\\
\enddata
\end{deluxetable}

\begin{deluxetable}{lcccccccccc}
\tablewidth{0pc}
\tablecaption{Incidence of LLS \label{tab:incidence}}
\tabletypesize{\footnotesize}
\tablehead{\colhead{z} & \colhead{$\Delta X$} & \colhead{$\Delta z^a$} & \colhead{$m_{\rm LLS}^{b}$} & \colhead{$<z>^c$} & 
\colhead{\loz} &
\colhead{\lox} }
\startdata
\cutinhead{\hwfc; $\mtll \ge 0.5$}
$\lbrack$1.20,2.00]&   22.9&    8.0&  13&1.85&$ 1.62^{+ 0.44}_{- 0.59}$&$ 0.57^{+ 0.15}_{- 0.20}$\\
$\lbrack$2.00,2.60]&   46.7&   14.7&  29&2.22&$ 1.97^{+ 0.36}_{- 0.44}$&$ 0.62^{+ 0.11}_{- 0.14}$\\
\cutinhead{\hwfc; $\mtll \ge 1$}
$\lbrack$1.20,2.00]&   32.5&   11.4&  12&1.80&$ 1.05^{+ 0.30}_{- 0.40}$&$ 0.37^{+ 0.10}_{- 0.14}$\\
$\lbrack$2.00,2.60]&   52.5&   16.5&  25&2.19&$ 1.51^{+ 0.30}_{- 0.37}$&$ 0.48^{+ 0.09}_{- 0.12}$\\
\cutinhead{\hwfc; $\mtll \ge 2$}
$\lbrack$1.20,2.00]&   43.1&   15.1&  15&1.79&$ 0.99^{+ 0.25}_{- 0.33}$&$ 0.35^{+ 0.09}_{- 0.12}$\\
$\lbrack$2.00,2.60]&   56.2&   17.7&  17&2.20&$ 0.96^{+ 0.23}_{- 0.29}$&$ 0.30^{+ 0.07}_{- 0.09}$\\
\cutinhead{\hacs; $\mtll \ge 2$}
$\lbrack$1.20,2.00]&   23.3&    8.4&   4&1.70&$ 0.48^{+ 0.23}_{- 0.38}$&$ 0.17^{+ 0.08}_{- 0.14}$\\
$\lbrack$2.00,2.60]&   17.4&    5.5&   5&2.16&$ 0.91^{+ 0.39}_{- 0.61}$&$ 0.29^{+ 0.12}_{- 0.19}$\\
\cutinhead{Combined \cite[with][]{ribaudo11}; $\mtll \ge 2$}
$\lbrack$1.20,2.00]&  108.0&   39.0&  29&1.77&$ 0.69^{+ 0.15}_{- 0.15}$&$ 0.24^{+ 0.05}_{- 0.05}$\\
$\lbrack$2.00,2.60]&  107.8&   34.8&  32&2.21&$ 0.92^{+ 0.18}_{- 0.18}$&$ 0.28^{+ 0.06}_{- 0.06}$\\
\enddata
\tablecomments{The cosmology assumed has $\Omega_\Lambda = 0.74, \Omega_m = 0.26$, and $H_0 = 72 \mkms \rm Mpc^{-1}$.  For the combined
values, we adopted the measurements from Table~7 of \cite{ribaudo11}
for the $z=[1.544,1.947]$ and $z=[1.947,2.594]$ bins.  We also 
converted their \lox\ measurements to our cosmology with a simple scaling.}
\tablenotetext{a}{Total redshift survey path.}
\tablenotetext{b}{Number of LLS discovered in the survey path.  Note that two LLS ocurring within $\approx 10,000 \mkms$ of one another have been treated as one system.}
\tablenotetext{c}{Mean absorption redshift of the LLS.}
 
\end{deluxetable}

\begin{deluxetable}{lcccccc}
\tablewidth{0pc}
\tablecaption{\hwfc\ STACKED SPECTRUM\label{tab:wfc3_stack}}
\tabletypesize{\footnotesize}
\tablehead{\colhead{\lrest} &\colhead{$f_\lambda^a$} & 
\colhead{$\sigma(f_\lambda)^b$}\\
(\AA)  }
\startdata
 601.92&0.46&0.09\\
 608.11&0.44&0.09\\
 614.30&0.42&0.09\\
 620.49&0.42&0.08\\
 626.68&0.43&0.08\\
 632.87&0.42&0.08\\
 639.06&0.42&0.08\\
 645.25&0.42&0.08\\
 651.44&0.42&0.08\\
 657.63&0.42&0.08\\
 663.82&0.42&0.07\\
 670.01&0.43&0.08\\
 676.20&0.43&0.07\\
 682.39&0.44&0.07\\
 688.58&0.44&0.07\\
 694.77&0.45&0.07\\
 700.96&0.46&0.07\\
 707.14&0.46&0.07\\
 713.33&0.44&0.07\\
 719.52&0.45&0.07\\
\enddata
\tablenotetext{a}{Flux per \AA\ normalized to unity at $\mlrest = 1450$\AA.}
\tablenotetext{b}{RMS (with $3\sigma$ clipping) in the flux from a bootstrap analysis (see text).}
\tablecomments{[The complete version of this table is in the electronic edition of the Journal.  The printed edition contains only a sample.]}
\end{deluxetable}

\begin{deluxetable}{ccclccrrcc}
\tablewidth{0pc}
\tablecaption{Stacked Spectrum Model Parameters and \lmfp \label{tab:best_mfp}}
\tabletypesize{\footnotesize}
\tablehead{\colhead{Parameter} &\colhead{Mean} & \colhead{Median} & \colhead{$1\sigma$ c.l.} &\colhead{$2\sigma$ c.l.}}
\startdata
$C_{\rm T}$  & 1.03&1.04&0.99,1.05&0.95,1.05\\
\dat  &$-0.37$&$-0.33$&$-0.60,-0.20$&$-0.80,-0.20$\\
\tlyman  &$ 0.29$&$ 0.28$&$ 0.24, 0.35$&$ 0.21, 0.44$\\
\gtL  &$ 1.64$&$ 1.73$&$ 1.40, 1.80$&$ 1.20, 1.80$\\
$\mkppo$  &$33.64$&$33.93$&$28.12,39.07$&$24.42,47.15$\\
$\gamma_\kappa$  &$ 0.42$&$ 0.00$&$ 0.00, 1.12$&$ 0.00, 2.55$\\
\zll  &$1.877$&$1.890$&$1.791,1.958$&$1.713,2.018$\\
\lmfp\ ($\mhmpc$) & 252.1&243.4&207.3,303.7&177.3,353.1\\
\enddata
\end{deluxetable}
 
\clearpage

\begin{deluxetable}{ccclccrrcc}
\tablewidth{0pc}
\tablecaption{Revised \lmfp\ Values from SDSS\label{tab:sdssmfp}}
\tabletypesize{\footnotesize}
\tablehead{\colhead{$<z_q>$} & \colhead{\lmfp} & \colhead{$\sigma(\mlmfp)^a$}\\
& (Mpc$^{-1}$) & (Mpc$^{-1}$) }
\startdata
3.73&54.9&4.1\\
3.78&45.3&3.6\\
3.83&45.4&3.9\\
3.89&47.1&4.2\\
3.96&38.9&3.2\\
4.08&33.4&3.0\\
4.23&27.8&2.2\\
\enddata
\tablecomments{Reevaluation of the \lmfp\ values using the stacked SDSS quasar spectra from \cite{pwo09}, but update for cosmology ($H_0=72 \mkms \, \rm Mpc^{-1}$; $\Omega_m = 0.26, \Omega_\Lambda = 0.74$) and the defintion of the mean free path applied in this paper.}
\tablenotetext{a}{Based on a bootstrap analysis. There is additional systematic error not included in this estimate.}
\end{deluxetable}
 
\clearpage

%% f(N)
%\input{../Tables/tab_fn_constraints.tex}

\begin{deluxetable}{ccccll}
\rotate
\tablewidth{0pc}
\tablecaption{\fnhi\ Constraints\label{tab:fn_constraints}}
\tabletypesize{\footnotesize}
\tablehead{\colhead{Constraint} &\colhead{$z^a$} & \colhead{log \nhi} & \colhead{Value$^b$} & \colhead{Comment}
& \colhead{Reference} }
\startdata
\cutinhead{Constraints for \fnhi\ at $z \approx 2.4$}
Lya Forest&2.34&12.50--13.00&$-11.19^{+  0.05}_{-  0.04}$&Recalculated for our Cosmology                       &K02  \\
&&13.00--13.50&$-11.76^{+  0.05}_{-  0.05}$\\
&&13.50--14.00&$-12.54^{+  0.07}_{-  0.07}$\\
&&14.00--14.50&$-13.30^{+  0.10}_{-  0.09}$\\
SLLS      &2.51&19.00--19.60&$-20.63^{+  0.13}_{-  0.13}$&Only 30 systems total                                &OPB07\\
&&19.60--20.30&$-21.50^{+  0.16}_{-  0.15}$\\
DLA       &2.51&20.30--20.50&$-21.83^{+  0.06}_{-  0.06}$&$z=[2.3,2.7]$; modest SDSS bias \citep[see][]{np+09}?&PW09 \\
&&20.50--20.70&$-22.29^{+  0.08}_{-  0.08}$\\
&&20.70--20.90&$-22.41^{+  0.08}_{-  0.07}$\\
&&20.90--21.10&$-22.88^{+  0.11}_{-  0.10}$\\
&&21.10--21.30&$-23.36^{+  0.15}_{-  0.15}$\\
&&21.30--21.50&$-23.60^{+  0.16}_{-  0.15}$\\
&&21.50--21.70&$-24.33^{+  0.34}_{-  0.30}$\\
&&21.70--21.90&$-25.00^{+  0.76}_{-  0.52}$\\
&&21.90--22.10&$-99.00^{+-99.00}_{--24.63}$\\
&&22.10--22.30&$-99.00^{+-99.00}_{--24.83}$\\
\tlox&2.23&$> 17.49$&$  0.30\pm  0.07$& &This paper\\
\tlya&2.40&12.00--17.00&$ 0.198\pm 0.007$&Converted to \tlya\ from $D_A$.  No LLS, no metals.&K05\\
\lmfp&2.44&12 -- 22&$ 243\pm 43\, \mhmpc $& &This paper\\
\cutinhead{Constraints for \fnhi\ at $z \approx 3.7$}
Lya Forest&3.75&13.65--14.05&$-12.42^{+  0.10}_{-  0.10}$&Read from Figure 4                                  &K01  \\
SLLS      &3.58&19.00--19.60&$-20.37^{+  0.12}_{-  0.12}$&$z=[3.1,4.5]$                                       &OPB07\\
&&19.60--20.30&$-21.22^{+  0.14}_{-  0.13}$\\
DLA       &3.61&20.30--20.50&$-21.68^{+  0.06}_{-  0.06}$&$z=[3.3,4.2]$; modest SDSS bias \citep[see][]{pwo09}&PW09 \\
&&20.50--20.70&$-21.98^{+  0.07}_{-  0.07}$\\
&&20.70--20.90&$-22.45^{+  0.10}_{-  0.10}$\\
&&20.90--21.10&$-22.71^{+  0.11}_{-  0.10}$\\
&&21.10--21.30&$-23.28^{+  0.17}_{-  0.16}$\\
&&21.30--21.50&$-23.53^{+  0.18}_{-  0.17}$\\
&&21.50--21.70&$-23.93^{+  0.24}_{-  0.22}$\\
&&21.70--21.90&$-24.53^{+  0.45}_{-  0.37}$\\
&&21.90--22.10&$-99.00^{+-99.00}_{--24.45}$\\
&&22.10--22.30&$-99.00^{+-99.00}_{--24.65}$\\
$\beta$&3.70&12.50--14.00&$-1.30\pm 0.15$&Shallower than z=2, but very uncertain.&See text\\
\tlox&3.76&$> 17.49$&$  0.52\pm  0.07$& &POW10\\
\tlya&3.70&12.00--19.00&$ 0.795\pm 0.062$&Higher than other evaluations.&FG08\\
\lmfp&3.69&12 -- 22&$  51\pm  5\, \mhmpc $& &PWO09\\
\enddata
\tablenotetext{a}{Effective redshift where the constraint was determined.}
\tablenotetext{b}{\fnhi\ constraints are given in log.}
\tablerefs{KT97: \cite{kt97}; K01: \cite{kim+01}; K02: \cite{kim02}; K05: \cite{kts+05}; OPB07: \cite{opb+07}; PW09: \cite{pwo09}}
\end{deluxetable}

\begin{deluxetable}{cccccccccccccc}
\tablewidth{0pc}
\tablecaption{Model and Correlation Matrix for the 5-Parameter $f(\mnhi,X)$\label{tab:corrmat}}
\tabletypesize{\footnotesize}
\tablehead{\colhead{Param} &\colhead{Best} & \colhead{$\sigma^a$}
& \colhead{$\delta k_{12}$}
& \colhead{$\delta \beta_{12}$}
& \colhead{$\delta \beta_{17.5}$}
& \colhead{$\delta \beta_{20.3}$}
& \colhead{$\delta \beta_{21.5}$}
}
\startdata
\cutinhead{Model for \fnhi\ at $z \approx 2.4$}
$k_{12}$&$ -9.52$&0.07&$  1.00$&$ -0.91$&$  0.48$&$ -0.06$&$  0.01$\\
$\beta_{12}$&$ -1.67$&0.02&$ -0.91$&$  1.00$&$ -0.70$&$  0.09$&$ -0.02$\\
$\beta_{17.5}$&$ -1.07$&0.03&$  0.48$&$ -0.70$&$  1.00$&$ -0.53$&$  0.11$\\
$\beta_{20.3}$&$ -1.71$&0.12&$ -0.06$&$  0.09$&$ -0.53$&$  1.00$&$ -0.30$\\
$\beta_{21.5}$&$-11.10$&7.54&$  0.01$&$ -0.02$&$  0.11$&$ -0.30$&$  1.00$\\
\enddata
\tablenotetext{a}{Evaluated from the diagonal of the covariance matrix.}
\end{deluxetable}

%%%%%%%%%%%%%%%%%%%%%%%%%%%%%%%%%%%%%%%%%%%%%%%%%%%%%%%%%
%%%%%%%%%%%%%%%%%%%%%%%%%%%%%%%%%%%%%%%%%%%%%%%%%%%%%%%%%
%%%%%%%%%%%%%%%%%%%%%%%%%%%%%%%%%%%%%%%%%%%%%%%%%%%%%%%%%

\begin{figure}
\includegraphics[width=5in]{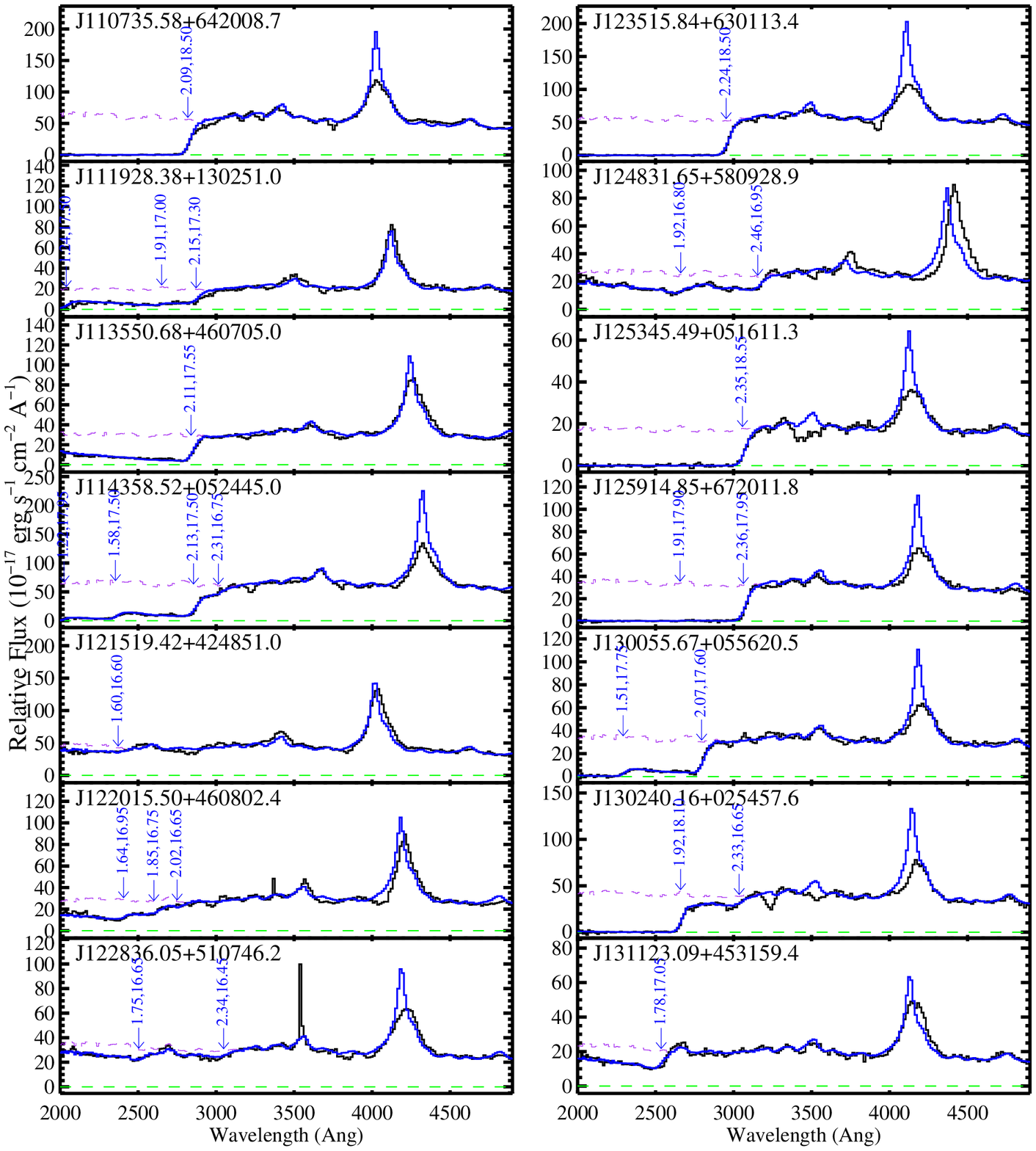}
\end{figure}

\begin{figure}
\includegraphics[width=5in]{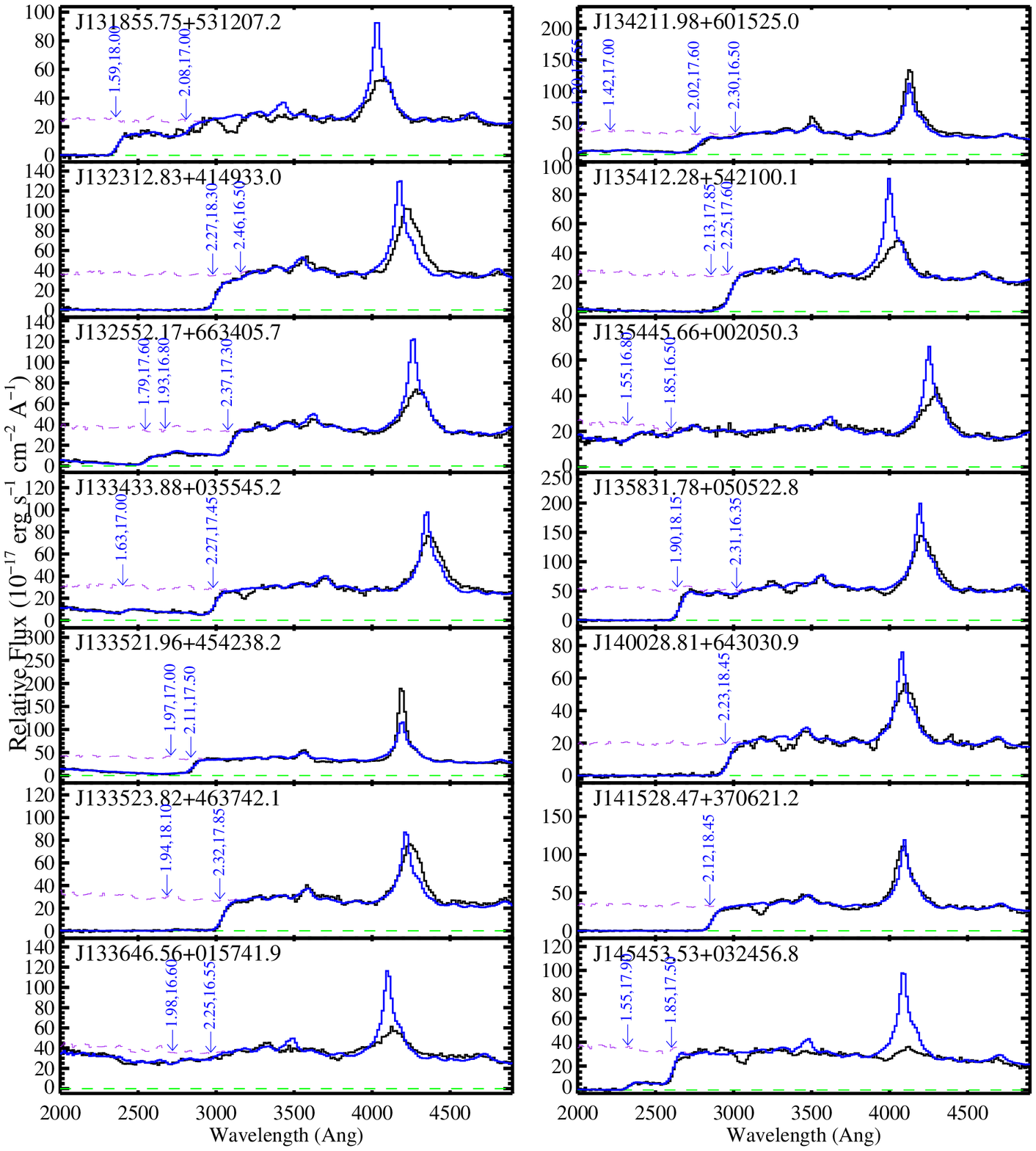}
\end{figure}

\begin{figure}
\includegraphics[width=5in]{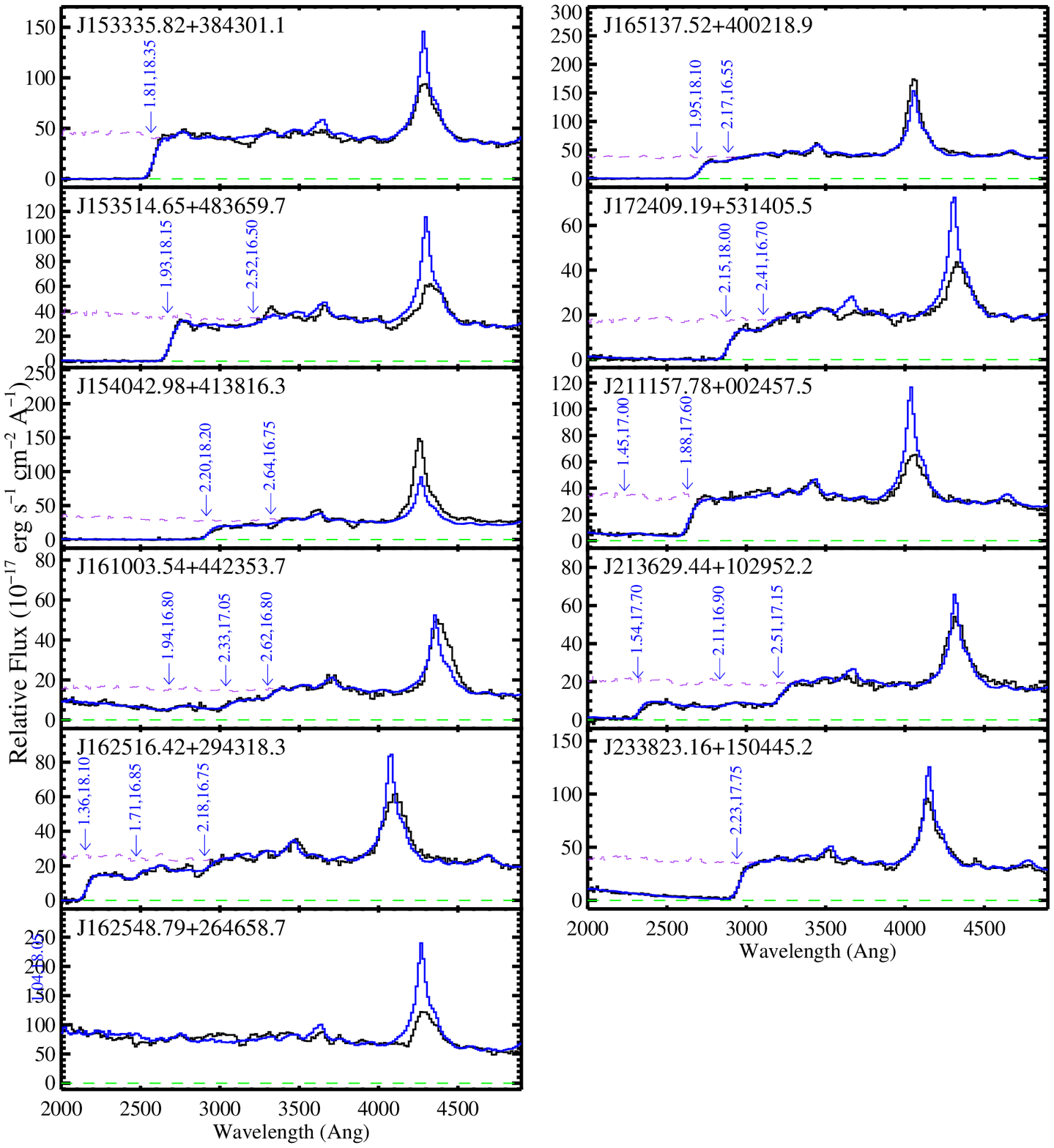}
\end{figure}

\begin{figure}
\includegraphics[width=5in]{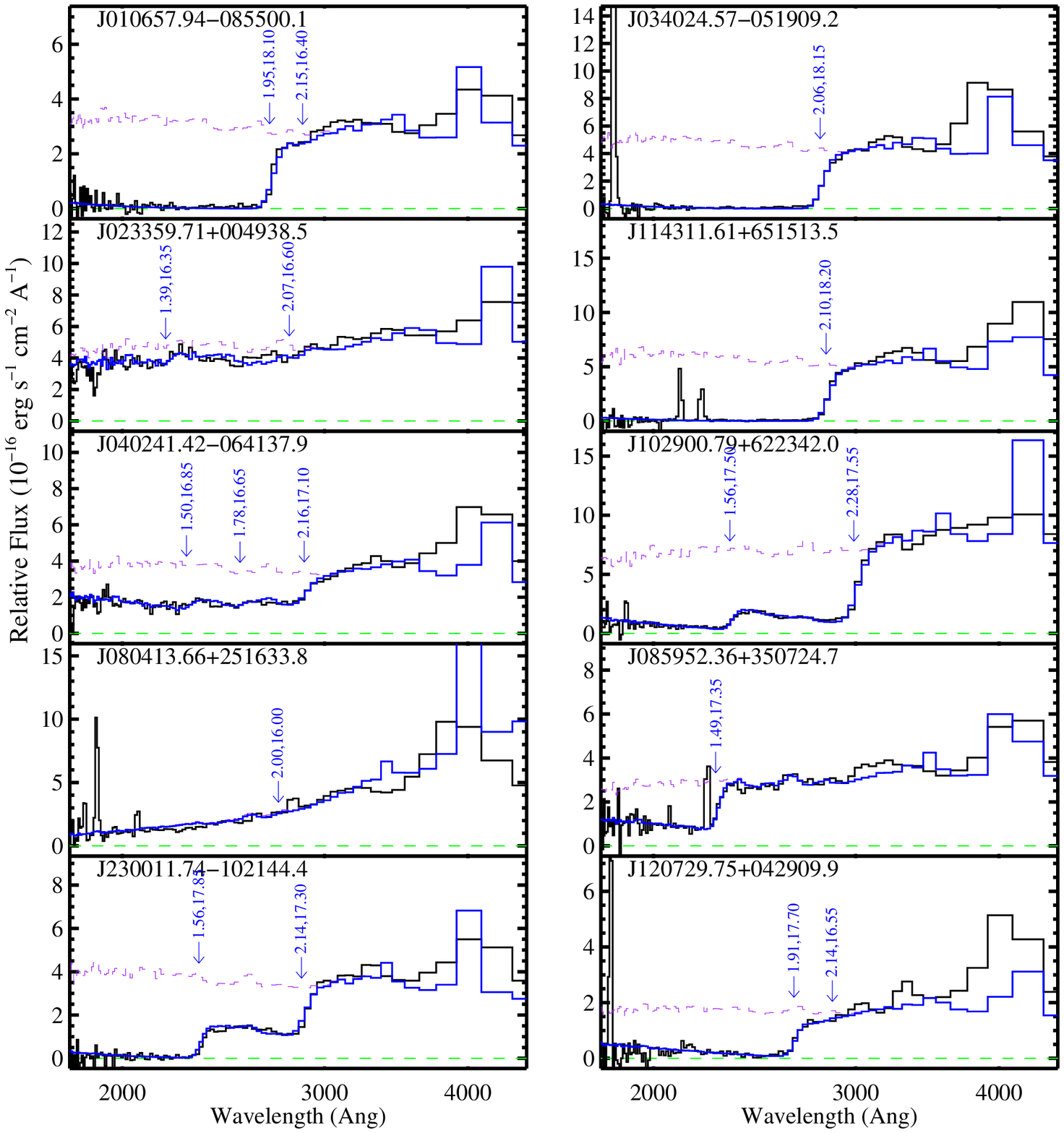}
\caption{Same as for Figure~\ref{fig:wfc3_models} but for the \acs\
  observations. Note that the x-axis has a logarithmic scaling.
}
\label{fig:acs_models}
\end{figure}

\begin{figure}
\includegraphics[width=5in]{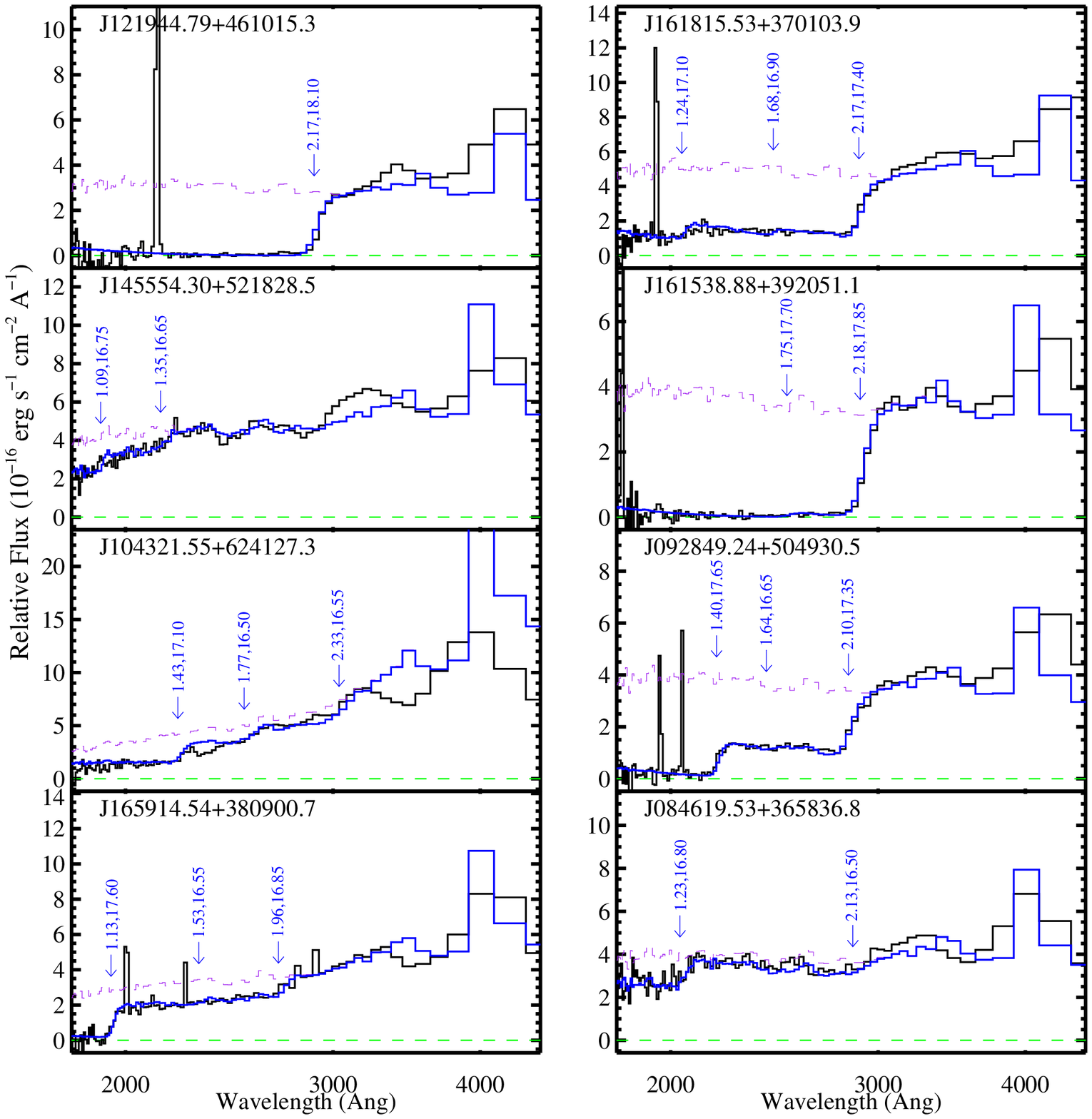}
\end{figure}

\clearpage
\begin{figure}
\includegraphics[width=5in]{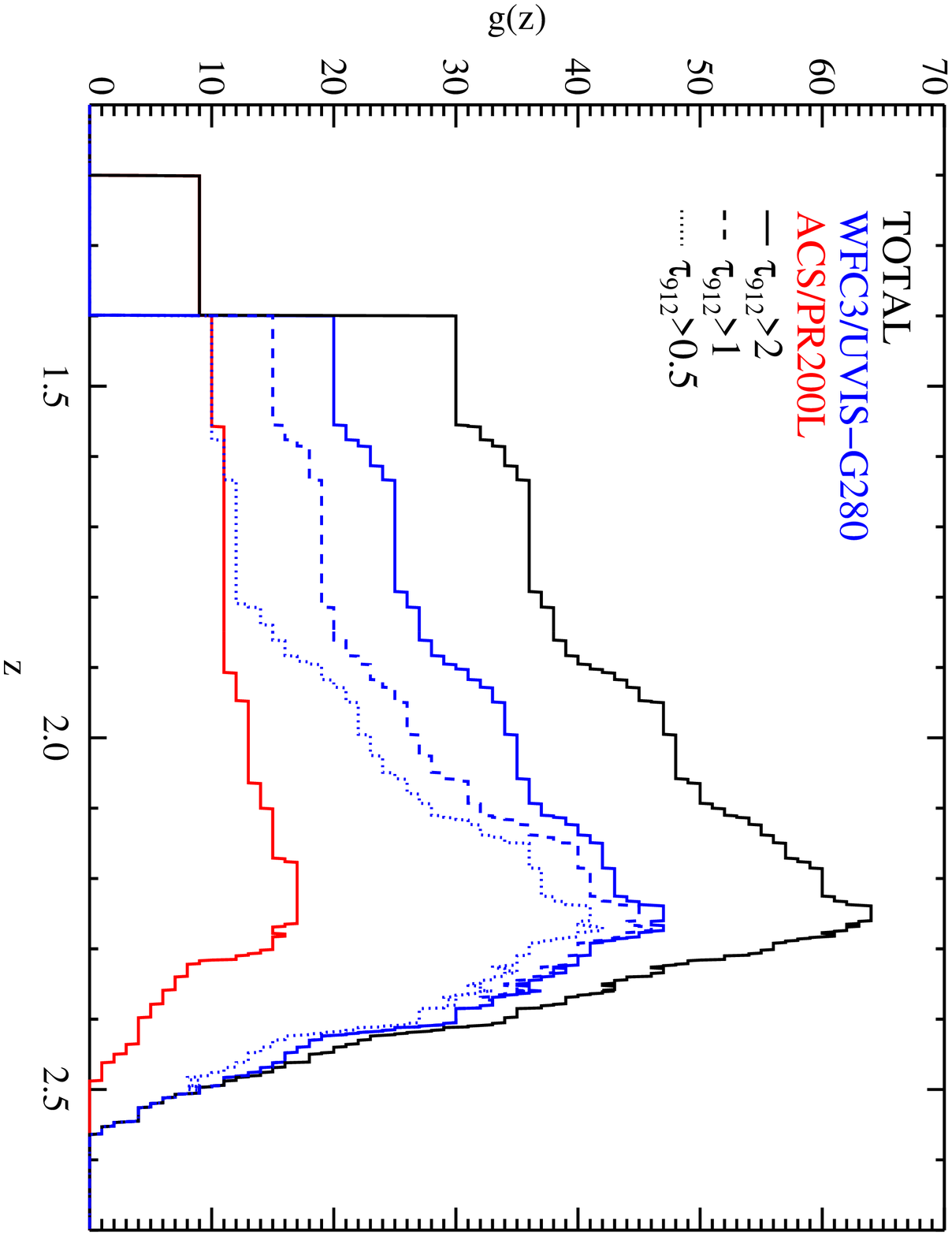}
\caption{These $g(z)$ curves describe the survey size as a function of
  redshift.  Specifically, $g(z)$ expresses the number of quasars at a
  given $z$ that
  were surveyed for LLS absorption to a limiting optical depth at
  $\lambda_r = 912$\AA.  They decrease with decreasing redshift
  because the presence of an LLS along a sightline often
  precludes the search for additional LLSs.
  The solid black curve shows the combined
  $g(z)$ functions for the \hacs\ and \hwfc\ datasets for $\tau_{912} \ge 2$.
}
\label{fig:goz}
\end{figure}

\begin{figure}
\includegraphics[width=5in]{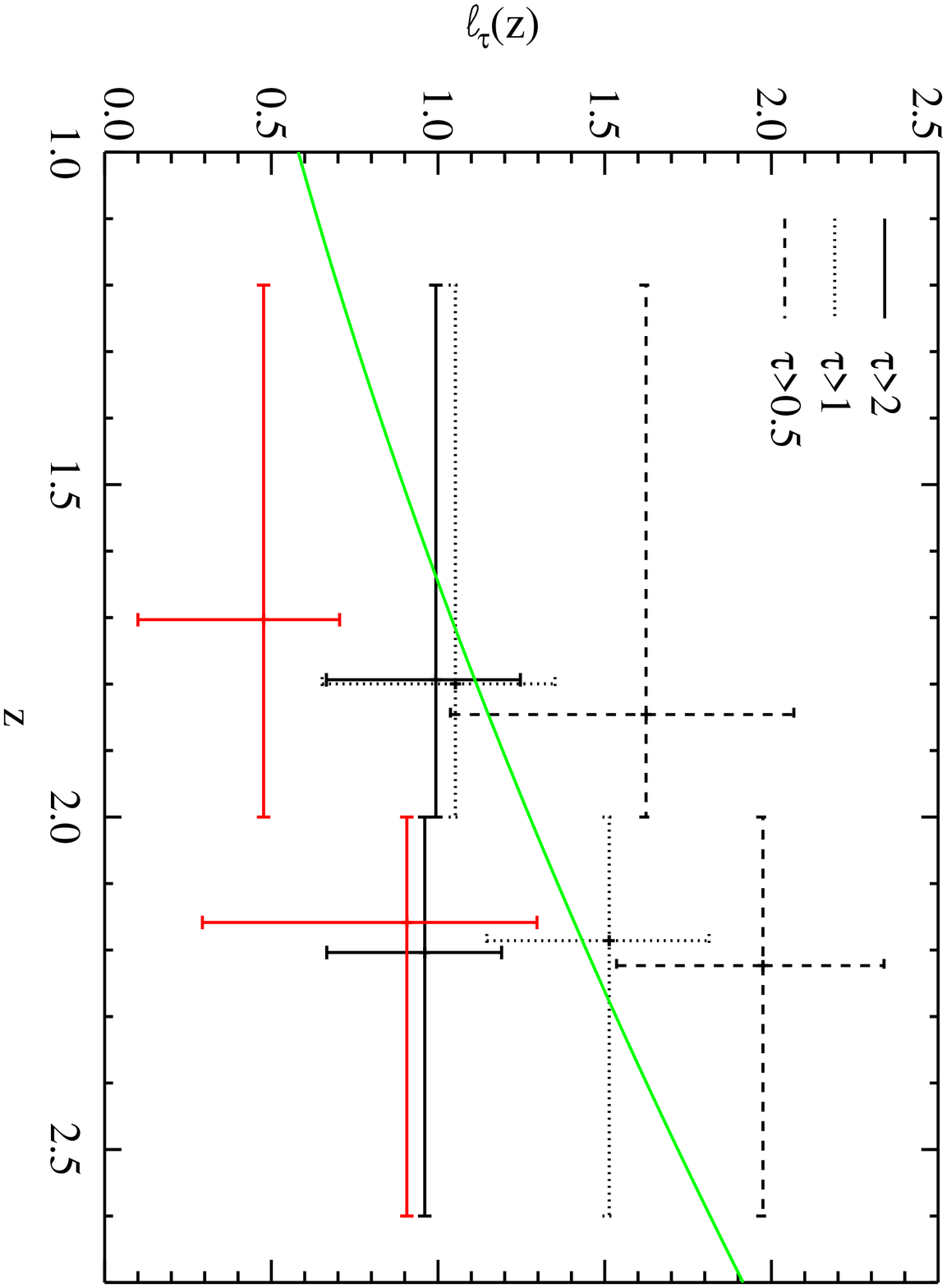}
\caption{Incidence of LLSs per unit redshift \loz, as estimated from
  our {\it HST} datasets.  The \wfc3\ data (black points)
  are shown for 3 limiting 
  optical depths \tlim.  The red (solid) points are for the \acs\ sample
  and $\mtlim \ge 2$.  For each limiting optical depth, the data
  suggest only a modest (if any) evolution in \loz\ with redshift.
  The solid green line represents to the fit by \citet{songaila10} for
  $\mtll \ge 1$.
}
\label{fig:loz}
\end{figure}

\begin{figure}
\includegraphics[width=5in]{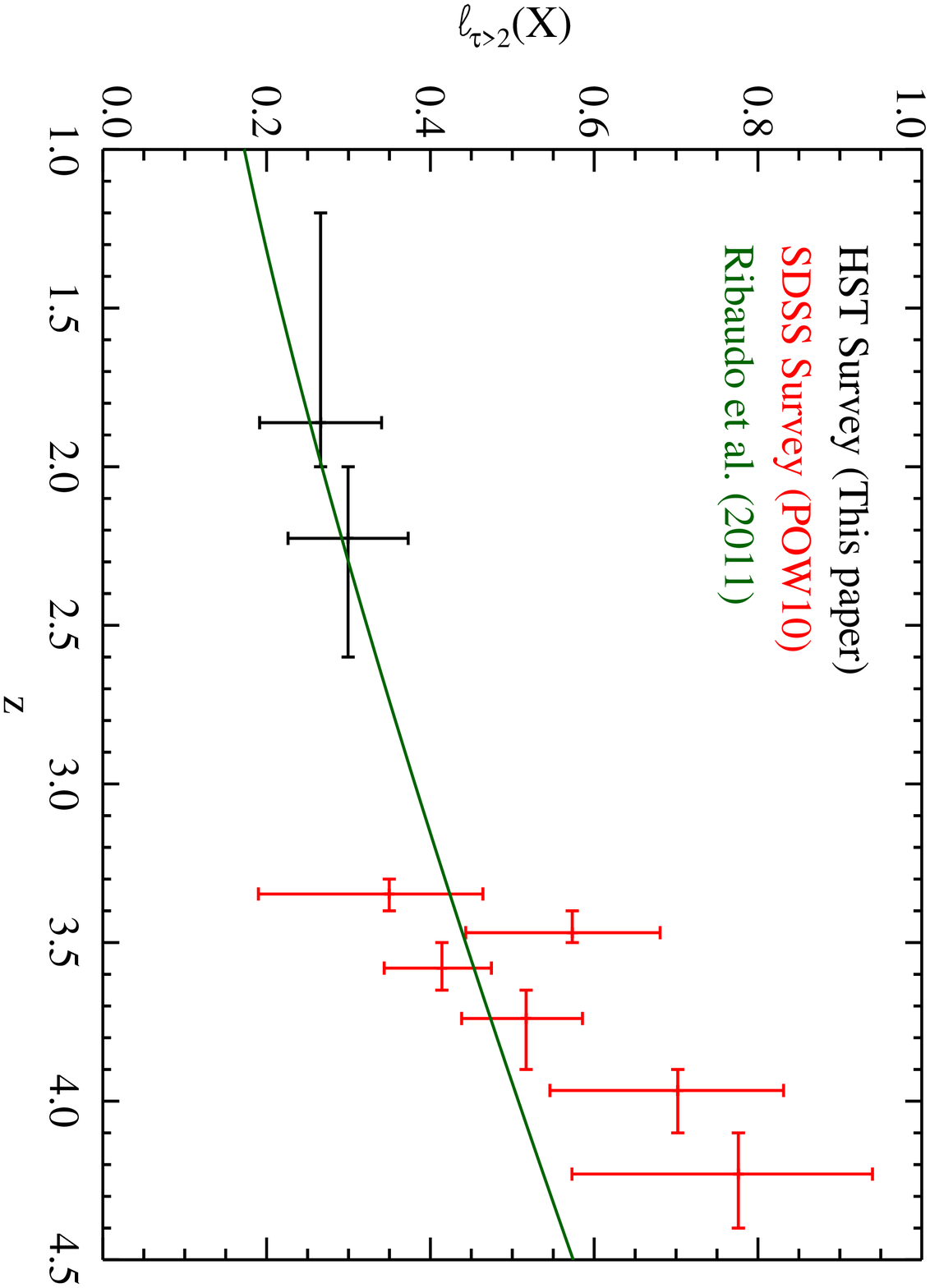}
\caption{Incidence of LLSs per unit
  path length \lox\ restricted to $\mtlim > 2$, as estimated
  from our combined \hwfc\ and \hacs\ surveys.  Shown in
  solid green is the \lox\ curve (converted to the cosmology adopted
  in this paper) of \cite{ribaudo11} 
  derived from a single power-law
  fit to archival analysis of lower $z$ {\it HST} data and high
  $z$ observations \citep{pow10,songaila10}. 
  Their results are in good agreement with our
  estimates at $z<2.5$ and together the data indicate mild
  evolution in \lox\ at low $z$.
}
\label{fig:lox}
\end{figure}

\begin{figure}
\includegraphics[width=5in]{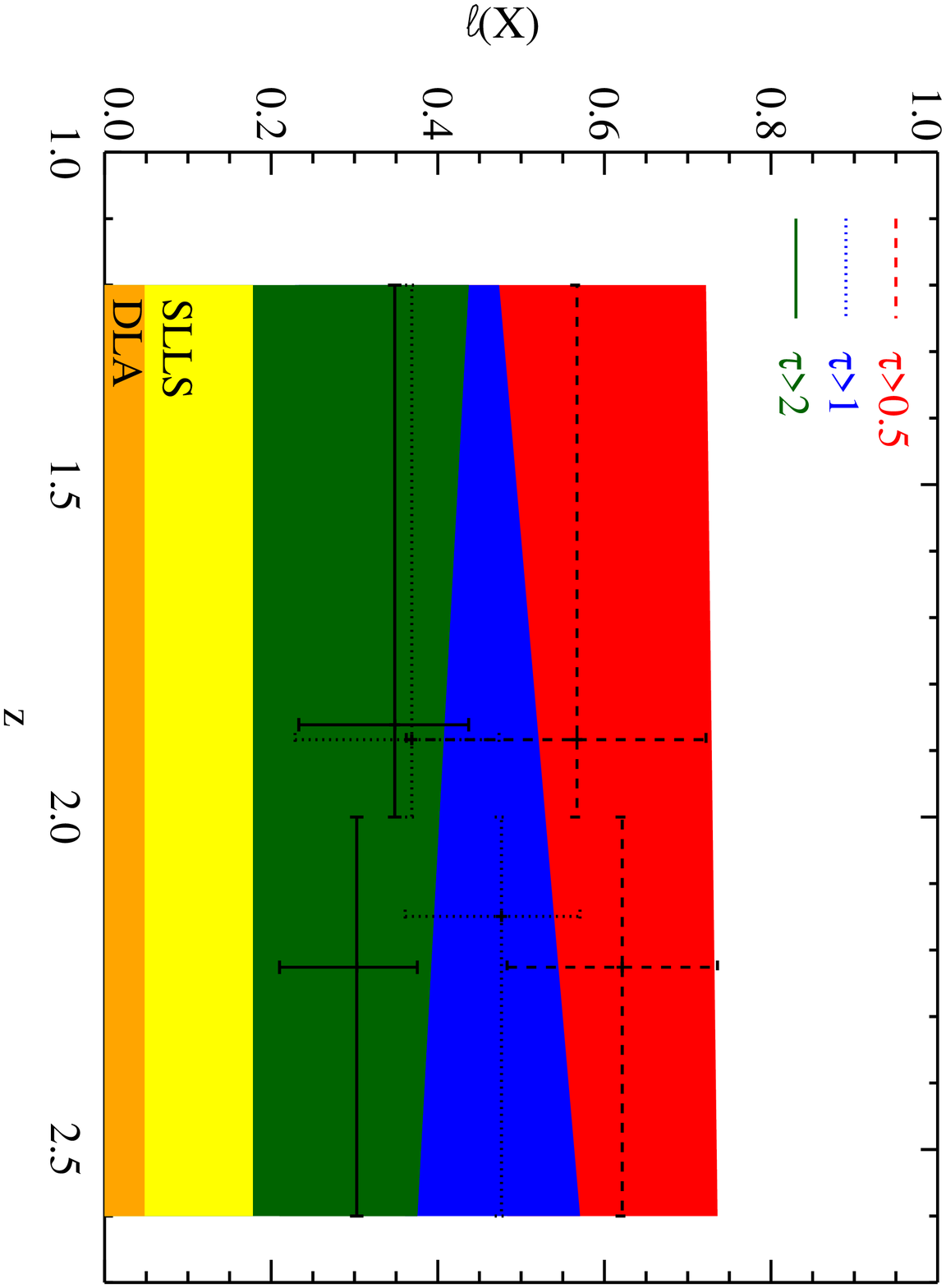}
\caption{The points (and error bars) show the incidence per unit
  absorption length \lox\ for LLSs with limiting optical depths $\mtlim
  = 0.5 - 2$, for the \hwfc\ sample only.  The red, blue, and green
  shaded regions express these same quantities.  The orange shaded
  region shows an estimate for the incidence of DLAs taken from \cite{pw09}.
  Similarly, the yellow shaded region gives the 
  contribution of super Lyman limit systems (SLLSs) to \lox\
  \citep{opb+07}.  Together, systems with $\mnhi > 10^{19} \cm{-2}$
  have $\mlox \approx 0.18$, implying a contribution of $50-80\%$ to
  \tlox.
}
\label{fig:complox}
\end{figure}

\begin{figure}
\includegraphics[width=5in]{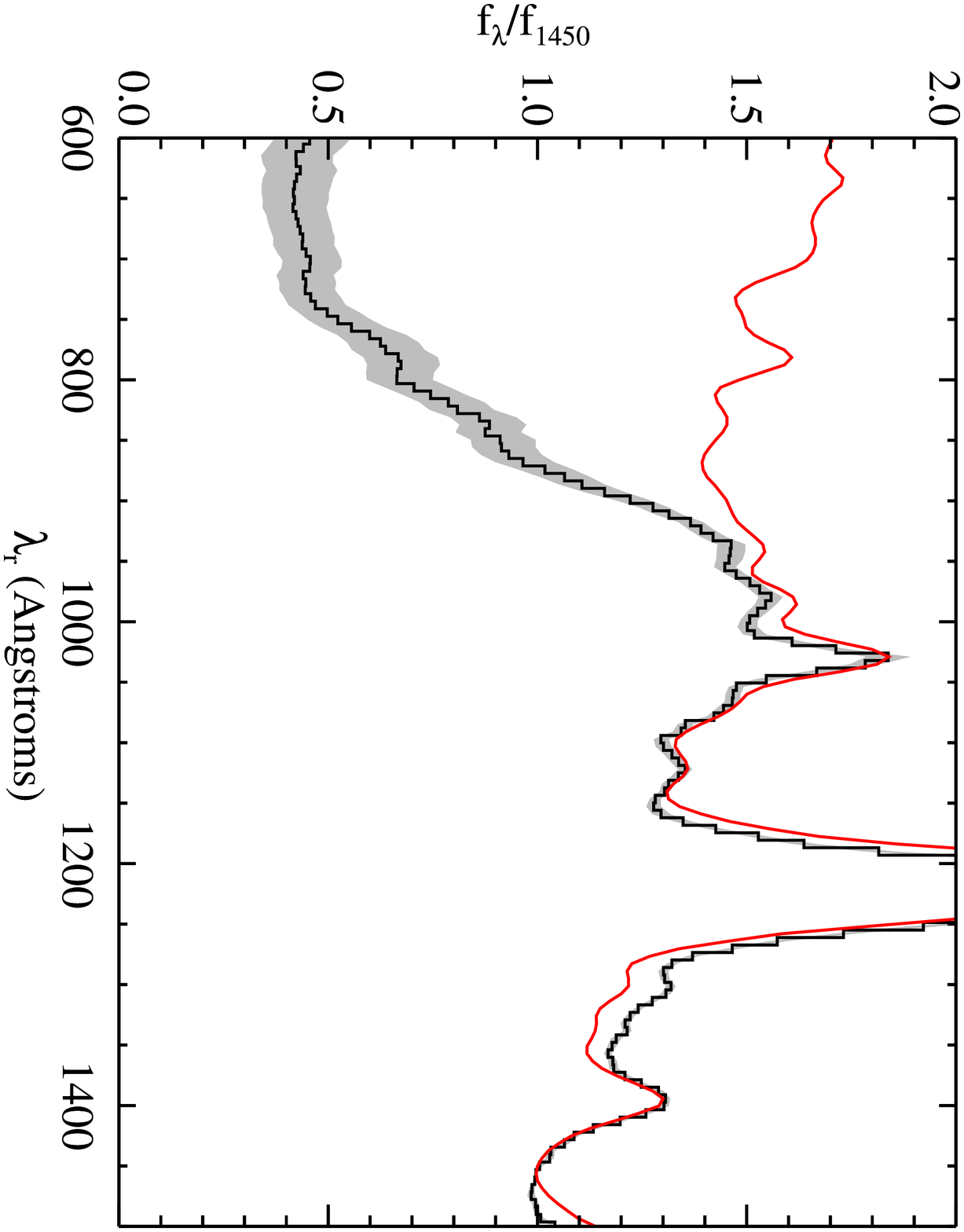}
\caption{The black histogram shows the average quasar spectrum at
  $z=\zmedian$ as estimated from a stack of the \wfc\ spectra, each
  normalized to unit flux at $\mlrest = 1450$\AA.  Note the strong
  decrease in flux at $\mlrest < 900$\AA\ due to the onset and cumulative
  effect of Lyman limit opacity.  The gray shaded region shows the RMS
  at each pixel as measured from standard bootstrap analysis.  The
  solid red curve shows the radio-quiet average quasar spectrum of
  \cite{telfer02}, smoothed to the resolution and dispersion of the
  \hwfc\ stack.  Their estimate of the intrinsic quasar flux, when
  normalized at $\mlrest = 1450$\AA, lies below the \hwfc\ stack at
  $\mlrest \approx 1300$\AA\ and would imply an effective opacity at \lya\
  \tlya\ that is significantly smaller than previous estimates.  We
  infer, therefore, that our \hwfc\ quasar cohort has an average SED
  that is bluer (harder) than the \cite{telfer02} sample at far-UV
  wavelengths.  
}
\label{fig:wfc3_stack}
\end{figure}

\begin{figure}
\includegraphics[width=5in]{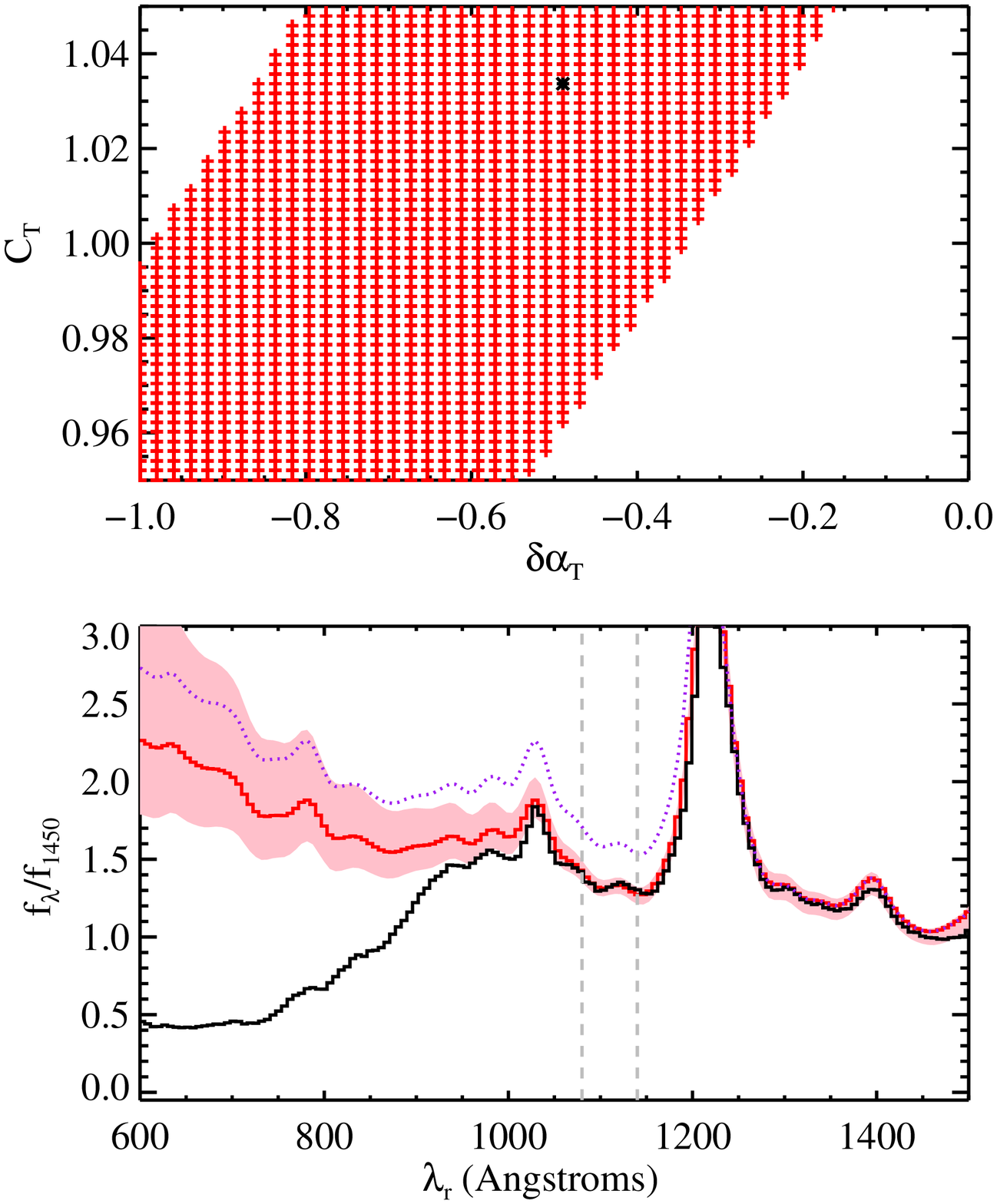}
\caption{Upper panel: Allowed values for the normalization $C_{\rm T}$
  and tilt \dat\ of the \cite{telfer02} quasar SED (relative to the
  Telfer et al.\ 2002 spectrum) that give the
  effective \lya\ opacity \tlya\ measured at $z=2.14$ by
  \cite{tytler04}.  
  The black asterisk denotes the model that minimizes $\chi^2$ from
  $\lambda_r = 1080-1140$\AA; this
  model of the intrinsic spectrum is shown in the lower panel.
  We find that the \hwfc\ stack favors a bluer SED ($\mdat < 0$) than 
  the Telfer spectrum. 
  Lower panel:  The black histogram is the \hwfc\ stacked spectrum and
  the dotted curve is the Telfer spectrum scaled and tilted with the
  `best' $C_{\rm T}$ and \dat\ values (black asterisk in the upper panel).
  The solid red line shows this intrinsic SED attenuated by a
  constant \tlya=0.19 for $\lambda < \lambda_{\rm Ly\alpha}$. % Checked on 8/9/2011 by JXP  
  The shaded region shows the range of models
  corresponding to the range of $C_{\rm T}$ and \dat\ values in the
  upper panel.
  The gray dashed lines indicate the region where $D_A$ was evaluated.
}
\label{fig:DA}
\end{figure}

%\begin{figure}
%%\includegraphics[width=5in,angle=90]{Figures/fig_cumul_tlyman.ps}
%\caption{The effective Lyman series opacity \tlyman, plotted in
%  cumulative form as a function of the Lyman series transition (n=2
%  corresponds to \lya).  The curves correspond to the set of \fnhi\
%  models list in Table~\ref{tab:mocks}.  In each case, the circles
%  indicate where one achieves 50\%\ of the total.  The results
%  indicate the Lyman series transitions with $n \gtrsim 4$ tend to
%  contribute a significant fraction of \tlyman.
%}
%\label{fig:cumul_tau_lyman}
%\end{figure}

\begin{figure}
\includegraphics[width=5in]{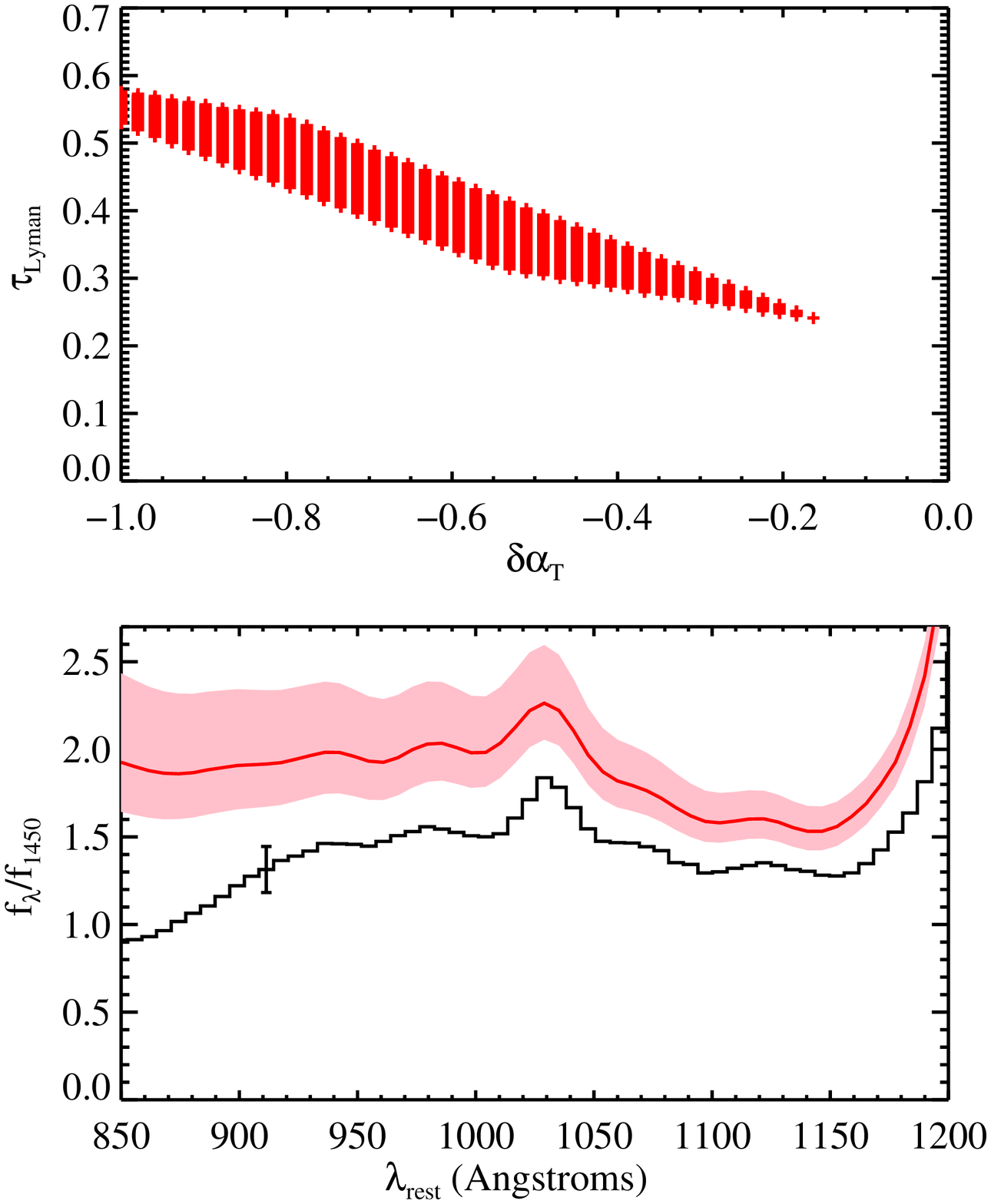}
\caption{Upper panel: Estimations of the effective Lyman series
  opacity \tlyman\ as a function of the tilt in the Telfer quasar
  spectrum \dat.  Naturally, one estimates larger \tlyman\ values for
  steeper quasar SEDs.  We estimate $\mtlyman \approx 0.4$ at $z=\zmedian$.  
  Lower panel: The black histogram shows the stacked \hwfc\ quasar
  spectrum with the error bar at $\mlrest = 912$\AA\ showing a 10\%\ error
  estimate in the average observed flux at that wavelength.  
  The solid red curves and the shaded region show the range
  of quasar SEDs that satisfy the \tlya\ constraint
  (Figure~\ref{fig:DA}).  These are used in the estimates of \tlyman\
  shown in the upper panel.
}
\label{fig:wfc3_tlyman}
\end{figure}

\begin{figure}
\includegraphics[width=5in]{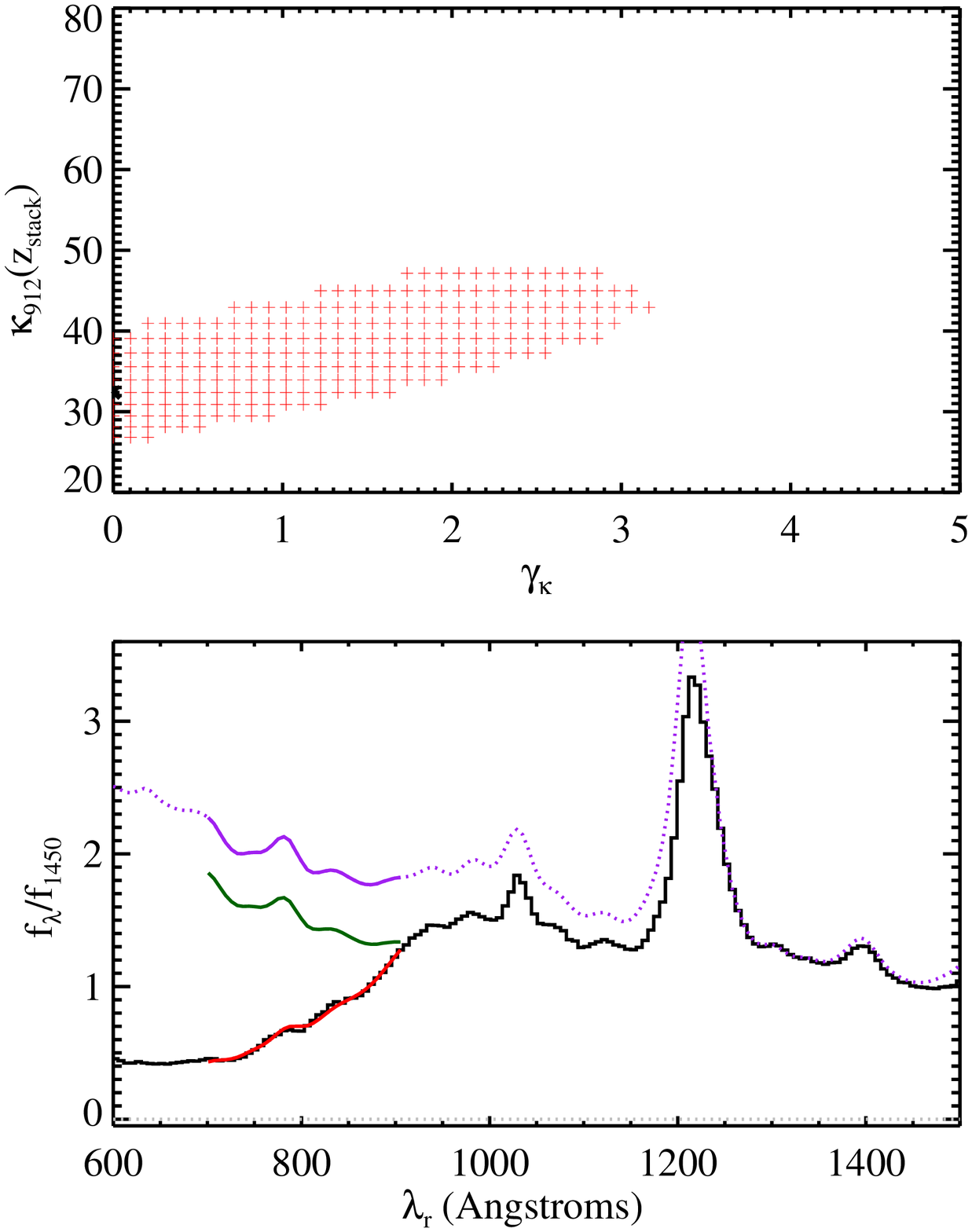}
\caption{Upper panel: Estimated constraints on \ztkll($z_{\rm stack}$)
  and $\gamma_\kappa$, our model parameters for the opacity of the
  universe to ionizing radiation at $z_{\rm stack} = \zmedian$
  (Equation~\ref{eqn:kLL}).  These parameters are highly
  degenerate with one another.
  Lower panel:  The black histogram is the \hwfc\ stacked quasar
  spectrum, normalized at $\mlrest = 1450$\AA.  The purple dotted curve
  shows the best estimate of the intrinsic quasar continuum, i.e.\ the
  scaled and tilted Telfer spectrum as given in
  Table~\ref{tab:best_mfp}.  The solid green curve gives the intrinsic
  SED attenuated by Lyman series opacity \tlyman, constrained to match
  the stacked spectrum at 912\AA\ to within 10\%\ and assumed to
  decrease as $(1+z)^{\gamma_\tau}$.  Lastly, the solid red curve is
  the complete model that also includes the effective Lyman limit
  opacity \tll.
}
\label{fig:wfc3_kappa}
\end{figure}

\begin{figure}
\includegraphics[width=5in]{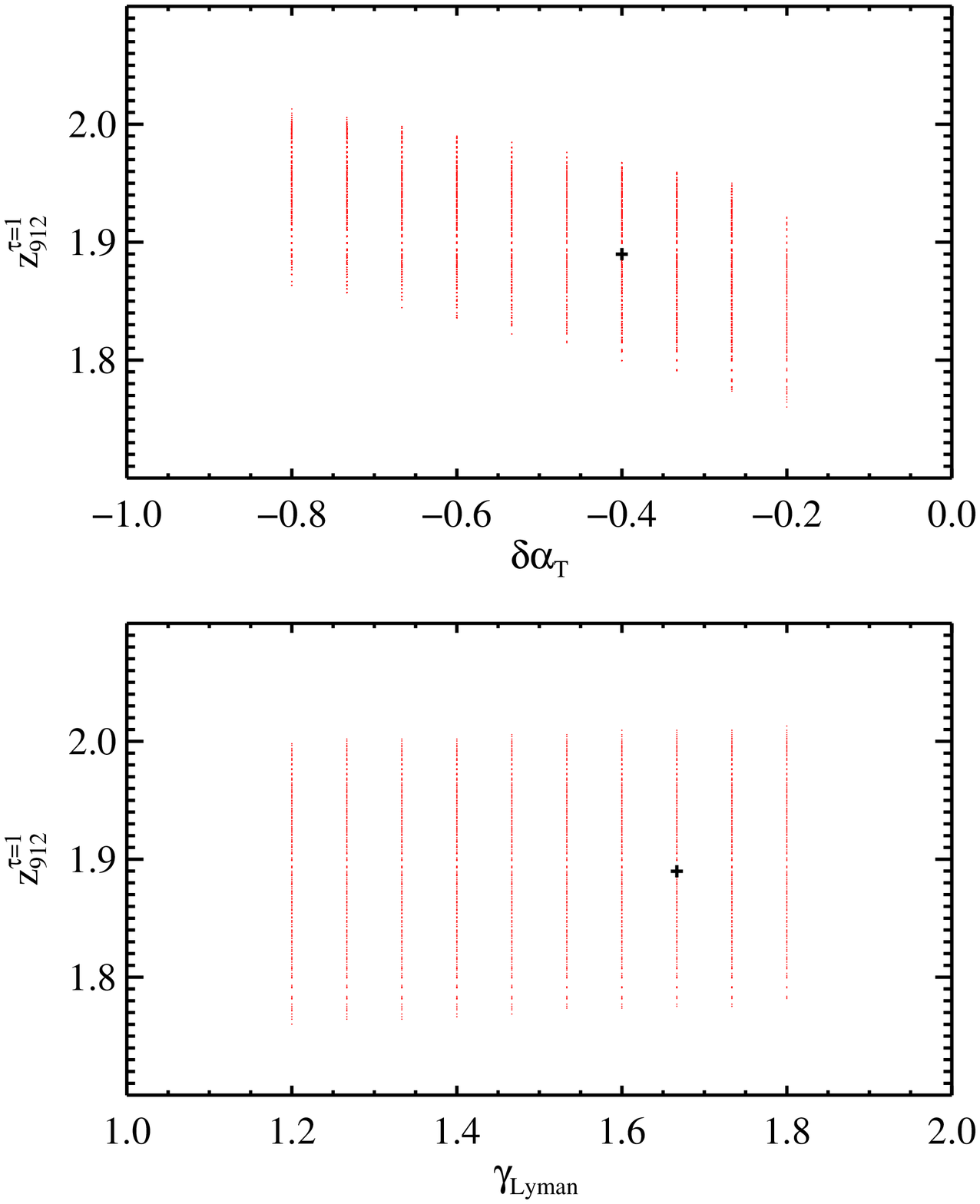}
\caption{Each panel shows a crude estimate of the allowed \ozll\
  values corresponding to \tll=1 from our analysis of the \hwfc\
  stacked quasar spectrum, as a function of (upper) the quasar tilt
  \dat\ and (lower) the assumed redshift evolution for \tlyman.  The
  plus sign indicates values from the `best fit' model shown in
  Figure~\ref{fig:wfc3_kappa}.  Aside from \dat\ values near zero, we
  find $z_{912}=1.85-2.1$ implying \lmfp$\approx 150 - 300 \mhmpc$.
}
\label{fig:wfc3_z912}
\end{figure}

\begin{figure}
\includegraphics[width=5in]{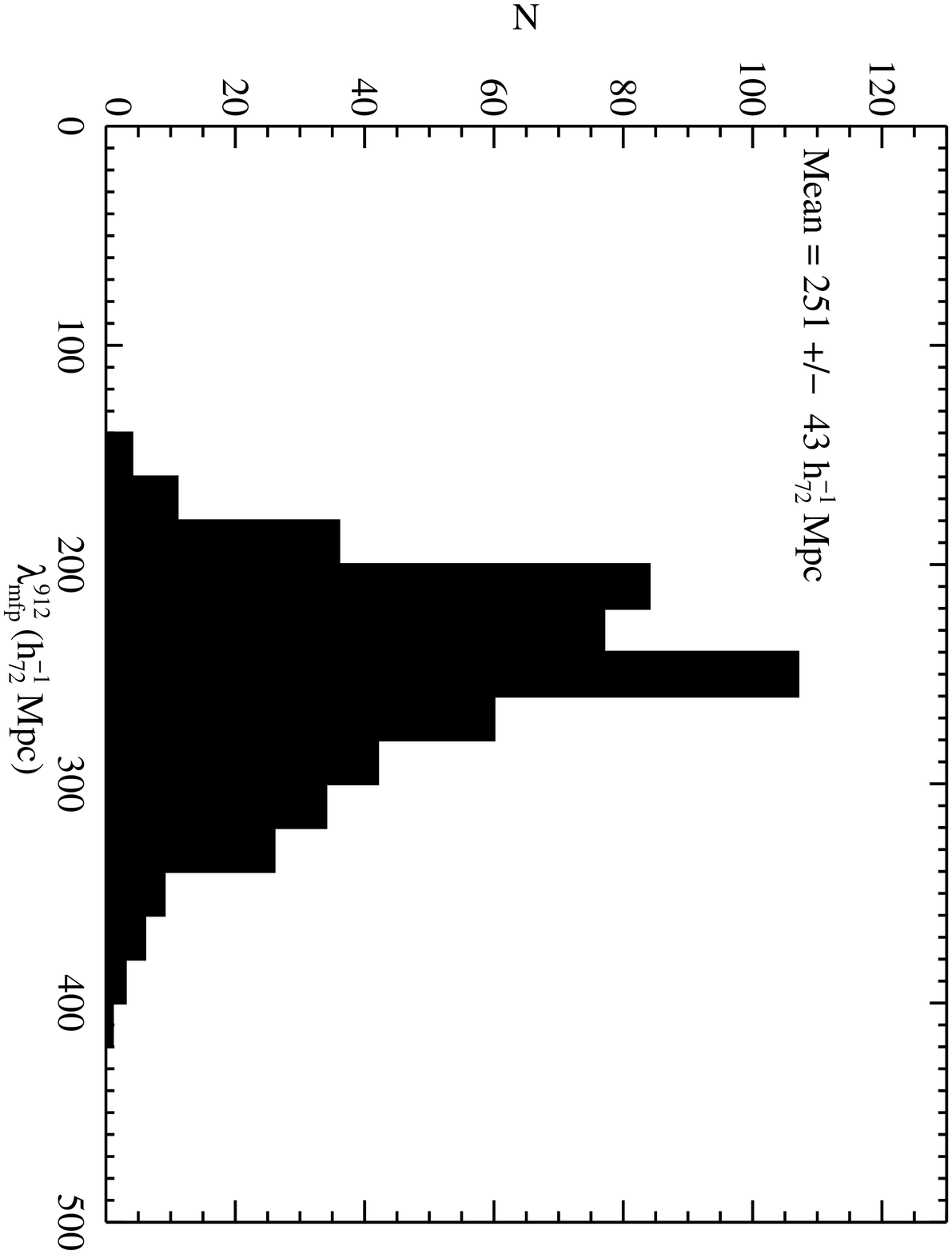}
\caption{Bootstrap analysis of \lmfp\ measured from the \hwfc\ stacked
  quasar spectrum (\zem=2.44) and using the
  techniques describe in $\S$~\ref{sec:stack}.  The values peak at
  $\approx 250 \mhmpc$ with a non-Gaussian tail extending to larger
  values. 
}
\label{fig:mfp_boot}
\end{figure}

%%%%%%%%%%%%%%%%%%%%%%%%%%%%%%%%%
\begin{figure}
\includegraphics[width=5in]{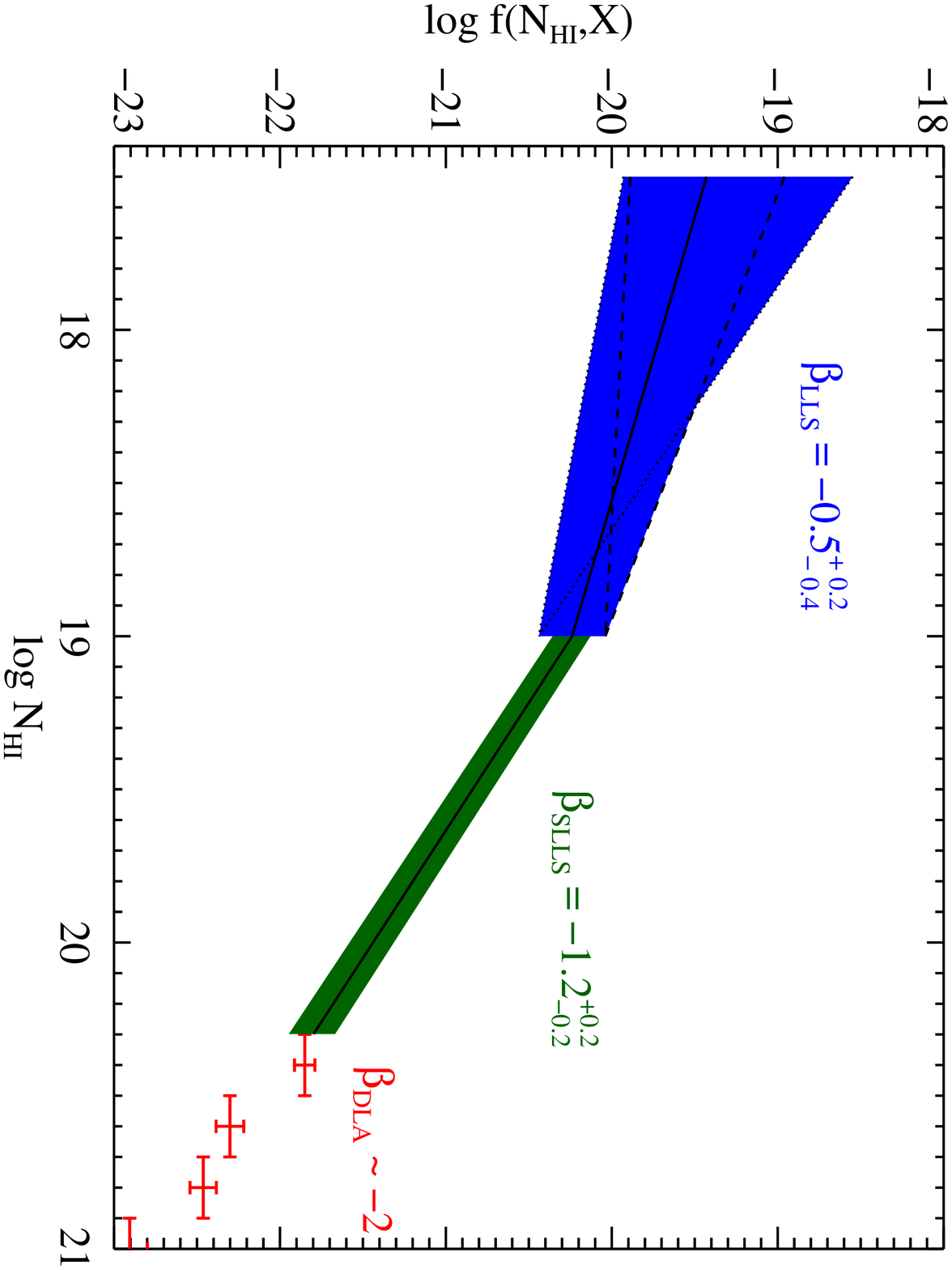}
\caption{Estimates of the \nhi\ frequency distribution \fnhi\ for
  systems with $\mnhi > 10^{17.5} \cm{-2}$ at $z\approx 2.4$.  The
  measurements for the super Lyman limit systems (SLLSs) and damped \lya\
  systems (DLAs) were taken from \cite{opb+07} and \cite{pw09}
  respectively.  
  The blue band is an estimate of \fnhi\ for LLSs having 
  $\mnhi = 10^{17.5} - 10^{19} \cm{-2}$ for an assumed power-law
  $f(\mnhi,X) \propto \mnhi^{\beta_{\rm LLS}}$
  and constrained by the observed incidence of SLLSs and \tlox\ (this
  paper). We find $\mbtlls = -0.6^{+0.2}_{-0.3}$ (68\%\ c.l.) for
  conservative estimates on the value of \fnhi\ at $\mnhi = 10^{19} \cm{-2}$
  and allowing for the uncertainty in \tlox.
  The dashed and dotted curves indicate the range of power-laws that
  satisfy the observations.
}
\label{fig:fn_high}
\end{figure}

\begin{figure}
\includegraphics[width=5in]{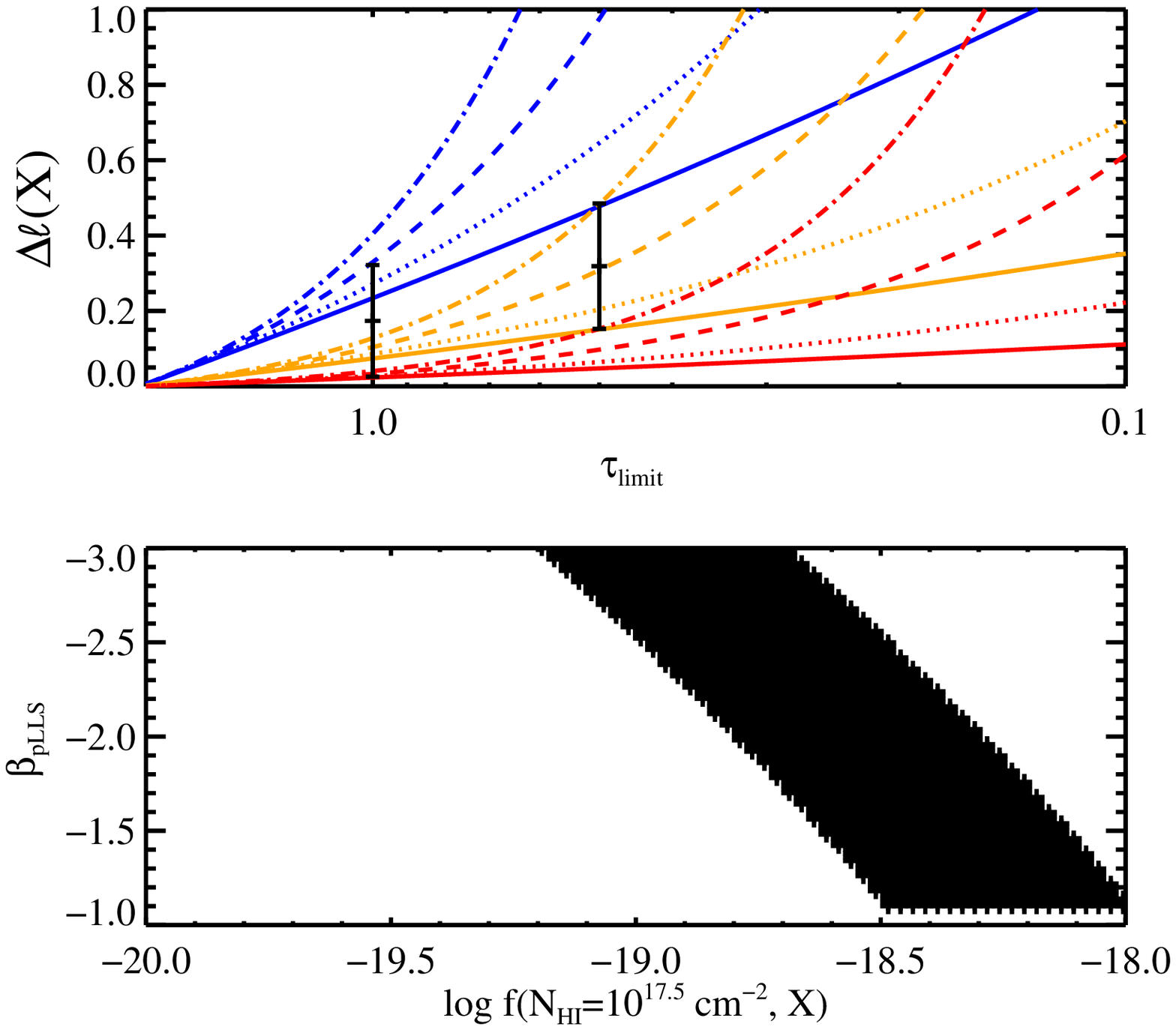}
\caption{Upper panel: The points with error-bars show the measured
  offset $\Delta \mlox$ in incidence of LLS \lox\ from the $\mtlim >2$
  estimate plotted at the limiting optical depth.  The curves show a
  series of model estimates for $\Delta \mlox$ as a function of \tlim\
  for an assumed $f(\mnhi=10^{17.5} \cm{-2}, X)$ of 
  $10^{-19} \cm{2}$ (red), 
  $10^{-18.5} \cm{2}$ (orange),  and
  $10^{-18} \cm{2}$ (blue). 
  We also assume a range of slopes \bplls\ for an assumed single
  power-law of $-1$ (solid), $-1.5$ (dotted), $-2$ (dashed) and $-2.5$
  (dash-dot).  Models with $\mfnlls < 10^{-19} \cm{2}$ are ruled out
  by the observations.  Similarly, models with $\mfnlls \approx
  10^{-18.5} \cm{2}$ prefer $\mbplls < -1.5$.  
  Lower panel: Permitted values for \bplls\ and $\mfnlls$ for an
  assumed single power-law (Equation~\ref{eqn:power_pLLS}) that match
  the observations within the $1\sigma$ uncertainty.  Unless one
  adopts a large $\mfnlls$ value, the observations require a very
  steep power-law ($\mbplls < -1.5$).
}
\label{fig:fn_pLLS}
\end{figure}

\begin{figure}
\includegraphics[width=5in]{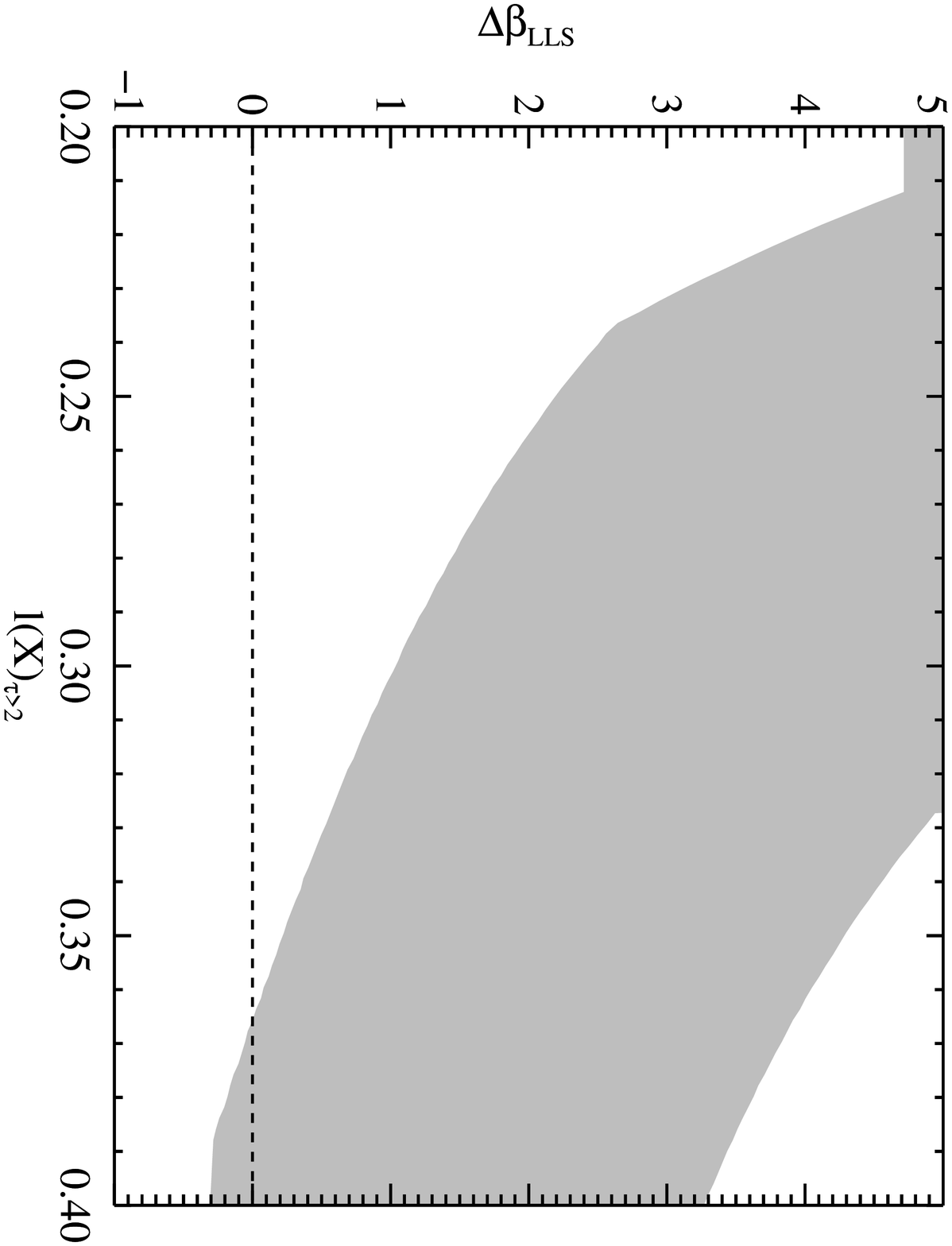}
\caption{Estimates for the change in slope \dbeta\ of the \nhi\
  frequency distribution from $\mnhi \approx 10^{18} \cm{-2}$ to
  $10^{17} \cm{-2}$ as inferred from the incidence of LLS at $z
  \approx 2$ in our {\it HST} survey.  Results are plotted as a
  function of the incidence of $\tau > 2$ LLS, \tlox.   For our
  preferred value $\mtlox = 0.29 \pm 0.05$, the data indicate a
  significant steepening of \fnhi\, $\mdbeta > 1$, as one transitions
  from optically thick gas to optically thin regions. 
}
\label{fig:fn_dbeta}
\end{figure}

\begin{figure}
\includegraphics[width=5in]{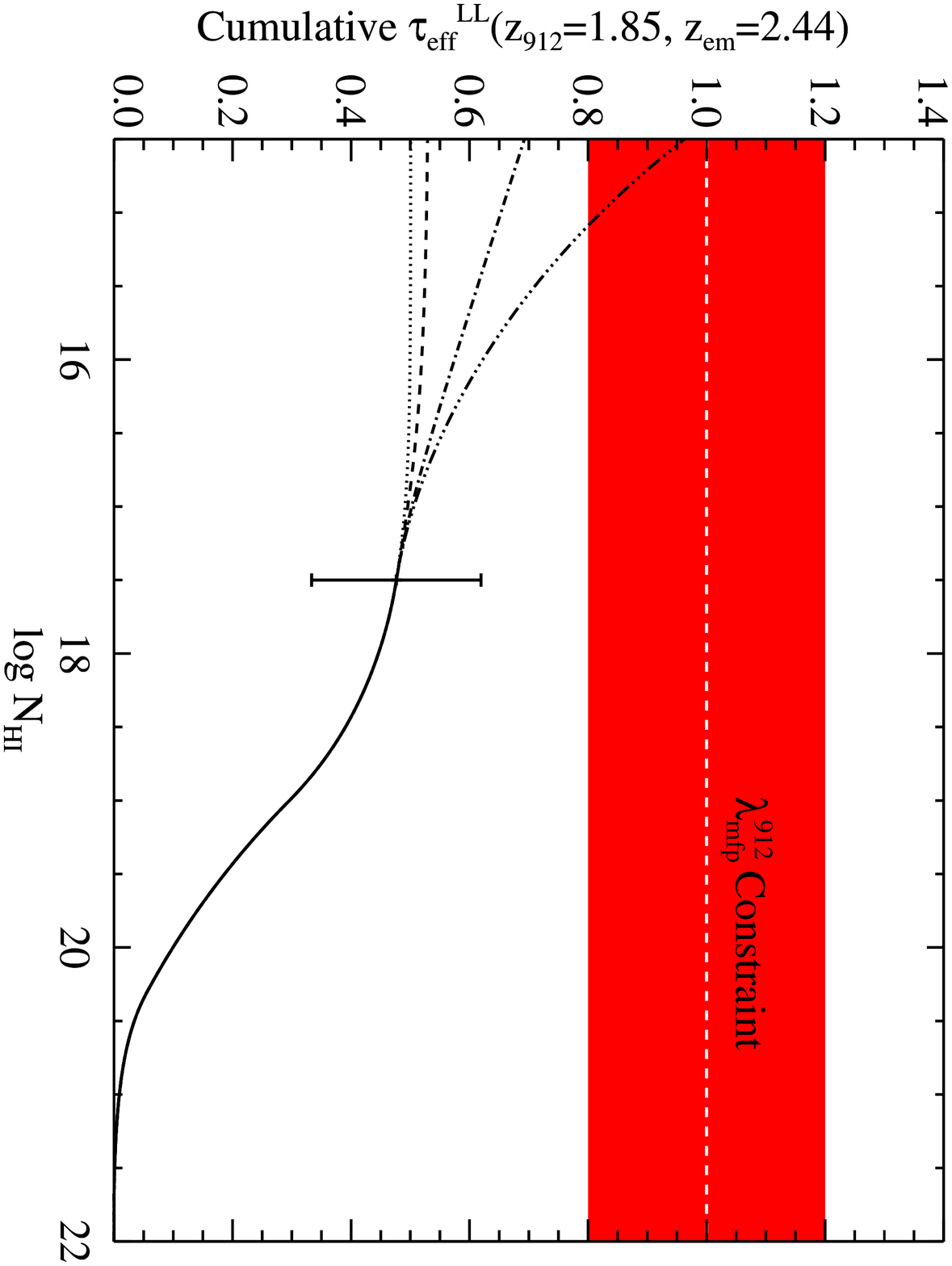}
\caption{The curves trace the cumulative effective opacity for Lyman
  limit absorption \tll\ evaluated from $z_{912} = 1.85$ to $\mzem =
  2.44$ (Equation~\ref{eqn:teff}).   Each curve assumes the identical
  \fnhi\ distribution for $\mnhi \ge 10^{17.5} \cm{-2}$ as described
  by Figure~\ref{fig:fn_high}.
  For $\mnhi < 10^{17.5} \cm{-2}$, the curves adopt a fixed power-law
  with $\beta = -1.0$ (dotted), $-1.5$ (dashed), $-2.0$ (dash-dot) and
  $-2.5$ (dash-dots).  The red shaded region shows our best estimate
  for \tll\ over this redshift interval, as assessed from our \lmfp\
  analysis ($\S$~\ref{sec:mfp}).  This constraint strongly prefers a
  steep \nhi\ frequency distribution at these intermediate column
  densities, consistent with our analysis of the pLLS
  (Figure~\ref{fig:fn_pLLS}).
}
\label{fig:cumul_tau}
\end{figure}

\begin{figure}
\includegraphics[width=5in]{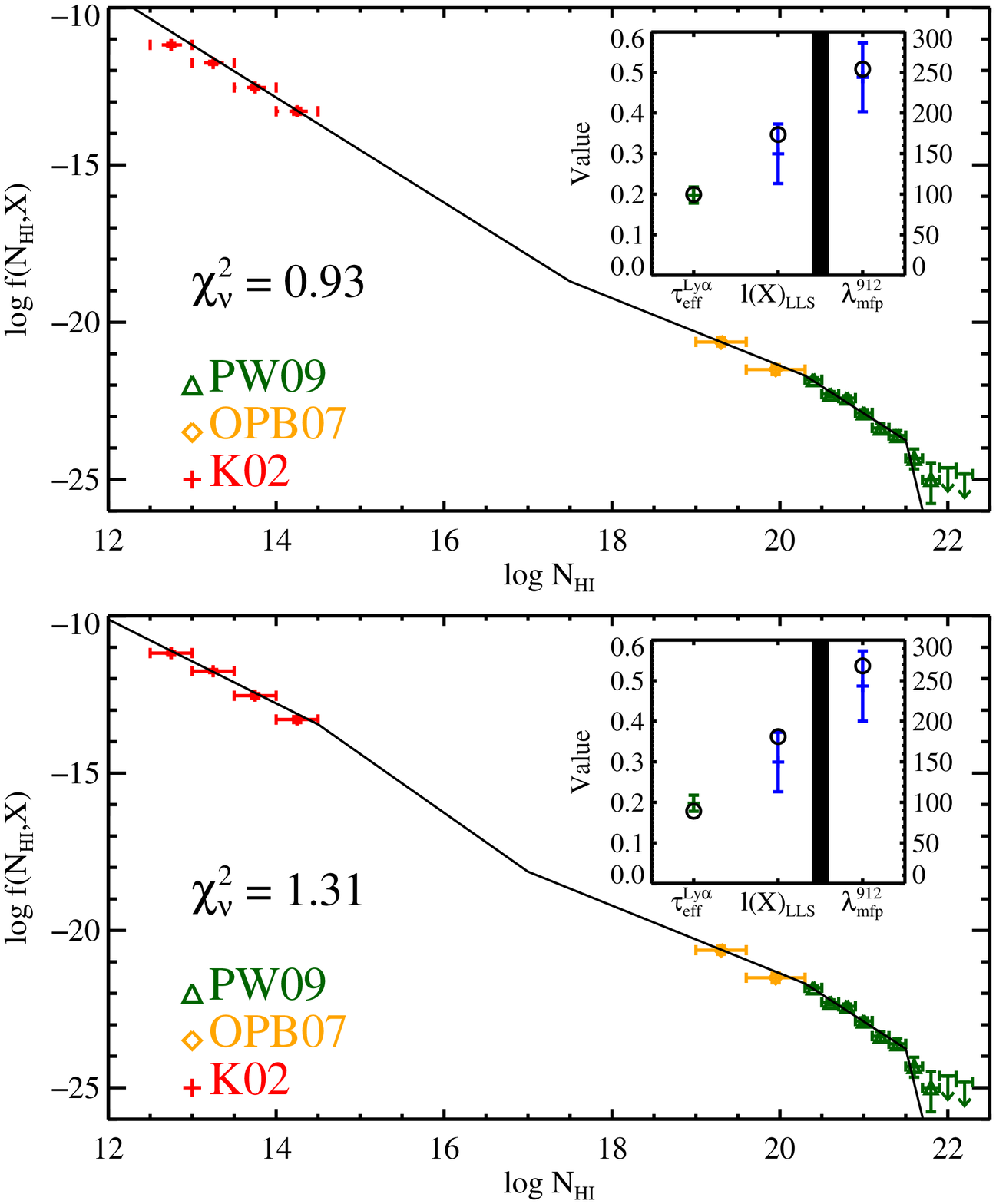}
\caption{The \nhi\ frequency distribution function at $z = 2.4$, as
  modeled by a series of broken power-laws.  The black curve shows the
  best-fit model, with parameters listed in
  Table~\ref{tab:corrmat}.  This model was derived by minimizing
  the reduced $\chi^2$ calculated against the observational
  constraints shown in the Figure (see also
  Table~\ref{tab:fn_constraints}).  This includes estimates of \fnhi\
  for the \lya\ forest \citep[K01;][]{kim+01}, the SLLS
  \citep[OPB07][]{opb+07}, and the DLAs \citep[PW09][]{pw09}.
  Furthermore, we adopted a model constraint on the power-law index in
  the \lya\ forest \citep{kim02} and integral constraints on the
  opacity of the \lya\ forest \citep{kts+05}.  Lastly, we also adopted
  results on \lmfp\ and \lox\ estimated in this paper.  We find that
  this simple (and assuredly non-physical) model for \fnhi\ provides a
  good description of all these observations.
}
\label{fig:bothfn_z25}
\end{figure}

%\begin{figure}
%\includegraphics[width=5in]{f17.eps}
%%\includegraphics[width=5in,angle=90]{Figures/fig_fn_z25.ps}
%\caption{The \nhi\ frequency distribution function at $z = 2.4$, as
%  modeled by a series of broken power-laws.  The black curve shows the
%  best-fit model, with parameters listed in
%  Table~\ref{tab:corrmat}.  This model was derived by minimizing
%  the reduced $\chi^2$ calculated against the observational
%  constraints shown in the Figure (see also
%  Table~\ref{tab:fn_constraints}).  This includes estimates of \fnhi\
%  for the \lya\ forest \citep[K01;][]{kim+01}, the SLLS
%  \citep[OPB07][]{opb+07}, and the DLAs \citep[PW09][]{pw09}.
%  Furthermore, we adopted a model constraint on the power-law index in
%  the \lya\ forest \citep{kim02} and integral constraints on the
%  opacity of the \lya\ forest \citep{kts+05}.  Lastly, we also adopted
%  results on \lmfp\ and \lox\ estimated in this paper.  We find that
%  this simple (and assuredly non-physical) model for \fnhi\ provides a
%  good description of all these observations.
%}
%\label{fig:fn_z25}
%\end{figure}

\begin{figure}
\includegraphics[width=5in]{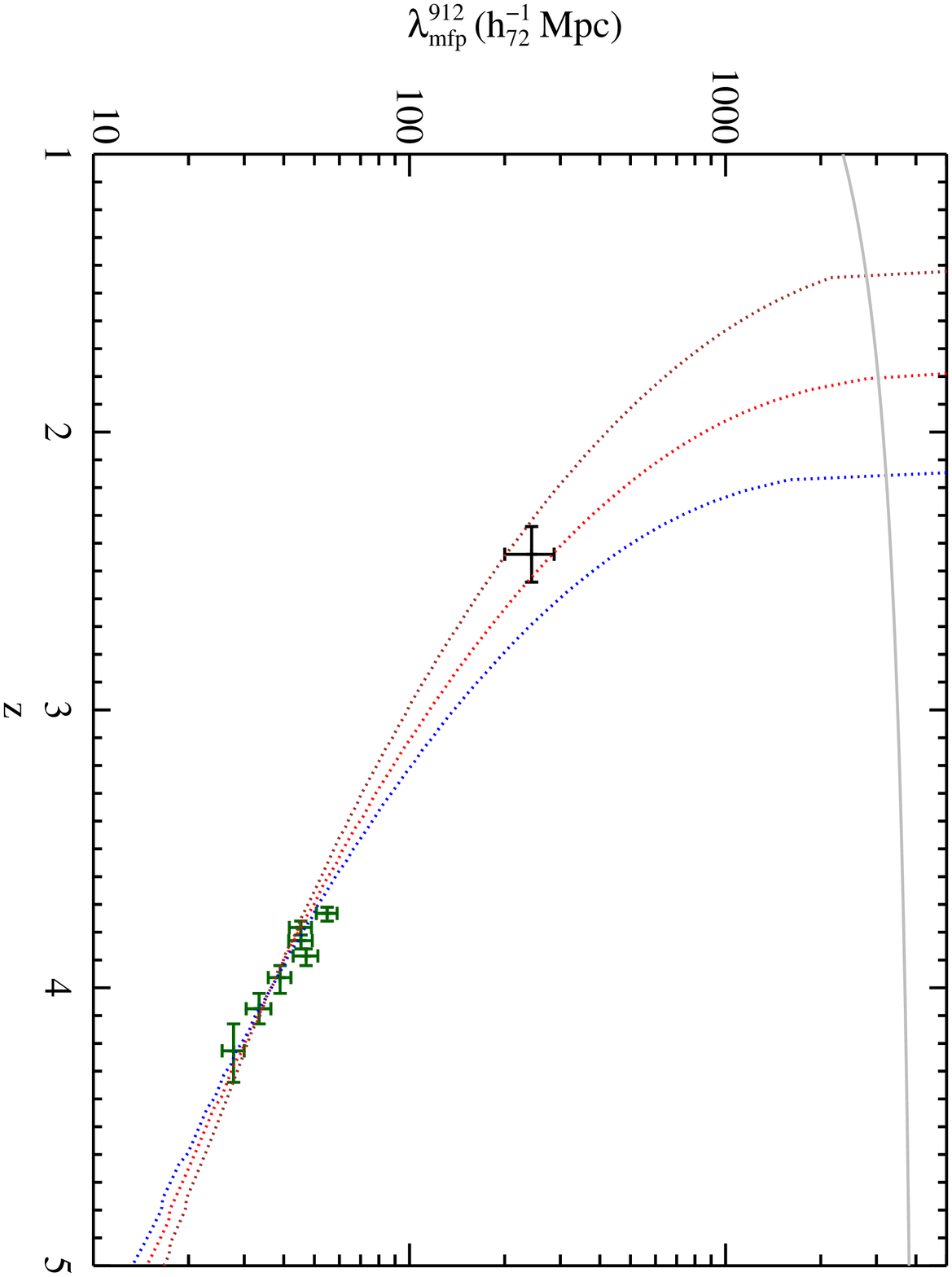}
\caption{The data points trace the \lmfp\ measurements as a function
  of redshift from this paper ($z \approx 2.44$; black) and the SDSS
  analysis of PWO09 ($z \sim 4$; green).  The colored curves
  show the predicted evolution in \lmfp\
  for a series of assumed \fnhi\ models.  Each has the functional form
  dervied for the IGM at $z=2.4$ (the 5 parameter model given in
  Table~\ref{tab:corrmat}).  We have allowed, however, the incidence
  of absorbers \loz\ to evolve
  as $(1+z)^\gamma$ with $\gamma = 1.5, 2.0, 2.5$
  for line colors brown, red, and blue
  respectively.   Furthermore, each 
  of the curves was forced to intersect the SDSS measurements at
  $z=3.9$.  Our \hwfc\ measurement rules out $\gamma = 1.5$ and favors
  $\gamma \approx 2$ which coincides well with the redshift evolution
  in \loz\ reported by \cite{ribaudo11}.  The gray curve in the figure
  traces the Horizon of the universe with redshift.  An extrapolation
  of the \lmfp\ curves predicts that the `breakthrough' epoch occurs
  at $z \approx 1.5-2$.
}
\label{fig:mfp_vs_z}
\end{figure}

\begin{figure}
\includegraphics[width=5in]{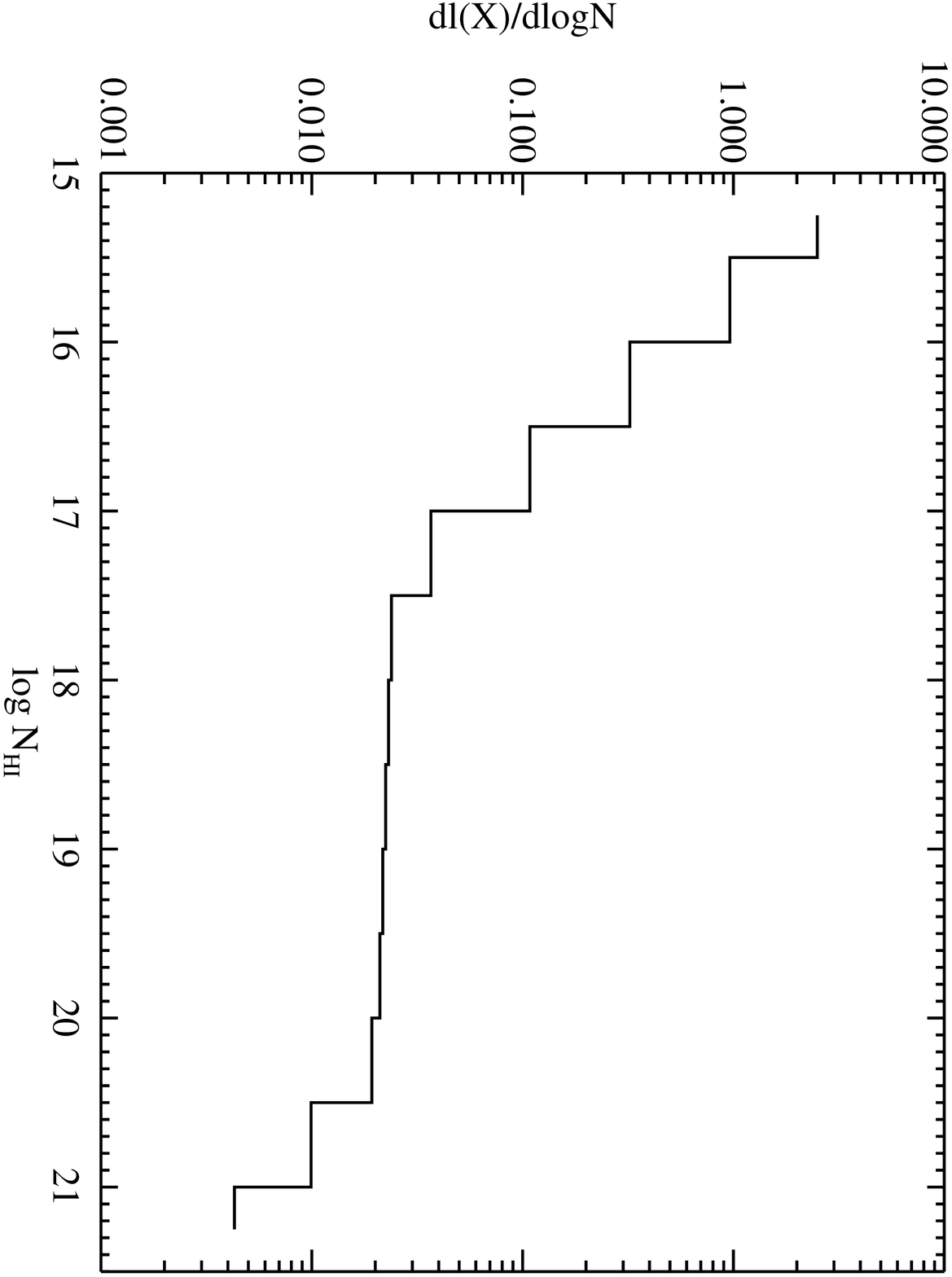}
\caption{Differential contribution of absorption systems to \lox as a
  function of \nhi,$d\ell(X)/d\ln\mnhi$. Abrupt changes in this
  quantity mirror inflections in \fnhi .  The transitions away from the
near constancy of this quantity across the LLS regime are likely
linked to ionization properties of the gas, either with the onset of
the gas being optically thick (\nhi $= 10^{17.2}$\cmm), or the onset
of self-shielding (\nhi $= 10^{20.5}$\cmm). 
}
\label{fig:dldN}
\end{figure}

\begin{figure}
\includegraphics[width=5in]{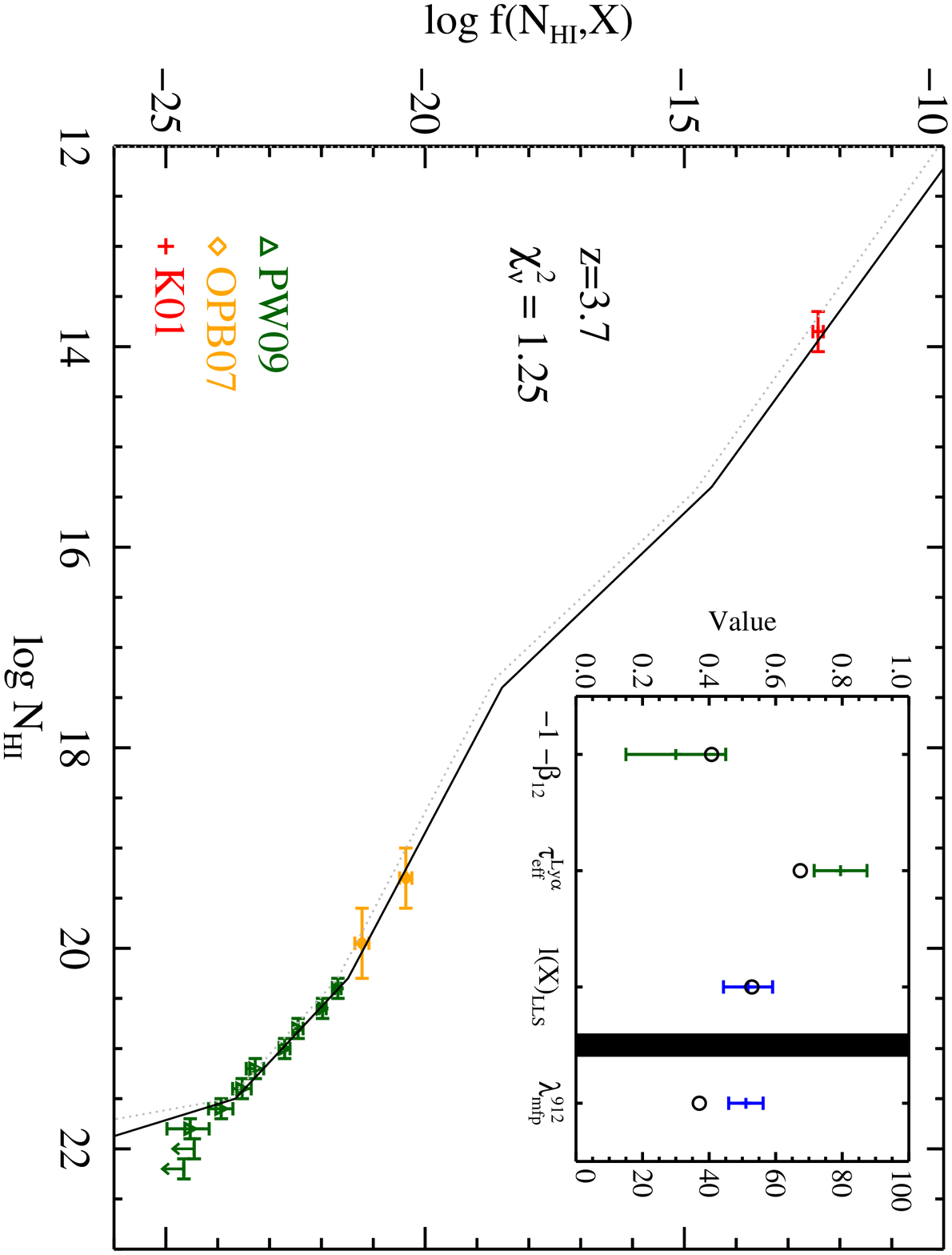}
\caption{Similar to Figure~\ref{fig:bothfn_z25} but for $z=3.7$ and,
  therefore, for a different set of observational constraints
  (Table~\ref{tab:fn_constraints}).  For \tlya, we adopted the
  estimate of \cite{fpl+08} and the \lox\ and \lmfp\ measurements from
  POW10 and PWO09 respectively.  The dotted gray line
  shows the best-fit model of \fnhi\ at $z=2.4$.  It has nearly
  identical shape but is offset to lower normalization by $\approx
  0.3$\,dex.  
}
\label{fig:fn_z37}
\end{figure}

\end{document}